\documentclass[12pt]{article}
\usepackage[utf8]{inputenc}
\usepackage{extsizes}
\usepackage{amsmath}
\usepackage[left=0.65in,right=0.65in,top = 0.7in, bottom = 0.7in]{geometry}
\usepackage{graphicx}
\usepackage[toc,page]{appendix}
\usepackage[inline]{enumitem}
\usepackage{multicol}
\usepackage[T1]{fontenc}
\usepackage{multirow}
\usepackage{amssymb}
\usepackage{setspace}
\usepackage{titling}
\usepackage{bigints}
\usepackage{bibentry}
\usepackage{import}
\usepackage{subcaption}
\usepackage{upgreek}
\usepackage{longtable}
\usepackage{bm}
\usepackage{bbm}
\usepackage{algorithm}
\usepackage{siunitx}
\usepackage{algpseudocode}
\usepackage{mathtools, nccmath}
\usepackage{setspace}
\usepackage[mathlines]{lineno}
\usepackage[most]{tcolorbox}
\usepackage{tabularray}
\usepackage{esint}
\usepackage[unicode=true]{hyperref}
\usepackage[nameinlink]{cleveref}
\usepackage{nameref}
\usepackage{booktabs}
\usepackage{collcell}
\usepackage{listings}
\usepackage{pgf,tikz,pgfplots}
\usepackage{asymptote}
\usepackage{fancyhdr}
\usepackage{lscape}
\usepackage{authblk}
\usepackage{pdfpages}
\usepackage{soul}
\pgfplotsset{compat=1.15}
\usepackage{amsthm}
\usepackage{mathrsfs}
\numberwithin{equation}{section}
\usepackage{float}
\usepackage{longfigure}
\usepackage{adjustbox}
\usepackage[english]{babel}
\usetikzlibrary{arrows}
\usetikzlibrary{shapes}
\newcommand*{\currentname}{\@currentlabelname}
\makeatother

\usepackage{mathtools}
\usepackage{upquote}
\usepackage{amsthm}
\usepackage{chngcntr}
\usepackage{apptools}
\usepackage{etoolbox}
\AtAppendix{\counterwithin{theorem}{section}}

\newcommand\mydots{\ifmmode\ldots\else\makebox[1em][c]{.\hfil.\hfil.}\thinspace\fi}

\usepackage{scalerel,stackengine}
\stackMath
\newcommand\reallywidehat[1]{%
\savestack{\tmpbox}{\stretchto{%
  \scaleto{%
    \scalerel*[\widthof{\ensuremath{#1}}]{\kern-.6pt\bigwedge\kern-.6pt}%
    {\rule[-\textheight/2]{1ex}{\textheight}}
  }{\textheight}%
}{0.5ex}}%
\stackon[1pt]{#1}{\tmpbox}%
}
\definecolor{qqwuqq}{rgb}{0.,0.39215686274509803,0.}
\definecolor{ududff}{rgb}{0.30196078431372547,0.30196078431372547,1.}
\newcommand{\splitfigurealt}[2]{%
  \ifnum#1=6
    \adjincludegraphics[trim={0cm {0.8333333333333334\height} 0cm {0.0\height}},clip,width=0.95\textwidth]{#2}\\
    \adjincludegraphics[trim={0cm {0.6666666666666666\height} 0cm {0.16666666666666666\height}},clip,width=0.95\textwidth]{#2}\\
    \adjincludegraphics[trim={0cm {0.5\height} 0cm {0.3333333333333333\height}},clip,width=0.95\textwidth]{#2}\\
    \adjincludegraphics[trim={0cm {0.33\height} 0cm {0.5\height}},clip,width=0.95\textwidth]{#2}\\
    \adjincludegraphics[trim={0cm {0.16\height} 0cm {0.68\height}},clip,width=0.95\textwidth]{#2}\\
    \adjincludegraphics[trim={0cm {0.0\height} 0cm {0.85\height}},clip,width=0.95\textwidth]{#2}\\
  \fi
}
\newcommand{\splitfigure}[2]{%
  \ifnum#1=3
    \adjincludegraphics[trim={0cm {0.6666666666666666\height} 0cm {0.0\height}},clip,width=0.95\textwidth]{#2}\\
    \adjincludegraphics[trim={0cm {0.3333333333333333\height} 0cm {0.3333333333333333\height}},clip,width=0.95\textwidth]{#2}\\
    \adjincludegraphics[trim={0cm {0.0\height} 0cm {0.6666666666666666\height}},clip,width=0.95\textwidth]{#2}\\
  \fi
  \ifnum#1=4
    \adjincludegraphics[trim={0cm {0.75\height} 0cm {0.0\height}},clip,width=0.95\textwidth]{#2}\\
    \adjincludegraphics[trim={0cm {0.5\height} 0cm {0.25\height}},clip,width=0.95\textwidth]{#2}\\
    \adjincludegraphics[trim={0cm {0.25\height} 0cm {0.5\height}},clip,width=0.95\textwidth]{#2}\\
    \adjincludegraphics[trim={0cm {0.0\height} 0cm {0.75\height}},clip,width=0.95\textwidth]{#2}\\
  \fi
  \ifnum#1=5
    \adjincludegraphics[trim={0cm {0.8\height} 0cm {0.0\height}},clip,width=0.95\textwidth]{#2}\\
    \adjincludegraphics[trim={0cm {0.6\height} 0cm {0.2\height}},clip,width=0.95\textwidth]{#2}\\
    \adjincludegraphics[trim={0cm {0.4\height} 0cm {0.4\height}},clip,width=0.95\textwidth]{#2}\\
    \adjincludegraphics[trim={0cm {0.2\height} 0cm {0.6\height}},clip,width=0.95\textwidth]{#2}\\
    \adjincludegraphics[trim={0cm {0.0\height} 0cm {0.8\height}},clip,width=0.95\textwidth]{#2}\\
  \fi
  \ifnum#1=6
    \adjincludegraphics[trim={0cm {0.8333333333333334\height} 0cm {0.0\height}},clip,width=0.95\textwidth]{#2}\\
    \adjincludegraphics[trim={0cm {0.6666666666666666\height} 0cm {0.16666666666666666\height}},clip,width=0.95\textwidth]{#2}\\
    \adjincludegraphics[trim={0cm {0.5\height} 0cm {0.3333333333333333\height}},clip,width=0.95\textwidth]{#2}\\
    \adjincludegraphics[trim={0cm {0.3333333333333333\height} 0cm {0.5\height}},clip,width=0.95\textwidth]{#2}\\
    \adjincludegraphics[trim={0cm {0.16666666666666666\height} 0cm {0.6666666666666666\height}},clip,width=0.95\textwidth]{#2}\\
    \adjincludegraphics[trim={0cm {0.0\height} 0cm {0.8333333333333334\height}},clip,width=0.95\textwidth]{#2}\\
  \fi
}
\newcommand{\splitatcommas}[1]{%
  \begingroup
  \begingroup\lccode`~=`, \lowercase{\endgroup
    \edef~{\mathchar\the\mathcode`, \penalty0 \noexpand\hspace{0pt plus 1em}}%
  }\mathcode`,="8000 #1%
  \endgroup
}
\newcommand{\ignore}[1]{}
\newcommand{\nobibentry}[1]{{\let\nocite\ignore\bibentry{#1}}}

\newtcbtheorem[auto counter]{theo}%
  {Theorem}{fonttitle=\bfseries\upshape,
     arc=0mm, colback=black!5!white,colframe=black!75!black}{theorem}
\hypersetup{
    colorlinks=true,
    linkcolor=blue, 
    citecolor=blue,        
    filecolor=magenta,      
    urlcolor=cyan           
    }

\date{}
\makeatother
\title{Optimisation of neoadjuvant pembrolizumab therapy for locally advanced MSI-H/dMMR colorectal cancer using data-driven delay integro-differential equations}
\author[1, *]{Georgio Hawi}
\author[1, $\dagger$]{Peter S. Kim}
\author[2, $\dagger$]{Peter P. Lee}
\affil[1]{School of Mathematics and Statistics, University of Sydney, Sydney, Australia}
\affil[2]{Department of Immuno-Oncology, Beckman Research Institute, City of Hope, Duarte, California, USA}
\affil[*]{Corresponding author: \href{mailto:georgio.hawi@sydney.edu.au}{georgio.hawi@sydney.edu.au}}
\affil[$\dagger$]{These authors contributed comparably to this work}
\usepackage{float}
\floatstyle{plaintop}
\restylefloat{table}

\theoremstyle{definition}

\newcommand{\set}[1]{\left \{ #1 \right \}}

\allowdisplaybreaks
\usepackage{longtable}
\usepackage[numbers,square,sort&compress]{natbib}
\usepackage{bibunits}
\usepackage{hypernat}
\makeatletter
\pretocmd{\@startbibunit}{%
  \begingroup
    \count@\@bibunitauxcnt
    \advance\count@\@ne
    \xdef\@extra@binfo{.bu\the\count@}%
    \xdef\@extra@b@citeb{.bu\the\count@}%
  \endgroup
}{}{}
\apptocmd{\endbibunit}{%
  \gdef\@extra@binfo{}%
  \gdef\@extra@b@citeb{}%
}{}{}
\makeatother
\begin{document}
\maketitle
\begin{bibunit}[vancouver]
\vspace{-20mm}
\begin{abstract}
Colorectal cancer (CRC) poses a major public health challenge due to its increasing prevalence, particularly among younger populations. Microsatellite instability-high (MSI-H) CRC and deficient mismatch repair (dMMR) CRC constitute 15\% of all CRC and exhibit remarkable responsiveness to immunotherapy, especially with PD-1 inhibitors. Despite this, there is a significant need to optimise immunotherapeutic regimens to maximise clinical efficacy and patient quality of life. To address this, we employ a novel framework driven by delay integro-differential equations to model the interactions among cancer cells, immune cells, and immune checkpoints in locally advanced MSI-H/dMMR CRC (laMCRC). Several of these components are being modelled deterministically for the first time in cancer, paving the way for a deeper understanding of the complex underlying immune dynamics. We consider two compartments---the tumour site and the tumour-draining lymph node (TDLN)---taking into account phenomena such as DC migration, T cell proliferation, and CD8+ T cell exhaustion and reinvigoration. Parameter values and initial conditions are derived from experimental data, integrating various pharmacokinetic, bioanalytical, and radiographic studies, along with deconvolution of bulk RNA-sequencing data from the TCGA COADREAD and GSE26571 datasets. We finally optimised neoadjuvant treatment with pembrolizumab, a widely used PD-1 inhibitor, to balance efficacy, efficiency, and toxicity in laMCRC patients. We mechanistically analysed factors influencing treatment success and improved upon currently FDA-approved therapeutic regimens for metastatic MSI-H/dMMR CRC, demonstrating that a single medium-to-high dose of pembrolizumab may be sufficient for effective tumour eradication while being efficient, safe, and practical.
\end{abstract}
\section{Introduction}
Colorectal cancer (CRC) is the third most common cancer worldwide, accounting for approximately $10\%$ of all cancer cases \citep{Klimeck2023}, with more than 1.85 million cases and 850,000 deaths annually \citep{Biller2021}. The American Cancer Society estimates that in the United States, there will be 152,810 new cases of CRC diagnosed and 53,010 deaths due to CRC \citep{Siegel2024}, with individual risk factors including a family history of CRC, inflammatory bowel disease, and type 2 diabetes \citep{Gausman2020}. Despite CRC being diagnosed mostly in adults aged 65 and older, there has been an increase in the incidence rate of CRC amongst younger populations \citep{SifakiPistolla2022, Siegel2023, Siegel2024} since the mid-1990s, with CRC being the leading cause of cancer-related deaths in adults under 55 \citep{Siegel2024}. In particular, since many people will not experience symptoms in the early stages of CRC, diagnoses often occur at a later stage when the disease is more advanced, at which point treatment is significantly less effective and survival is much worse \citep{Andrew2018}. Of new CRC diagnoses, 20\% of patients present with metastatic disease, while an additional 25\% who initially have localised disease eventually develop metastases \citep{Biller2021}. In the United States, the 5-year survival rates for stage IIIA, stage IIIB, and stage IIIC colon cancer are 90\%, 72\%, and 53\%, respectively, whilst stage IV CRC has a 5-year survival of only 12\% \citep{Rawla2019}. \\~\\
Whilst many systemic therapies are available for advanced CRC, chemotherapy has been the main treatment approach, with fluoropyrimidine 5-fluorouracil being the only Food and Drug Administration (FDA) agent approved for metastatic CRC treatment for nearly 40 years \citep{Atreya2017}. Noting that folinic acid (leucovorin), a vitamin B derivative, increases the cytotoxicity of 5-fluorouracil \citep{moran1989leucovorin}, and with the approval of the topoisomerase I inhibitor irinotecan in 1996 and the platinum-based agent oxaliplatin, mainstay chemotherapy regimens such as FOLFOX (folinic acid, 5-fluorouracil, oxaliplatin) and FOLFIRI (folinic acid, 5-fluorouracil, irinotecan) have become integral to the treatment of advanced CRC \citep{MohelnikovaDuchonova2014}. However, the response rate of advanced CRC patients with 5-fluorouracil monotherapy remains at only 10--15\%, with the addition of other anti-cancer drugs increasing response rates to only 40--50\% \citep{Gu2019}.\\~\\
Moreover, patients with the hypermutant microsatellite instability-high (MSI-H) phenotype who have reached metastasis are less responsive to conventional chemotherapy and have a poorer prognosis compared to patients with microsatellite stable (MSS) CRC \citep{Shulman2018}. MSI-H CRC is associated with the inactivation of mismatch repair (MMR) genes, including \textit{MLH1}, \textit{MSH2}, \textit{MSH6}, and \textit{PMS2}, leading to deficient MMR (dMMR) and impaired recognition and correction of spontaneous mutations by cells \citep{MuletMargalef2023}. In particular, we note that in CRC, MSI-H and dMMR tumours are equivalent \citep{Boland1998-nd}, and we denote these tumours as MSI-H/dMMR for the remainder of this work. Approximately 20\% of stage II, 12\% of stage III, and 4\% of stage IV CRC tumours are diagnosed as MSI-H/dMMR \citep{Sinicrope2011, Venderbosch2014, Fan2024}, with approximately 80\% of sporadic MSI-H/dMMR CRC caused by MLH1 promoter hypermethylation \citep{Andr2020}. This leads to a highly increased mutational rate, with MSI-H/dMMR CRC tumours having 10--100 times more somatic mutations compared to microsatellite stable (MSS) CRC tumours \citep{MuletMargalef2023}, resulting in increased tumour mutation burden (TMB) and neoantigen load, and an immunogenic tumour microenvironment (TME) with dense immune cell infiltration \citep{Llosa2015, Giannakis2016}. This immunogenicity results in patients with MSI-H/dMMR CRC having a good prognosis for immunotherapy treatment, in particular to immune checkpoint inhibitors (ICIs) \citep{Ciardiello2019}.\\~\\
Immune checkpoints, such as programmed cell death protein 1 (PD-1), cytotoxic T-lymphocyte-associated antigen 4 (CTLA-4), and lymphocyte-activation gene 3 (LAG-3), normally downregulate immune responses after antigen activation \citep{Topalian2012}. CTLA-4 is expressed on activated T and B cells and plays a major role in downmodulating the initial stages of T cell activation and proliferation \citep{Buchbinder2016}. PD-1, a cell membrane receptor that is expressed on a variety of cell types, including activated T cells, activated B cells and monocytes, has been extensively researched in the context of cancer such as MSI-H/dMMR CRC \citep{Sarshekeh2018, Yaghoubi2019}. When PD-1 interacts with its ligands (PD-L1 and PD-L2), effector T cell activity is inhibited, resulting in the downregulation of pro-inflammatory cytokine secretion and the upregulation of immunosuppressive regulatory T cells (Tregs) \citep{Lin2024, han2020pd}. Cancers can exploit this by expressing PD-L1 themselves, evading immunosurveillance, and impairing the proliferation and activity of cytotoxic T lymphocytes (CTLs) \citep{Oliveira2019}. Blockade of PD-1/PD-L1 complex formation reinvigorates effector T cell activity, resulting in enhanced anti-tumour immunity and responses, leading to improved clinical outcomes in cancer patients \citep{Lee2015, Zhang2020nature}. \\~\\
The KEYNOTE-177 phase III trial, NCT02563002, aimed to evaluate the efficacy of first-line pembrolizumab, an anti-PD-1 antibody, in metastatic MSI-H/dMMR CRC (mMCRC) \citep{Andr2020}. In the trial, 307 treatment-naive mMCRC patients were randomly assigned to receive pembrolizumab at a dose of 200 mg every 3 weeks or 5-fluorouracil-based chemotherapy every 2 weeks. A partial or complete response was observed in 43.8\% of patients allocated to pembrolizumab therapy, compared with 33.1\% of patients participating in 5-fluorouracil-based therapy. Furthermore, among patients who responded, 83\% in the pembrolizumab group maintained a response at 24 months, compared with 35\% of patients receiving chemotherapy. These results motivated the FDA to approve pembrolizumab for the first-line treatment of unresectable or metastatic MSI-H/dMMR CRC on June 29, 2020 \citep{Casak2021}. \\~\\
In the past couple of years, there has been a surge in research into the efficacy of neoadjuvant pembrolizumab in the treatment of high-risk stage II and stage III MSI-H/dMMR CRC \citep{Zhang2022}. One such phase II study is NEOPRISM-CRC (NCT05197322), in which 31 patients with a high TMB and high-risk stage II or stage III MSI-H/dMMR CRC were given three cycles of pembrolizumab, at a dose of 200 mg every 3 weeks via IV infusion, and underwent surgery 4--6 weeks after the last dose was administered \citep{Shiu2024}. Seventeen patients exhibited pathologic complete responses (pCRs) (55\%, 95\% CI 36\%--73\%), with the remaining patients having their tumours removed after surgery. After a median follow-up time of 6 months, recurrence was found in no patients, and the median cancer-free period was 9.7 months. Another phase II trial, NCT04082572, aimed to evaluate the efficacy of neoadjuvant pembrolizumab on localised MSI-H/dMMR solid tumours \citep{Ludford2023}. As part of this, 27 patients with locally advanced MSI-H/dMMR CRC (laMCRC) were either given 200 mg pembrolizumab via IV infusion every 3 weeks for eight treatments, followed by surgical resection or 200 mg pembrolizumab via IV infusion every 3 weeks for 16 treatments. Overall, 21 patients exhibited pCRs, and among the 14 patients in the resection group, 11 exhibited pCRs. Additionally, after a median follow-up time of 9.5 months, only two patients who underwent surgical resection experienced recurrence or progression. \\~\\
In the IMHOTEP Phase II trial, patients with localised, resectable MSI-H/dMMR CRC received neoadjuvant pembrolizumab at a dose of 400 mg per cycle every 6 weeks, with one (67.1\%) or two (32.9\%) cycles administered \citep{delaFouchardiere2024}. Surgery was performed after the last dose, with a pCR observed in 53.8\% of patients, including 47.1\% of those who received one cycle and 68.0\% of those who received two cycles. However, in the RESET-C study, 84 patients with resectable stage I-III dMMR colon cancer received a single neoadjuvant cycle of pembrolizumab at 4 mg/kg (up to a maximum of 400 mg), followed by surgical resection within three to five weeks and CT scans at one and three years for follow-up \citep{Qvortrup2025}. This regimen resulted in a major pathological response in 57\% of patients and a pCR rate of 44\%, including 33\% for stage III cancer, suggesting that a single cycle of neoadjuvant pembrolizumab may be sufficient to achieve pCR in some laMCRC patients, particularly those with earlier-stage disease. \\~\\
An important question to consider is the appropriate dosing and spacing of ICI therapies to balance tumour reduction with factors such as monetary cost, toxicity, and side effects \citep{Centanni2019, LeLouedec2020}. A retrospective study by Dub\'e-Pelletier et al.\ of 80 patients with advanced non-small cell lung cancer (NSCLC) who received 4 mg/kg pembrolizumab every 6 weeks and 80 NSCLC patients who received 2 mg/kg pembrolizumab every 3 weeks, revealed that both therapies were comparable in terms of OS, toxicity, and progression-free survival \citep{DubPelletier2023}, despite the less frequent therapy being more cost-effective. Various pharmacokinetic models have been developed to optimise ICI therapy \citep{Lindauer2016, Shang2022, Yan2023, Wang2020}, with \citep{Puszkiel2024} showing that tripling the dosing interval of nivolumab, another anti-PD-1 antibody, from 240 mg every 2 weeks to 240 mg every 6 weeks leads to comparable efficacy despite financial costs decreasing threefold. Mathematical models provide a powerful framework for optimising treatment regimens, and in this work, we construct a comprehensive data-driven model of the immunobiology of MSI-H/dMMR CRC using delay integro-differential equations and use this to evaluate and optimise neoadjuvant pembrolizumab therapy in laMCRC.\\~\\
To date, there are no existing mathematical models in the literature for ICI therapy in laMCRC; however, there are numerous models of CRC. Kirshtein et al.\ developed an ordinary differential equation (ODE) model of CRC progression, incorporating immunological components such as T helper cells, Tregs, dendritic cells (DCs), and macrophages, and they considered the effects of carcinogenic cytokines and immunosuppressive agents \citep{Kirshtein2020}. They used data from the TCGA COADREAD database \citep{COADREAD2012} to perform estimates for the steady states and initial conditions of model variables and considered data from all patients, regardless of the TNM stage. Moreover, this model was extended in \citep{Budithi2021} to include FOLFIRI treatment. Bozkurt et al.\ presented a relatively simple ODE model of CRC treatment with anthracycline doxorubicin and IL-2 immunotherapy, modelling cancer cells, natural killer (NK) cells, CD8+ T cells, and other lymphocytes \citep{Bozkurt2023}. De Pillis et al.\ developed an ODE model of CRC with irinotecan and monoclonal antibody therapies, in particular cetuximab and panitumumab, modelling similar quantities \citep{dePillis2014}.\\~\\
ICI therapy has been modelled extensively in other cancers, and Butner et al.\ provide a comprehensive review of the merits and weaknesses of various modelling approaches, including partial differential equations (PDEs), ODEs, agent-based modelling (ABM), and hybrid modelling in \citep{Butner2022}. We now summarise a few pre-existing differential-equation-based models of PD-1 blockade therapies. Lai et al.\ modelled the effects of anti-PD-1 and vaccines on cancer, taking into account DC maturation by high mobility group box 1 (HMGB1) and interleukin-2 (IL-2) and interleukin-12 (IL-12) in \citep{Lai2017}. This model was adapted in \citep{Lai2019} to optimise combination PD-1 and vascular endothelial growth factor (VEGF) inhibitor therapies in cancer. Siewe et al.\ modelled how transforming growth factor-beta (TGF-$\upbeta$) can be used to overcome resistance to PD-1 blockade and also incorporated macrophages, Tregs, IL-2, IL-12, TGF-$\upbeta$, interleukin-10 (IL-10), and chemokine ligand 2 (CCL-2) \citep{Siewe2021}. This model was extended in \citep{Siewe2023} to model cancer therapy with PD-1 inhibitors with CSF-1 blockade, also including the cytokine tumour necrosis factor (TNF). Additionally, Liao et al.\ constructed a mathematical model that demonstrated the pro-cancer or anti-cancer nature of interleukin-27 (IL-27) in combination with anti-PD-1, incorporating the following cytokines: IL-27, TGF-$\upbeta$, IL-2, interferon-gamma (IFN-$\upgamma$), and IL-10 \citep{Liao2024}. Quantitative systems pharmacology (QSP) models have also been used to model clinical responses to ICI therapy, with a QSP model by Milberg et al.\ simulating the dynamics of immune cell interactions and PD-1 binding to its ligands, PD-L1 and PD-L2, across multiple physiological compartments, including the spleen, tumour microenvironment, and lymph nodes \citep{Milberg2019}. \\~\\
There are, however, a multitude of limitations and drawbacks to these pre-existing models of CRC and ICI therapy. One of the biggest issues is that mature DC migration to the tumour-draining lymph node (TDLN) to activate naive T cells is not addressed, with T cell proliferation and migration to the tumour site (TS) also not being addressed. Kumbhari et al.\ attempted to address this in \citep{Kumbhari2020} and \citep{Kumbhari2020n2} in the context of optimising cancer vaccine therapy; however, activation is treated as occurring instantaneously, which has been shown to be false experimentally \citep{Plambeck2022}. Moreover, since T cell activation and proliferation take a non-negligible amount of time to occur, immune checkpoint inhibition of these processes must take this into account. No papers to date have properly considered this inhibition deterministically throughout the whole proliferation and activation programs, since examining inhibition at a single moment in time is insufficient to characterise this properly. Additionally, damage-associated molecular patterns (DAMPs), released by necrotic cancer cells, induce DC maturation. To date, DC maturation is considered only by HMGB1 in some models, with other important DAMPs, such as calreticulin, not being taken into account \citep{DelPrete2023}. Furthermore, no deterministic model to date has adequately addressed CD8+ T cell exhaustion due to prolonged antigen exposure \citep{Blank2019} nor their potential reinvigoration through immune checkpoint blockade. Likewise, no immunobiological model has accurately represented the interactions between PD-1 and pembrolizumab. Whilst models such as \citep{Zhang2024model} by Zhang et al.\ consider PD-1 on PD-1-expressing cells separately, immunobiological models universally treat the effect of PD-1 as a mass-action depletion rather than accounting for the biologically more accurate processes of formation, dissociation, and internalisation of the PD-1/pembrolizumab complex. Whilst QSP models take these into account and integrate numerous immune processes and interactions, their large scale and complexity often pose a major drawback by limiting the ability to derive clear mechanistic insights. \\~\\
One thing to note is that pre-existing deterministic models mostly estimate cytokine production parameters via biologically informed assumptions, which can lead to inaccuracies and is a somewhat ad hoc approach. In this work, we construct a mathematical model using data-driven delay integro-differential equations that addresses these drawbacks, incorporating all of the aforementioned processes and species, and use this to optimise neoadjuvant pembrolizumab therapy for laMCRC.\\~\\
It is prudent for us to briefly outline the functions and processes of some immune cells in the TME since their interaction with cancer cells directly or through chemokine/cytokine signalling significantly influences the efficacy of therapeutic regimens \citep{Wu2017tme}. T cell activation occurs in the lymph node through T cell receptor (TCR) recognition of cancer antigen presented by major histocompatibility complex (MHC) class I molecules, in the case of CD8+ T cells, and MHC class II molecules, in the case of CD4+ T cells, expressed on the surfaces of mature DCs \citep{Shah2021}. CTLs recognise cancer cells through TCR detection of peptide major histocompatibility complexes (pMHCs) on cancer cell surfaces via MHC class I \citep{Sugiyarto2023}. CD8+ cells, as well as NK cells, are amongst the most cytotoxic and important cells in cancer cell lysis \citep{Maimela2019}, in addition to secreting pro-inflammatory cytokines such as IL-2, IFN-$\upgamma$, and TNF \citep{Hoekstra2021}. These are also secreted by CD4+ T helper 1 (Th1) cells and are an important part of cell-mediated immunity, allowing for neutrophil chemotaxis and macrophage activation \citep{Caza2015}. Furthermore, we must also consider Tregs, which are vital in immune tissue homeostasis since they are able to suppress the synthesis of pro-inflammatory cytokines and control intestinal inflammatory processes \citep{Furiati2019}. This is done in a variety of ways, including the production of immunomodulatory and immunosuppressive cytokines such as TGF-$\upbeta$, IL-10, and interleukin-35 (IL-35) \citep{Jarnicki2006, Turnis2016}. We note that naive CD4+ T cells can differentiate towards multiple additional phenotypes such as Th2, Th9, Th22, Tfh, and Th17 cells, each involved in the pathogenesis of cancer \citep{Cui2019n2, Hetta2020}.\\~\\
Also of importance in CRC are macrophages, which, like T cells, are able to produce pro-inflammatory and anti-inflammatory cytokines \citep{Kerneur2022}. Naive macrophages, denoted M0 macrophages, can differentiate into two main phenotypes: classically activated M1 macrophages and alternatively activated M2 macrophages. These names were given since M1 macrophages promote Th1 cell responses, and M2 macrophages promote Th2 responses, with Th1-associated cytokines downregulating M2 activity, and vice versa \citep{Mills2012}. M1 macrophages contribute to the inflammatory response by activating endothelial cells, promoting the induction of nitric oxide synthase, and producing large amounts of pro-inflammatory cytokines such as TNF, interleukin-$1\upbeta$ (IL-1$\upbeta$), and IL-12 \citep{Viola2019}. On the other hand, M2 macrophages are responsible for wound healing and the resolution of inflammation by phagocytosing apoptotic cells and releasing anti-inflammatory mediators such as IL-10, interleukin-13 (IL-13), and CC Motif Chemokine Ligand 17 (CCL17) \citep{Han2021}.\\~\\
It is important to note that the M1/M2 macrophage dichotomy is somewhat of a simplification. Macrophages are highly plastic and have been demonstrated to integrate environmental signals to change their phenotype and physiology \citep{Xue2014}. To account for this, in the model, we incorporate macrophage polarisation and repolarisation between their anti-tumour and immunosuppressive phenotypes in response to various cytokines.
\section{Mathematical Model}
\subsection{Model Variables and Assumptions}
The variables and their units in the model are shown in \autoref{modelvars}. For simplicity, we ignore spatial effects in the model, including diffusion, advection, and chemotaxis for all species. We assume the system has two compartments: one at the TS, located in the colon or rectum, and one at the tumour-draining lymph node (TDLN). This is a simplification since locally advanced CRC typically involves multiple tumour-draining lymph nodes \citep{Wang2024}; however, for simplicity, we focus on the sentinel node and refer to it as the TDLN for the purposes of the model. We assume that cytokines in the TS are produced only by effector or activated cells and that DAMPs in the TS are only produced by necrotic cancer cells. We assume that all mature DCs in the TDLN are cancer-antigen-bearing and that all T cells in the TS are primed with cancer antigens. Furthermore, we assume that all activated T cells in the TDLN are activated with cancer antigens and that T cell proliferation/division follows a deterministic program. We ignore CD4+ and CD8+ memory T cells and assume that naive CD4+ T cells differentiate immediately upon activation. We also assume that all Tregs in the TS are natural Tregs (nTregs), ignoring induced Tregs (iTregs). We assume, for simplicity, that activated macrophages polarise into the M1/M2 dichotomy. We also assume that the duration of pembrolizumab infusion is negligible compared to the timescale of the model. Therefore, we treat its infusion as an intravenous bolus so that drug absorption occurs immediately after infusion. \\~\\
We assume that all species, $X_i$, degrade/die at a rate proportional to their concentration, with decay constant $d_{X_i}$. We assume that the rate of activation/polarisation of a species $X_i$ by a species $X_j$ follows the Michaelis-Menten kinetic law $\lambda_{X_i X_j}X_i\frac{X_j}{K_{X_i X_j}+X_j}$, for rate constant $\lambda_{X_i X_j}$, and half-saturation constant $K_{X_i X_j}$. Similarly, we model the rate of inhibition of a species $X_i$ by a species $X_j$ using a term of the form $\lambda_{X_i X_j}\frac{X_i}{1+X_j/K_{X_i X_j}}$ for rate constant $\lambda_{X_i X_j}$, and inhibition constant $K_{X_i X_j}$. Production of $X_i$ by $X_j$ is modelled using mass-action kinetics unless otherwise specified, so that the rate at which $X_i$ is formed is given by $\lambda_{X_iX_j}X_j$ for some positive constant $\lambda_{X_iX_j}$. Finally, we assume that the rate of lysis of $X_i$ by $X_j$ follows mass-action kinetics in the case where $X_j$ is a cell and follows Michaelis-Menten kinetics in the case where $X_j$ is a cytokine.
\begin{table}[H]
\centering
\resizebox{\columnwidth}{!}{%
\begin{tabular}{|lp{7.7cm}|lp{7.7cm}|}
\hline
\textbf{Var} & \textbf{Description} & \textbf{Var} & \textbf{Description} \\ 
\hline
$C$ & Viable cancer cell density & $N_c$ & Necrotic cell density \\
$D_0$ & Immature DC density & $D$ & Mature DC density in the TS \\
$D^\mathrm{LN}$ & Mature DC density in the TDLN & $T_0^8$ & Naive CD8+ T cell density in the TDLN \\
$T_A^8$ & Effector CD8+ T cell density in the TDLN & $T_8$ & Effector CD8+ T cell density in the TS \\
$T_{\mathrm{ex}}$ & Exhausted CD8+ T cell density in the TS & $T_0^4$ & Naive CD4+ T cell density in the TDLN \\
$T_A^1$ & Effector Th1 cell density in the TDLN & $T_1$ & Effector Th1 cell density in the TS \\
$T_0^r$ & Naive Treg density in the TDLN & $T_A^r$ & Effector Treg density in the TDLN \\ 
$T_r$ & Effector Treg density in the TS & $M_0$ & Naive macrophage density \\
$M_1$ & M1 macrophage density & $M_2$ & M2 macrophage density \\
$K_0$ & Resting NK cell density & $K$ & Activated NK cell density\\
\hline
$H$ & HMGB1 concentration & $S$ & Calreticulin concentration \\
$I_2$ & IL-2 concentration & $I_\upgamma$ & IFN-$\upgamma$ concentration \\
$I_\upalpha$ & TNF concentration & $I_\upbeta$ & TGF-$\upbeta$ concentration \\
$I_{10}$ & IL-10 concentration & & \\
\hline
$P_D^{T_8}$ & Unbound PD-1 receptor concentration on effector CD8+ T cells in the TS & $P_D^{T_1}$ & Unbound PD-1 receptor concentration on effector Th1 cells in the TS \\
$P_D^{K}$ & Unbound PD-1 receptor concentration on activated NK cells & $Q_A^{T_8}$ & PD-1/pembrolizumab complex concentration on effector CD8+ T cells in the TS \\
$Q_A^{T_1}$ & PD-1/pembrolizumab complex concentration on effector Th1 cells in the TS & $Q_A^{K}$ & PD-1/pembrolizumab complex concentration on activated NK cells \\
$P_L$ & Unbound PD-L1 concentration in the TS & $Q^{T_8}$ & PD-1/PD-L1 complex concentration on effector CD8+ T cells in the TS \\
$Q^{T_1}$ & PD-1/PD-L1 complex concentration on effector Th1 cells in the TS & $Q^{K}$ & PD-1/PD-L1 complex concentration on activated NK cells \\
$A_{1}$ & Concentration of pembrolizumab in the TS & & \\
\hline 
$P_D^{8\mathrm{LN}}$ & Unbound PD-1 receptor concentration on effector CD8+ T cells in the TDLN & $P_D^{1\mathrm{LN}}$ & Unbound PD-1 receptor concentration on effector Th1 cells in the TDLN \\
$Q_A^{8\mathrm{LN}}$ & PD-1/pembrolizumab complex concentration on effector CD8+ T cells in the TDLN & $Q_A^{1\mathrm{LN}}$ & PD-1/pembrolizumab complex concentration on effector Th1 cells in the TDLN \\
$P_L^\mathrm{LN}$ & Unbound PD-L1 concentration in the TDLN & $Q^{8\mathrm{LN}}$ & PD-1/PD-L1 complex concentration on effector CD8+ T cells in the TDLN \\
$Q^{1\mathrm{LN}}$ & PD-1/PD-L1 complex concentration on effector Th1 cells in the TDLN & $A_{1}^\mathrm{LN}$ & Concentration of pembrolizumab in the TDLN \\
\hline
\end{tabular}%
}
\caption{\label{modelvars}Variables used in the model. Quantities in the top box are in units of $\mathrm{cell/{cm}^3}$, quantities in the second box are in units of $\mathrm{g/{cm}^3}$, and all other quantities are in units of $\mathrm{molec/{cm}^3}$. All quantities pertain to the tumour site unless otherwise specified. TDLN denotes the tumour-draining lymph node, whilst TS denotes the tumour site.}
\end{table}
\subsection{Model Summary}
We now outline some of the main processes accounted for in the model, with all processes and equations being explained in \Cref{modeleqnssection}. 
\begin{enumerate}
    \item Effector CD8+ T cells and NK cells induce apoptosis of cancer cells, with this being inhibited by TGF-$\upbeta$ and the PD-1/PD-L1 complex. However, TNF and IFN-$\upgamma$ induce necroptosis of cancer cells, causing them to become necrotic before they are removed.
    \item Necrotic cancer cells release DAMPs such as HMGB1 and calreticulin, which stimulate immature DCs to mature.
    \item Some mature DCs migrate to the T cell zone of the TDLN and activate naive CD8+ and CD4+ T cells (including Tregs), with CD8+ T cell and Th1 cell activation being inhibited by Tregs and the PD-1/PD-L1 complex.
    \item Activated T cells undergo clonal expansion and proliferate rapidly in the TDLN, with CD8+ T cell and Th1 cell proliferation being inhibited by Tregs and the PD-1/PD-L1 complex.
    \item T cells that have completed proliferation migrate to the TS and perform effector functions including the production of pro-inflammatory (IL-2, IFN-$\upgamma$, TNF) and immunosuppressive (TGF-$\upbeta$, IL-10) cytokines. Extended exposure to the cancer antigen can lead CD8+ T cells to become exhausted; however, this exhaustion can be reversed by pembrolizumab.
    \item In addition, mature DCs, NK cells and macrophages secrete cytokines that can activate NK cells and polarise and repolarise macrophages into pro-inflammatory and immunosuppressive phenotypes.
    \item Pembrolizumab infusion promotes the binding of unbound PD-1 receptors to pembrolizumab, forming the PD-1/pembrolizumab complex instead of the PD-1/PD-L1 complex. This reduces the inhibition of pro-inflammatory CD8+ and Th1 cell activation and proliferation while also reducing the inhibition of cancer cell lysis.
\end{enumerate}
\subsection{Model Equations\label{modeleqnssection}}
\subsubsection{Equations for Cancer Cells ($C$ and $N_c$)}
Viable cancer cells are killed by effector CD8+ T cells \citep{Raskov2020} and activated NK cells \citep{Zhang2020} through direct contact, whilst TNF and IFN-$\upgamma$ indirectly eliminate cancer cells via activating cell death pathways \citep{Josephs2018, Wang2008, Jorgovanovic2020}. In particular, TNF and IFN-$\upgamma$ induce the necroptosis, programmed necrotic cell death, of cancer cells \citep{Wang2008, Castro2018}. We note that TGF-$\upbeta$ and the PD-1/PD-L1 complex inhibit cancer cell lysis by CD8+ T cells \citep{Thomas2005, Juneja2017, Azuma2008}, and that TGF-$\upbeta$ and PD-1/PD-L1 have been shown to inhibit NK cell cytotoxicity \citep{Regis2020, Batlle2019, Hsu2018, Lin2024, Quatrini2020, Liu2017}. We assume that viable cancer cells grow logistically, as is done in many CRC models \citep{Kirshtein2020, dePillis2014, Budithi2021}, due to space and resource competition in the TME. Combining these, we have
\begin{align}
  \begin{split}
  \frac{dC}{dt} &= \underbrace{\lambda_{C}C\left(1-\frac{C}{C_0}\right)}_{\text{growth}} - \underbrace{\lambda_{CT_8}T_8 \frac{1}{1+I_{\upbeta}/K_{CI_{\upbeta}}}\frac{1}{1+Q^{T_8}/K_{CQ^{T_8}}}C}_{\substack{\text{elimination by $T_8$} \\ \text{inhibited by $I_{\upbeta}$ and $Q^{T_8}$}}} - \underbrace{\lambda_{CK}K \frac{1}{1+I_{\upbeta}/K_{CI_{\upbeta}}}\frac{1}{1+Q^K/K_{CQ^K}}C}_{\substack{\text{elimination by $K$} \\ \text{inhibited by $I_\upbeta$ and $Q^K$}}} \\
  &- \underbrace{\lambda_{CI_{\upalpha}}\frac{I_{\upalpha}}{K_{CI_{\upalpha}}+I_{\upalpha}}C}_{\text{elimination by $I_{\upalpha}$}}- \underbrace{\lambda_{CI_{\upgamma}}\frac{I_{\upgamma}}{K_{CI_{\upgamma}}+I_{\upgamma}}C}_{\text{elimination by $I_{\upgamma}$}},
  \end{split} \label{cancereqn} \\
  \begin{split}
    \frac{dN_c}{dt}&= \underbrace{\lambda_{CI_{\upalpha}}\frac{I_{\upalpha}}{K_{CI_{\upalpha}}+I_{\upalpha}}C}_{\text{elimination by $I_{\upalpha}$}} + \underbrace{\lambda_{CI_{\upgamma}}\frac{I_{\upgamma}}{K_{CI_{\upgamma}}+I_{\upgamma}}C}_{\text{elimination by $I_{\upgamma}$}} -\underbrace{d_{N_c}N_c}_{\text{removal}}.
    \end{split}\label{necroticcelleqn}
\end{align}
\subsubsection{Equation for HMGB1 ($H$)}
The molecule HMGB1 is released by necrotic cancer cells \citep{Fucikova2020} so that
\begin{equation}
 \frac{dH}{dt} = \underbrace{\lambda_{HN_c} N_c}_{\text{production by $N_c$}} - \underbrace{d_{H} H}_{\text{degradation}}. \label{Heqn}
\end{equation}
\subsubsection{Equation for Calreticulin ($S$)}
Necrotic cancer cells release calreticulin \citep{Ahmed2020} so that
\begin{equation}
\frac{dS}{dt} = \underbrace{\lambda_{SN_c} N_c}_{\text{production by $N_c$}}- \underbrace{d_{S} S}_{\text{degradation}}. \label{Seqn}
\end{equation}
\subsubsection{Equations for Immature and Mature DCs in the TS ($D_0$ and $D$)}
Immature DCs are stimulated to mature via DAMPs such as HMGB1 and calreticulin \citep{DelPrete2023}; however, we employ Michaelis-Menten kinetics to account for the limited rate of receptor recycling time \citep{Lai2017}. In addition, activated NK cells have been shown to efficiently kill immature DCs but not mature DCs; however, this is inhibited by TGF-$\upbeta$ \citep{Morandi2012, Vivier2008, Castriconi2003}. We also need to consider that some mature DCs migrate into the T cell zone of the TDLN and stimulate naive T cells, causing them to be activated \citep{Ruhland2020, Choi2017}. Assuming that immature DCs are supplied at a rate $\mathcal{A}_{D_0}$, we have that
\begin{align}
\begin{split}
    \frac{dD_0}{dt} &= \underbrace{\mathcal{A}_{D_0}}_{\text{source}} - \underbrace{\lambda_{DH}D_0\frac{H}{K_{DH}+H}}_{\text{$D_0 \to D$ by $H$}} - \underbrace{\lambda_{DS}D_0\frac{S}{K_{DS}+S}}_{\text{$D_0 \to D$ by $S$}}-\underbrace{\lambda_{D_0K} D_0K\frac{1}{1+I_\upbeta/K_{D_0I_\upbeta}}}_{\substack{\text{elimination by $K$} \\ \text{inhibited by $I_\upbeta$}}} - \underbrace{d_{D_0}D_0}_{\text{death}},
    \end{split} \label{D0eqn} \\
  \frac{dD}{dt} &= \underbrace{\lambda_{DH}D_0\frac{H}{K_{DH}+H}}_{\text{$D_0 \to D$ by $H$}} +   \underbrace{\lambda_{DS}D_0\frac{S}{K_{DS}+S}}_{\text{$D_0 \to D$ by $S$}} - \underbrace{\lambda_{DD^\mathrm{LN}}D}_{\substack{\text{$D$ migration} \\ \text{to TDLN}}} - \underbrace{d_{D}D}_{\text{death}}. \label{Deqn}
 \end{align}
\subsubsection{Equation for Mature DCs in the TDLN ($D^\mathrm{LN}$)}
\begin{figure}[H]
    \centering
    \fbox{\includegraphics[width=0.75\textwidth]{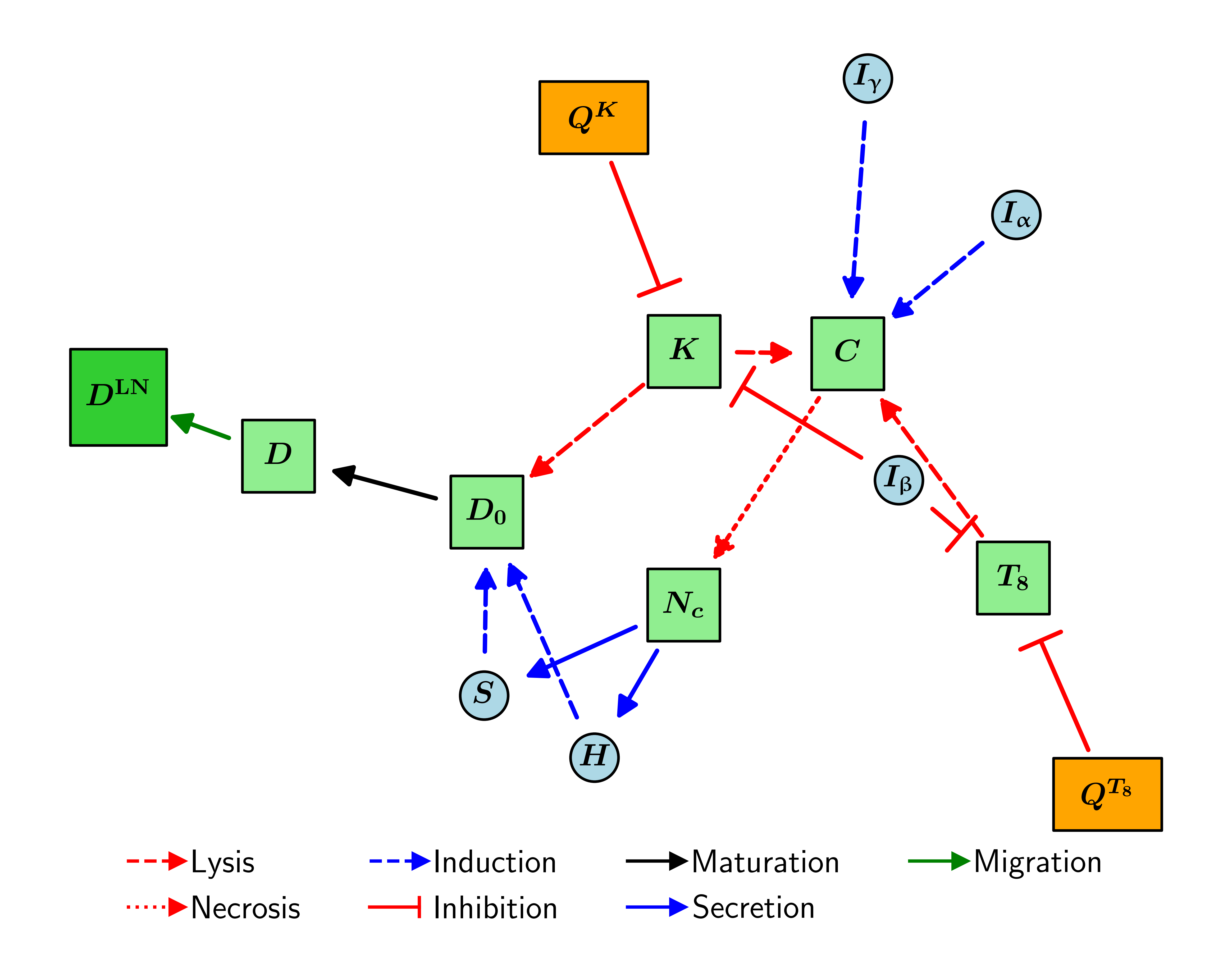}}
    \caption{\label{modeldiagramcancer}Schematic diagram of the interactions of cancer cells, DAMPs, and DCs in the model.}
\end{figure}

We assume a fixed DC migration time of $\tau_m$ and also assume that only $\exp\left(-d_{D}\tau_m\right)$ of the mature DCs that leave the TS survive migration. Taking into account the volume change between the TS and the TDLN, we have that
\begin{equation}
\frac{dD^\mathrm{LN}}{dt} = \frac{V_\mathrm{TS}}{V_\mathrm{LN}}\underbrace{\lambda_{DD^\mathrm{LN}}\exp\left(-d_D \tau_m\right)D(t-\tau_m)}_{\text{$D$ migration to TDLN}} - \underbrace{d_{D}D^\mathrm{LN}}_{\text{death}}. \label{DLNeqn}
 \end{equation}
A diagram encompassing the interactions of cancer cells, DAMPs, and DCs is shown in \autoref{modeldiagramcancer}.
\subsubsection{Equation for Naive CD8+ T Cells in the TDLN ($T_0^8$)}
We assume that naive CD8+ T cells come into the TDLN at a constant rate and that they have not undergone cell division, nor will they until their activation. For simplicity, we do not consider cytokines in the TDLN, absorbing their influence into $\lambda_{T_0^8 T_A^8}$. We do, however, explicitly take into account the influence of effector Tregs and the PD-1/PD-L1 complex in the TDLN, which have been shown to inhibit T cell activation via mechanisms including limiting naive T cells from binding to mature DCs \citep{Tadokoro2006,Chen2022treg, Li2020, Sakaguchi2008, BrunnerWeinzierl2018, Mizuno2019, Chen2023immun, Peng2020, Arasanz2017}. Recalling that T cells that have become activated by mature DCs are no longer naive, and taking this all into account, leads to
\begin{equation}
  \frac{dT_0^8}{dt}=\underbrace{\mathcal{A}_{T_0^8}}_{\text{source}} -\underbrace{R^8(t)}_{\substack{\text{CD8+ T cell} \\ \text{activation}}} - \underbrace{d_{T_0^8}T_0^8}_{\text{death}}, \label{naivecd8eqn}
\end{equation}
where $R^8(t)$ is defined as
\begin{equation}
    R^8(t) := \underbrace{\frac{\lambda_{T_0^8 T_A^8}\exp\left(-d_{T_0^8}\tau_8^\mathrm{act}\right)D^\mathrm{LN}(t-\tau_8^\mathrm{act})T_0^8(t-\tau_8^\mathrm{act})}{\left(1+\int_{t-\tau_8^\mathrm{act}}^{t} T_A^r(s) \ ds/K_{T_0^8T_A^r}\right)\left(1+\int_{t-\tau_8^\mathrm{act}}^{t} Q^{8\mathrm{LN}}(s) \ ds/K_{T_0^8Q^{8\mathrm{LN}}}\right)}}_{\text{CD8+ T cell activation inhibited by $T_A^r$ and $Q^{8\mathrm{LN}}$}}.
\end{equation}
In particular, since effector Tregs and the PD-1/PD-L1 complex inhibit T cell activation during the whole activation process, it is not sufficient to consider point estimates of effector Treg and PD-1/PD-L1 concentration. Instead, we resort to considering the integrals of the concentrations of the relevant species throughout the entire $\tau_8^\mathrm{act}$ time that the CD8+ T cell takes to complete activation. This is because these integrals are proportional (with a proportionality constant of $1/{\tau_8^\mathrm{act}}$) to the average concentration of these species throughout activation, allowing us to properly incorporate their inhibition by effector Tregs and the PD-1/PD-L1 complex.
\subsubsection{Equation for Effector CD8+ T Cells in the TDLN ($T_A^8$)}
It is known that activated CD8+ T cells undergo clonal expansion in the TDLN and differentiate before they stop proliferating and migrate to the TS \citep{Harris2002, Catron2006}.\\~\\
We assume that activated CD8+ T cells proliferate up to $n^8_\mathrm{max}$ times, upon which they stop dividing. For simplicity, we assume that the death rate of CD8+ T cells that have not completed their division program is equal to $d_{T_0^8}$, the death rate of naive CD8+ T cells, regardless of the number of cell divisions previously undergone. We also assume that only activated CD8+ T cells that have undergone $n^8_\mathrm{max}$ divisions become effector CD8+ T cells, which will leave the TDLN and migrate to the TS. Furthermore, we assume a constant cell cycle time of $\Delta_8$, except for the first cell division, which has a cycle time of $\Delta_8^0$. Thus, the duration of the activated CD8+ T cell division program to $n^8_\mathrm{max}$ divisions is given by 
\begin{equation}
    \tau_{T_A^8} := \Delta_8^0 + (n^8_\mathrm{max}-1)\Delta_8.
\end{equation}
In particular, we must take into account that some T cells will die before the division program is complete, so we must introduce a shrinkage factor of $\exp\left(-d_{T_0^8}\tau_{T_A^8}\right)$. Furthermore, we must also take into account that effector Tregs and the PD-1/PD-L1 complex inhibit CD8+ T cell proliferation throughout the program \citep{Chen2022treg, Li2020, Sakaguchi2008, Riley2009, Buchbinder2016}. We must also consider that some of these effector CD8+ T cells will migrate to the TS to perform effector functions. We finally assume that the death rate of CD8+ T cells that have completed their division program is equal to the death rate of CD8+ T cells in the TS. Taking this all into account leads to
\begin{equation}
\begin{split}
    \frac{dT_A^8}{dt} &= \underbrace{\frac{2^{n^8_\mathrm{max}}\exp\left(-d_{T_0^8}\tau_{T_A^8}\right)R^8(t- \tau_{T_A^8})}{\left(1+\int_{t- \tau_{T_A^8}}^{t} T_A^r(s) \ ds/K_{T_A^8 T_A^r}\right)\left(1+\int_{t- \tau_{T_A^8}}^{t} Q^{8\mathrm{LN}}(s) \ ds/K_{T_A^8 Q^{8\mathrm{LN}}}\right)}}_{\text{CD8+ T cell proliferation inhibited by $T_A^r$ and $Q^{8\mathrm{LN}}$}} - \underbrace{\lambda_{T_A^8T_8}T_A^8}_{\substack{\text{$T_A^8$ migration} \\ \text{to the TS}}}-\underbrace{d_{T_8} T_A^8}_\text{death}.
\end{split}\label{TA8n8maxeqn}
\end{equation}
\subsubsection{Equations for Effector and Exhausted CD8+ T Cells in the TS ($T_8$ and $T_\mathrm{ex}$)}
We assume that it takes time $\tau_a$ for effector CD8+ T cells in the TDLN to migrate to the TS. We must also account for CTL expansion due to IL-2 \citep{Rosenberg2014}, noting that this proliferation is inhibited by effector Tregs \citep{Chen2022treg, Li2020, Sakaguchi2008}. Furthermore, the death of CD8+ T cells is resisted by IL-10 \citep{Oft2019, Qiu2017}. \\~\\
However, chronic antigen exposure can cause effector CD8+ T cells to enter a state of exhaustion, where they lose their ability to kill cancer cells, and the rate of cytokine secretion significantly decreases \citep{Blank2019, Lee2016, Shive2021}. We denote this exhausted CD8+ T cell population as $T_\mathrm{ex}(t)$. It has also been shown that pembrolizumab can ``reinvigorate'' these cells back into the effector state \citep{Pauken2015, Lee2015}. We model the reinvigoration and exhaustion using Michaelis-Menten terms in $A_1$ and $\int_{t-\tau_{l}}^t C(s) \ ds$ respectively, where $\tau_l$ is the median time that CD8+ T cells take to become exhausted after entering the TS. In particular, this has been shown to be more appropriate than simple mass-action kinetics as it accounts for extended antigen exposure \citep{DeBoer1995}. \\~\\
As such, remembering to take the volume change between the TDLN and the TS into account, this implies that
\begin{align}
\begin{split}
    \frac{d T_8}{dt} &= \frac{V_\mathrm{LN}}{V_\mathrm{TS}}\underbrace{\lambda_{T_A^8T_8}\exp\left(-d_{T_8} \tau_a\right)T_A^8(t - \tau_a)}_{\text{$T_A^8$ migration to the TS}} + \underbrace{\lambda_{T_8 I_{2}}\frac{T_8 I_{2}}{K_{T_8 I_{2}}+ I_{2}}\frac{1}{1+T_r/K_{T_8T_r}}}_{\text{growth by $I_2$ inhibited by $T_r$}} \\
    &- \underbrace{\lambda_{T_8C}\frac{T_8\int_{t-\tau_{l}}^t C(s) \ ds}{K_{T_8C}+\int_{t-\tau_{l}}^t C(s) \ ds}}_{\text{$T_8 \to T_\mathrm{ex}$ from $C$ exposure}} + \underbrace{\lambda_{T_\mathrm{ex}A_1}\frac{T_\mathrm{ex}A_1}{K_{T_\mathrm{ex}A_1} + A_1}}_{\text{$T_\mathrm{ex} \to T_8$ by $A_1$}} - \underbrace{\frac{d_{T_8} T_8}{1+I_{10}/K_{T_8I_{10}}}}_{\substack{\text{death} \\ \text{inhibited by $I_{10}$}}},
    \end{split}\label{t8eqn}\\
    \frac{dT_\mathrm{ex}}{dt} &= \underbrace{\lambda_{T_8C}\frac{T_8\int_{t-\tau_{l}}^t C(s) \ ds}{K_{T_8C}+\int_{t-\tau_{l}}^t C(s) \ ds}}_{\text{$T_8 \to T_\mathrm{ex}$ from $C$ exposure}} - \underbrace{\lambda_{T_\mathrm{ex}A_1}\frac{T_\mathrm{ex}A_1}{K_{T_\mathrm{ex}A_1} + A_1}}_{\text{$T_\mathrm{ex} \to T_8$ by $A_1$}} - \underbrace{\frac{d_{T_\mathrm{ex}} T_\mathrm{ex}}{1+I_{10}/K_{T_\mathrm{ex}I_{10}}}}_{\substack{\text{death} \\ \text{inhibited by $I_{10}$}}}. \label{Texeqn}
\end{align}
\subsubsection{Equation for Naive CD4+ T Cells in the TDLN ($T_0^4$)}
For simplicity, we consider only the Th1 subtype that naive CD4+ T cells differentiate into upon activation, absorbing the influence of cytokines via the kinetic rate constant $\lambda_{T_0^4 T_A^1}$. Taking into account that effector Tregs and the PD-1/PD-L1 complex inhibit Th1 cell activation and some mature DCs migrate into the TDLN and activate naive CD4+ T cells, causing them to no longer be naive, and assuming that naive CD4+ T cells come into the TDLN at a rate $\mathcal{A}_{T_0^4}$, we can write a similar equation to \eqref{naivecd8eqn}:
\begin{equation}
  \frac{dT_0^4}{dt} =\underbrace{\mathcal{A}_{T_0^4}}_{\text{source}} - \underbrace{R^1(t)}_{\text{Th1 cell activation}} - \underbrace{d_{T_0^4}T_0^4}_{\text{death}}, \label{naivecd4eqn}
\end{equation}
where $R^1(t)$ is defined as
\begin{equation}
  R^1(t) := \underbrace{\frac{\lambda_{T_0^4 T_A^1}\exp\left(-d_{T_0^4}\tau_4^\mathrm{act}\right) D^\mathrm{LN}(t-\tau_4^\mathrm{act})T_0^4(t-\tau_4^\mathrm{act})}{\left(1+\int_{t-\tau_4^\mathrm{act}}^{t} T_A^r(s) \ ds/K_{T_0^4 T_A^r}\right)\left(1+\int_{t-\tau_4^\mathrm{act}}^{t} Q^{1\mathrm{LN}}(s) \ ds/K_{T_0^4 Q^{1\mathrm{LN}}}\right)}}_{\text{Th1 cell activation inhibited by $T_A^r$ and $Q^{1\mathrm{LN}}$}}.
\end{equation}
\subsubsection{Equation for Effector Th1 Cells in the TDLN ($T_A^1$)}
We assume that Th1 cells proliferate up to $n^1_\mathrm{max}$ times, upon which they stop dividing and become effector cells. As before, we assume that the death rate of Th1 cells that have not completed their division program is equal to $d_{T_0^4}$, the death rate of naive CD4+ T cells, regardless of the number of cell divisions previously undergone. We assume a constant cell cycle time of $\Delta_1$, except for the first cell division, which has a cycle time of $\Delta_1^0$. Thus, the duration of the Th1 cell division program to $n^1_\mathrm{max}$ divisions is given by
\begin{equation}
    \tau_{T_A^1} := \Delta_1^0 + (n^1_\mathrm{max}-1)\Delta_1.
\end{equation}
In particular, we must take into account that some Th1 cells will die before the division program is complete, so we must introduce a shrinkage factor of $\exp\left(-d_{T_0^4}\tau_{T_A^1}\right)$. Furthermore, we must also take into account that effector Tregs and the PD-1/PD-L1 complex inhibit Th1 cell proliferation throughout their program. We also assume that the death rate of Th1 cells that have completed their division program is equal to the corresponding degradation rate in the TS. Taking this all into account, and incorporating effector Th1 cell migration to the TS, leads to
\begin{equation}
    \frac{dT_A^1}{dt} = \underbrace{\frac{2^{n^1_\mathrm{max}}\exp\left(-d_{T_0^4}\tau_{T_A^1}\right) R^1(t- \tau_{T_A^1})}{\left(1+\int_{t- \tau_{T_A^1}}^{t} Q^{1\mathrm{LN}}(s) \ ds/K_{T_A^1 Q^{1\mathrm{LN}}}\right)\left(1+\int_{t- \tau_{T_A^1}}^{t} T_A^r(s) \ ds/K_{T_A^1 T_A^r}\right)}}_{\text{Th1 cell proliferation inhibited by $T_A^r$ and $Q^{1\mathrm{LN}}$}} - \underbrace{\lambda_{T_A^1T_1}T_A^1}_{\substack{\text{$T_A^1$ migration} \\ \text{to the TS}}}-\underbrace{d_{T_1} T_A^1}_\text{death} \label{TA1n1maxeqn}.
\end{equation}
\subsubsection{Equation for Effector Th1 Cells in the TS ($T_1$)}
We assume that it takes time $\tau_a$ for these cells to migrate to the TS. We take into account the fact that IL-2 induces the growth of effector Th1 cells \citep{Choudhry2018}, noting that this proliferation is inhibited by effector Tregs \citep{Chen2022treg, Li2020, Sakaguchi2008}. Furthermore, the PD-1/PD-L1 axis converts Th1 cells to Tregs \citep{Amarnath2011, Cai2019}, a process we consider to be mediated by the PD-1/PD-L1 complex on Th1 cells. Thus, we have that
\begin{align}
\frac{dT_1}{dt} &= \frac{V_\mathrm{LN}}{V_\mathrm{TS}}\underbrace{\lambda_{T_A^1T_1}\exp\left(-d_{T_1} \tau_a\right)T_A^1(t-\tau_a)}_{\text{$T_A^1$ migration to the TS}} +\underbrace{\lambda_{T_1 I_{2}}\frac{T_1 I_{2}}{K_{T_1 I_{2}}+ I_{2}}\frac{1}{1+T_r/K_{T_1T_r}}}_{\text{growth by $I_2$ inhibited by $T_r$}} - \underbrace{\lambda_{T_1T_r}T_1\frac{Q^{T_1}}{K_{T_1Q^{T_1}} + Q^{T_1}}}_{\text{$T_1 \to T_r$ by $Q^{T_1}$}} - \underbrace{d_{T_1}T_1}_{\text{death}} \label{th1eqn}.
\end{align}
\subsubsection{Equation for Naive Tregs in the TDLN ($T_0^r$)}
Finally, we consider the concentration of naive Tregs in the TDLN, following the same procedure as for CD8+ T cells and Th1 cells. We absorb the influence of cytokines on Treg activation via the kinetic rate constant $\lambda_{T_0^r T_A^r}$. We also take into account that some mature DCs migrate into the TDLN and activate naive Tregs, causing them to no longer be naive. Assuming that naive Tregs come into the TDLN at a rate $\mathcal{A}_{T_0^r}$, we can write a similar equation to \eqref{naivecd8eqn} and \eqref{naivecd4eqn}: 
\begin{equation}
  \frac{dT_0^r}{dt} =\underbrace{\mathcal{A}_{T_0^r}}_{\text{source}} - \underbrace{R^r(t)}_{\text{Treg activation}} - \underbrace{d_{T_0^r}T_0^r}_{\text{death}}, \label{naivetregeqn}
\end{equation}
where $R^r(t)$ is defined as
\begin{equation}
  R^r(t) := \underbrace{\lambda_{T_0^r T_A^r}\exp\left(-d_{T_0^r}\tau_r^\mathrm{act}\right) D^\mathrm{LN}(t-\tau_r^\mathrm{act})T_0^r(t-\tau_r^\mathrm{act})}_{\text{Treg activation}}.
\end{equation}
\subsubsection{Equation for Effector Tregs in the TDLN ($T_A^r$)}
We assume that activated Tregs proliferate up to $n^r_\mathrm{max}$ times, upon which they stop dividing and become effector Tregs. As before, we assume that the death rate of Tregs that have not completed their division program is equal to $d_{T_0^r}$, the death rate of naive Tregs. We assume a constant cell cycle time of $\Delta_r$, except for the first cell division, which has a cycle time of $\Delta_r^0$. Thus, the duration of the activated Treg division program to $n^r_\mathrm{max}$ divisions is given by 
\begin{equation}
    \tau_{T_A^r} := \Delta_r^0 + (n^r_\mathrm{max}-1)\Delta_r.
\end{equation}
In particular, we must take into account that some T cells will die before the division program is complete, so we must introduce a shrinkage factor of $\exp\left(-d_{T_0^r}\tau_{T_A^r}\right)$. We also assume that the death rate of effector Tregs in the TDLN is equal to the corresponding degradation rate in the TS. Taking this all into account, and incorporating effector Treg migration to the TS, leads to
\begin{equation}
\begin{split}
    \frac{dT_A^r}{dt} &= \underbrace{2^{n^r_\mathrm{max}}\exp\left(-d_{T_0^r}\tau_{T_A^r}\right) R^r(t- \tau_{T_A^r})}_{\text{Treg proliferation}} - \underbrace{\lambda_{T_A^rT_r}T_A^r}_{\substack{\text{$T_A^r$ migration} \\ \text{to the TS}}}-\underbrace{d_{T_r} T_A^r}_\text{death}.
\end{split}\label{TArnrmaxeqn}
\end{equation}
\subsubsection{Equation for Effector Tregs in the TS ($T_r$)}
\begin{figure}[ht]
    \centering
    \fbox{\includegraphics[width=0.75\textwidth]{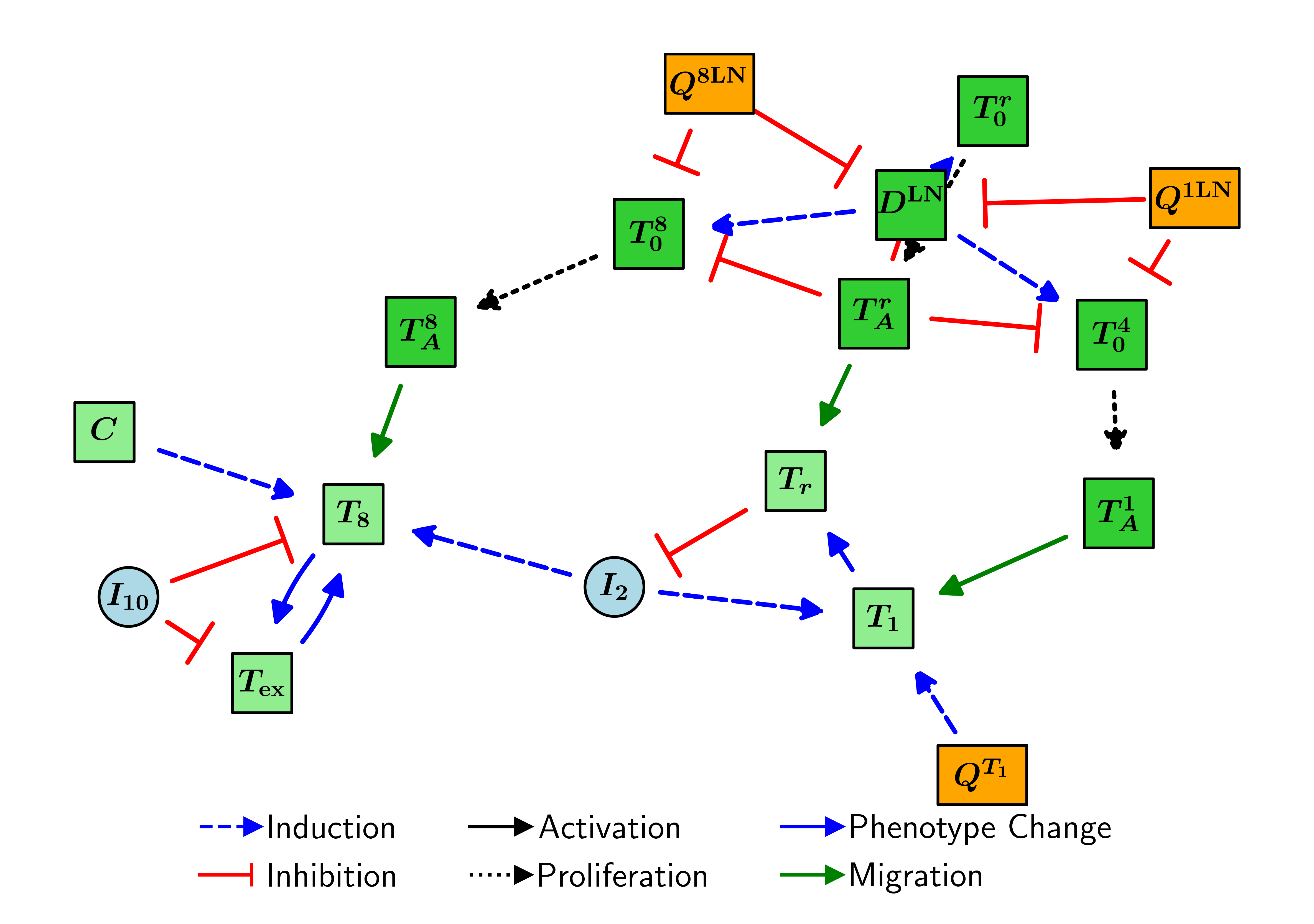}}
    \caption{\label{modeldiagramtcell}Schematic diagram of the interactions of T cells in the model.}
\end{figure}

Assuming that it also takes time $\tau_a$ for Tregs to migrate to the TS, we have that 
\begin{equation}
\frac{dT_r}{dt} = \frac{V_\mathrm{LN}}{V_\mathrm{TS}}\underbrace{\lambda_{T_A^rT_r}\exp\left(-d_{T_r} \tau_a\right)T_A^r(t-\tau_a)}_{\text{$T_A^r$ migration to the TS}} + \underbrace{\lambda_{T_1T_r}T_1\frac{Q^{T_1}}{K_{T_1Q^{T_1}} + Q^{T_1}}}_{\text{$T_1 \to T_r$ by $Q^{T_1}$}} - \underbrace{d_{T_r}T_r}_{\text{death}}. \label{tregeqn}
\end{equation}
A diagram encompassing the interactions of T cells is shown in \autoref{modeldiagramtcell}.
\subsubsection{Equations for Naive, M1, and M2 Macrophages ($M_0$, $M_1$, and $M_2$)}
TNF and IFN-$\upgamma$ polarise naive macrophages into M1 macrophages \citep{Kroner2014, Kratochvill2015, Nathan1983, Ivashkiv2018}, whilst IL-10 and TGF-$\upbeta$ polarise naive macrophages into the M2 phenotype \citep{Ambade2016, Chen2019TAM, Zhang2016}. In addition, TGF-$\upbeta$ induces M1 macrophages to convert into M2 macrophages \citep{Zhang2016}. Furthermore, M2 macrophages change phenotype to M1 under the influence of TNF \citep{Kroner2014} and IFN-$\upgamma$ \citep{Ye2021}. Assuming a production rate $\mathcal{A}_{M_0}$ of naive macrophages, we thus have that
\begin{align}
 \begin{split}
 \frac{dM_0}{dt} &=\underbrace{\mathcal{A}_{M_0}}_{\text{source}} - \underbrace{\lambda_{M_1I_\upalpha}M_0\frac{I_\upalpha}{K_{M_1 I_\upalpha}+I_\upalpha}}_{\text{$M_0 \to M_1$ by $I_{\upalpha}$}} - \underbrace{\lambda_{M_1I_\upgamma}M_0\frac{I_\upgamma}{K_{M_1 I_\upgamma}+I_\upgamma}}_{\text{$M_0 \to M_1$ by $I_\upgamma$}} - \underbrace{\lambda_{M_2I_{10}}M_0\frac{I_{10}}{K_{M_2 I_{10}}+I_{10}}}_{\text{$M_0 \to M_2$ by $I_{10}$}} \\
 &- \underbrace{\lambda_{M_2I_\upbeta}M_0\frac{I_\upbeta}{K_{M_2 I_\upbeta}+I_\upbeta}}_{\text{$M_0 \to M_2$ by $I_\upbeta$}} - \underbrace{d_{M_0}M_0}_{\text{degradation}},
 \end{split} \label{M0eqn} \\
 \begin{split}
 \frac{dM_1}{dt} &= \underbrace{\lambda_{M_1I_\upalpha}M_0\frac{I_\upalpha}{K_{M_1 I_\upalpha}+I_\upalpha}}_{\text{$M_0 \to M_1$ by $I_{\upalpha}$}} + \underbrace{\lambda_{M_1I_\upgamma}M_0\frac{I_\upgamma}{K_{M_1 I_\upgamma}+I_\upgamma}}_{\text{$M_0 \to M_1$ by $I_\upgamma$}} + \underbrace{\lambda_{MI_{\upgamma}}M_2\frac{I_{\upgamma}}{K_{MI_{\upgamma}}+I_{\upgamma}}}_{\text{$M_2 \to M_1$ by $I_{\upgamma}$}} + \underbrace{\lambda_{MI_{\upalpha}}M_2 \frac{I_{\upalpha}}{K_{MI_{\upalpha}}+I_{\upalpha}}}_{\text{$M_2 \to M_1$ by $I_{\upalpha}$}} \\
 &- \underbrace{\lambda_{MI_{\upbeta}}M_1 \frac{I_{\upbeta}}{K_{MI_{\upbeta}}+I_{\upbeta}}}_{\text{$M_1 \to M_2$ by $I_{\upbeta}$}} - \underbrace{d_{M_1}M_1}_{\text{degradation}},
 \end{split} \label{M1eqn} \\
\begin{split}
 \frac{dM_2}{dt} &= \underbrace{\lambda_{M_2I_{10}}M_0\frac{I_{10}}{K_{M_2 I_{10}}+I_{10}}}_{\text{$M_0 \to M_2$ by $I_{10}$}} + \underbrace{\lambda_{M_2I_\upbeta}M_0\frac{I_\upbeta}{K_{M_2 I_\upbeta}+I_\upbeta}}_{\text{$M_0 \to M_2$ by $I_\upbeta$}} - \underbrace{\lambda_{MI_{\upgamma}}M_2\frac{I_{\upgamma}}{K_{MI_{\upgamma}}+I_{\upgamma}}}_{\text{$M_2 \to M_1$ by $I_{\upgamma}$}} - \underbrace{\lambda_{MI_{\upalpha}}M_2 \frac{I_{\upalpha}}{K_{MI_{\upalpha}}+I_{\upalpha}}}_{\text{$M_2 \to M_1$ by $I_{\upalpha}$}} \\
 &+ \underbrace{\lambda_{MI_{\upbeta}}M_1 \frac{I_{\upbeta}}{K_{MI_{\upbeta}}+I_{\upbeta}}}_{\text{$M_1 \to M_2$ by $I_{\upbeta}$}} - \underbrace{d_{M_2}M_2}_{\text{degradation}}.
 \end{split} \label{M2eqn}
\end{align}
\subsubsection{Equations for Resting and Activated NK Cells ($K_0$ and $K$)}
\begin{figure}[ht]
    \centering
    \fbox{\includegraphics[width=0.75\textwidth]{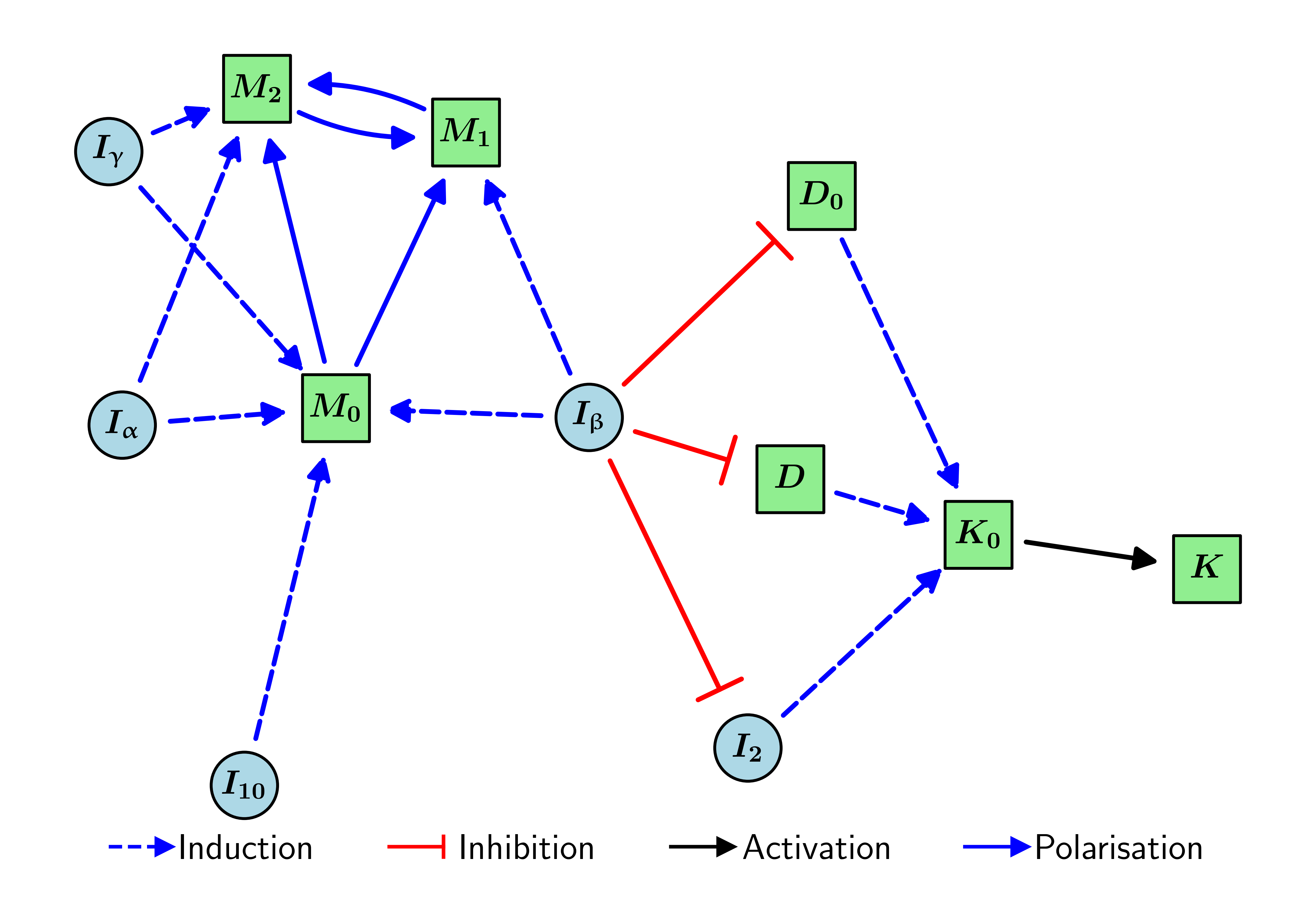}}
    \caption{\label{modeldiagramimmunecellTS}Schematic diagram of the interactions of macrophages and NK cells in the model.}
\end{figure}

Resting NK cells are activated by IL-2 \citep{Konjevi2019, Widowati2020} and immature and mature DCs \citep{Ferlazzo2002}. However, NK cell activation is inhibited by TGF-$\upbeta$ \citep{Viel2016}. Thus, assuming a supply rate $\mathcal{A}_{K_0}$ of resting NK cells, we have that
\begin{align}
    \frac{dK_0}{dt} &= \underbrace{\mathcal{A}_{K_0}}_{\text{source}} - \left(\underbrace{\lambda_{KI_2}K_0\frac{I_2}{K_{KI_2}+I_2}}_{\text{$K_0 \to K$ by $I_2$}} +\underbrace{\lambda_{KD_{0}}K_0\frac{D_{0}}{K_{KD_0}+D_0}}_{\text{$K_0 \to K$ by $D_0$}}+\underbrace{\lambda_{KD}K_0\frac{D}{K_{KD}+D}}_{\text{$K_0 \to K$ by $D$}}\right)\underbrace{\frac{1}{1+I_\upbeta/K_{KI_\upbeta}}}_{\substack{\text{activation}\\ \text{inhibited by $I_\upbeta$}}} - \underbrace{d_{K_0}K_0}_{\text{degradation}}, \label{K0eqn}\\
    \frac{dK}{dt} &= \left(\underbrace{\lambda_{KI_2}K_0\frac{I_2}{K_{KI_2}+I_2}}_{\text{$K_0 \to K$ by $I_2$}} + \underbrace{\lambda_{KD_{0}}K_0\frac{D_{0}}{K_{KD_0}+D_0}}_{\text{$K_0 \to K$ by $D_0$}}+\underbrace{\lambda_{KD}K_0\frac{D}{K_{KD}+D}}_{\text{$K_0 \to K$ by $D$}}\right)\underbrace{\frac{1}{1+I_\upbeta/K_{KI_\upbeta}}}_{\substack{\text{activation}\\ \text{inhibited by $I_\upbeta$}}} - \underbrace{d_{K}K}_{\text{degradation}}. \label{Keqn}
  \end{align}
A diagram encompassing the interactions of macrophages and NK cells is shown in \autoref{modeldiagramimmunecellTS}.
\subsubsection{Equation for IL-2 ($I_2$)}
IL-2 is produced by effector CD8+ T cells \citep{Kish2005,DSouza2004} and Th1 cells \citep{Hwang2005}, so that
\begin{equation}
 \frac{dI_2}{dt}= \underbrace{\lambda_{I_2 T_8}T_8}_{\text{production by $T_8$}} + \underbrace{\lambda_{I_2 T_1}T_1}_{\text{production by $T_1$}} - \underbrace{d_{I_2}I_2}_{\text{degradation}}. \label{il2eqn}
\end{equation}
\subsubsection{Equation for IFN-$\upgamma$ ($I_\upgamma$)}
IFN-$\upgamma$ is produced by effector CD8+ T cells \citep{Bhat2017} and Th1 cells \citep{Szabo2002, Castro2018}, with both expressions being inhibited by Tregs \citep{Sojka2011}. Furthermore, activated NK cells also produce IFN-$\upgamma$ \citep{Cui2019}. Thus,
\begin{equation}
 \frac{dI_{\upgamma}}{dt}=\left(\underbrace{\lambda_{I_{\upgamma} T_8}T_8}_{\text{production by $T_8$}} + \underbrace{\lambda_{I_{\upgamma} T_1}T_1}_{\text{production by $T_1$}}\right)\underbrace{\frac{1}{1+T_r/K_{I_{\upgamma}T_r}}}_{\text{inhibition by $T_r$}} + \underbrace{\lambda_{I_{\upgamma} K}K}_{\text{production by $K$}} - \underbrace{d_{I_\upgamma}I_\upgamma}_{\text{degradation}}. \label{ifngammaeqn}
\end{equation}
\subsubsection{Equation for TNF ($I_\upalpha$)}
TNF is produced by effector CD8+ T cells \citep{Hoekstra2021, Mehta2018} and Th1 cells \citep{Basu2021, Zhang2014}, M1 macrophages \citep{Chen2023}, and activated NK cells \citep{Fauriat2010, Wang2011}. Hence,
\begin{equation}
 \frac{dI_\upalpha}{dt}= \underbrace{\lambda_{I_{\upalpha}T_8}T_8}_{\text{production by $T_8$}} + \underbrace{\lambda_{I_{\upalpha}T_1}T_1}_{\text{production by $T_1$}} + \underbrace{\lambda_{I_{\upalpha}M_1}M_1}_{\text{production by $M_1$}} + \underbrace{\lambda_{I_{\upalpha}K}K}_{\text{production by $K$}}- \underbrace{d_{I_{\upalpha}}I_{\upalpha}}_{\text{degradation}}. \label{tnfeqn}
\end{equation}
\subsubsection{Equation for TGF-$\upbeta$ ($I_\upbeta$)}
TGF-$\upbeta$ is produced by viable cancer cells \citep{Massagu2008}, effector Tregs \citep{Tang2008} and M2 macrophages \citep{Chen2019TAM, Nuez2018}. Thus,
\begin{equation}
 \frac{dI_\upbeta}{dt}= \underbrace{\lambda_{I_{\upbeta}C}C}_{\text{production by $C$}} + \underbrace{\lambda_{I_{\upbeta}T_r}T_r}_{\text{production by $T_r$}} + \underbrace{\lambda_{I_{\upbeta}M_2}M_2}_{\text{production by $M_2$}} - \underbrace{d_{I_{\upbeta}}I_{\upbeta}}_{\text{degradation}}. \label{tgfbetaeqn}
\end{equation}
\subsubsection{Equation for IL-10 ($I_{10}$)}
\begin{figure}[ht]
    \centering
    \fbox{\includegraphics[width=0.75\textwidth]{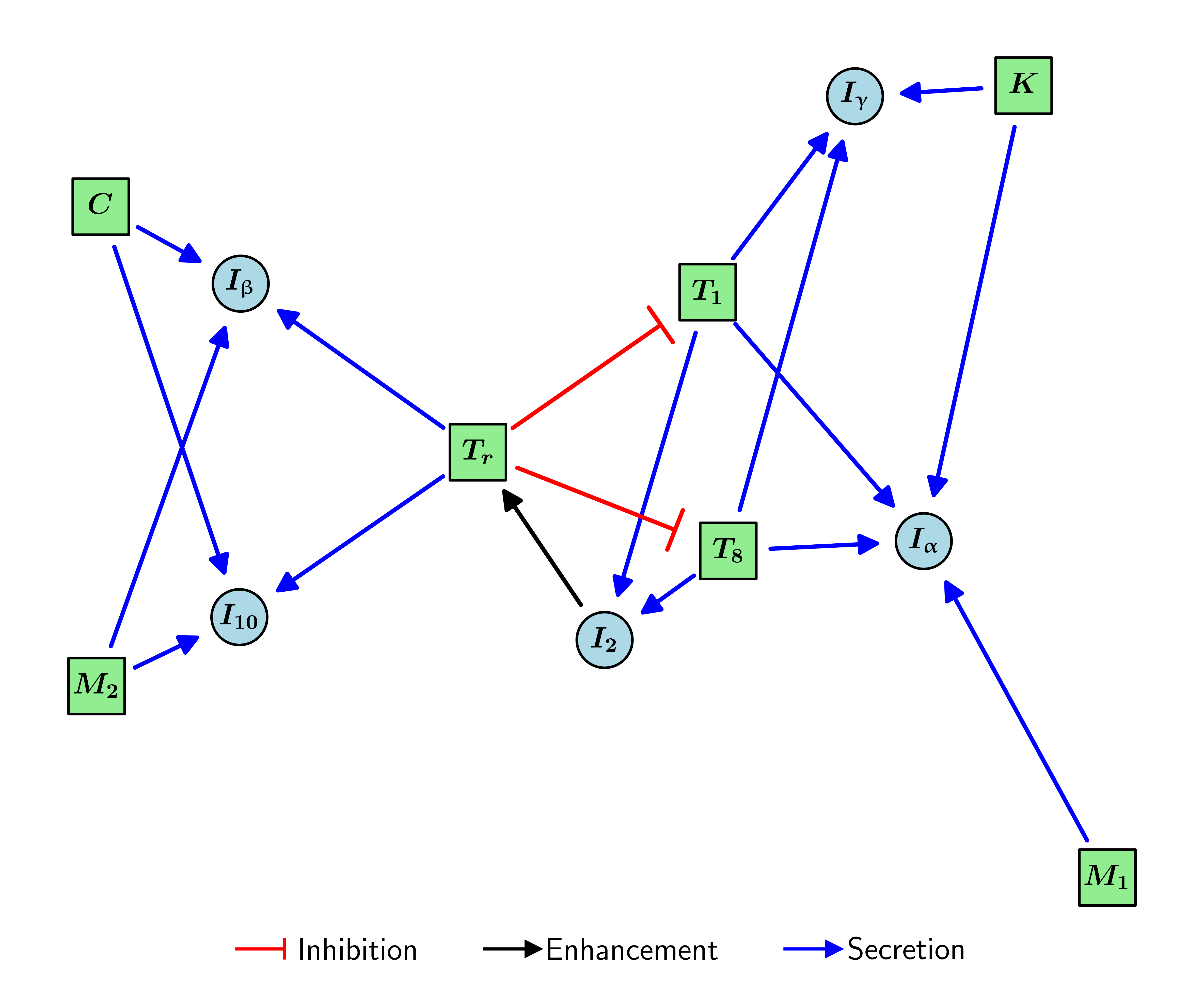}}
    \caption{\label{modeldiagramcytokines}Schematic diagram of the interactions of cytokines in the model.}
\end{figure}

IL-10 is produced by viable cancer cells \citep{Itakura2011, KrgerKrasagakes1994} and M2 macrophages \citep{Chen2019, Qi2016}. Additionally, effector Tregs secrete IL-10 \citep{Moore2001} with IL-2 enhancing this production \citep{TsujiTakayama2008, TsujiTakayama2008n2}. Hence,
\begin{equation}
\frac{dI_{10}}{dt}= \underbrace{\lambda_{I_{10}C}C}_{\text{production by $C$}} + \underbrace{\lambda_{I_{10}M_2}M_2}_{\text{production by $M_2$}} + \underbrace{\lambda_{I_{10}T_{r}}T_r\left(1+\lambda_{I_{10}I_{2}}\frac{I_{2}}{K_{I_{10}I_{2}}+I_{2}}\right)}_{\text{production by $T_r$ enhanced by $I_2$}} - \underbrace{d_{I_{10}}I_{10}}_{\text{degradation}}. \label{il10eqn}
\end{equation}
A diagram encompassing the interactions of cytokines is shown in \autoref{modeldiagramcytokines}.
\subsubsection{Equations for Unbound PD-1 receptors on Cells in the TS ($P_D^{T_8}$, $P_D^{T_1}$, $P_D^{K}$)}
It is known that PD-1 is expressed on the surface of effector CD8+ T cells \citep{Saito2012, Wu2014, Jiang2015}, effector Th1 cells \citep{LuzCrawford2012} and activated NK cells \citep{Liu2017, AlMterin2022, Hsu2018}. We assume that the rate of PD-1 synthesis is proportional to the concentration of the cell expressing it. 
However, unbound PD-1 receptors on these PD-1-expressing cells can bind to either pembrolizumab or PD-L1, forming the PD-1/pembrolizumab and PD-1/PD-L1 complexes, respectively, resulting in the depletion of unbound PD-1 molecules \citep{han2020pd, Lin2008}. For simplicity, we assume that the formation and dissociation rates of the PD-1/PD-L1 and PD-1/pembrolizumab complexes are invariant of the type of cell expressing PD-1. Considering unbound PD-1 receptors on effector CD8+ T cells in the TS at first, and taking into account the degradation of PD-1 receptors, this motivates the equation for $P_D^{T_8}$ to be
\begin{align}
    \frac{dP_D^{T_8}}{dt} &= \underbrace{\lambda_{P_D^{T_8}}T_8}_{\text{synthesis}} + \underbrace{\lambda_{Q_A}Q_A^{T_8}}_{\substack{\text{dissociation} \\ \text{of $Q_A^{T_8}$}}} + \underbrace{\lambda_{Q}Q^{T_8}}_{\substack{\text{dissociation} \\ \text{of $Q^{T_8}$}}} - \underbrace{\lambda_{P_DA_1}P_D^{T_8}A_1}_{\text{binding to $A_1$}} - \underbrace{\lambda_{P_DP_L}P_D^{T_8}P_L}_{\text{binding to $P_L$}} - \underbrace{d_{P_D}P_D^{T_8}}_{\text{degradation}}. \label{PD8eqntemp}
\intertext{Similarly, we have that}
    \frac{dP_D^{T_1}}{dt} &= \underbrace{\lambda_{P_D^{T_1}}T_1}_{\text{synthesis}} + \underbrace{\lambda_{Q_A}Q_A^{T_1}}_{\substack{\text{dissociation} \\ \text{of $Q_A^{T_1}$}}} + \underbrace{\lambda_{Q}Q^{T_1}}_{\substack{\text{dissociation} \\ \text{of $Q^{T_1}$}}} - \underbrace{\lambda_{P_DA_1}P_D^{T_1}A_1}_{\text{binding to $A_1$}} - \underbrace{\lambda_{P_DP_L}P_D^{T_1}P_L}_{\text{binding to $P_L$}} - \underbrace{d_{P_D}P_D^{T_1}}_{\text{degradation}}, \label{PD1eqntemp} \\
    \frac{dP_D^{K}}{dt} &= \underbrace{\lambda_{P_D^{K}}K}_{\text{synthesis}} + \underbrace{\lambda_{Q_A}Q_A^{K}}_{\substack{\text{dissociation} \\ \text{of $Q_A^{K}$}}} + \underbrace{\lambda_{Q}Q^{K}}_{\substack{\text{dissociation} \\ \text{of $Q^{K}$}}} - \underbrace{\lambda_{P_DA_1}P_D^{K}A_1}_{\text{binding to $A_1$}} - \underbrace{\lambda_{P_DP_L}P_D^{K}P_L}_{\text{binding to $P_L$}} - \underbrace{d_{P_D}P_D^{K}}_{\text{degradation}}. \label{PDKeqntemp}
\end{align}
\subsubsection{Equations for the PD-1/pembrolizumab Complex on Cells in the TS ($Q_A^{T_8}$, $Q_A^{T_1}$, $Q_A^{K}$)}
Pembrolizumab binds to unbound PD-1 on the surfaces of PD-1-expressing cells in a 1:1 ratio \citep{Na2016}, forming the PD-1/pembrolizumab complex in a reversible chemical process \citep{Tan2016, Wang2023pd}. We must also account for loss due to the endocytosis and internalisation of the PD-1/pembrolizumab complex from the surface of cells \citep{Cowles2022, BenSaad2024}. We assume that the rates of PD-1/pembrolizumab complex internalisation and dissociation are invariant of the type of cell expressing PD-1, so that
\begin{align}
    \frac{dQ_A^{T_8}}{dt} &= \underbrace{\lambda_{P_DA_1}P_D^{T_8}A_1}_{\text{formation of $Q_A^{T_8}$}} - \underbrace{\lambda_{Q_A}Q_A^{T_8}}_{\text{dissociation of $Q_A^{T_8}$}} - \underbrace{d_{Q_A}Q_A^{T_8}}_{\text{internalisation}}, \label{QA8eqn} \\
    \frac{dQ_A^{T_1}}{dt} &= \underbrace{\lambda_{P_DA_1}P_D^{T_1}A_1}_{\text{formation of $Q_A^{T_1}$}} - \underbrace{\lambda_{Q_A}Q_A^{T_1}}_{\text{dissociation of $Q_A^{T_1}$}} - \underbrace{d_{Q_A}Q_A^{T_1}}_{\text{internalisation}}, \label{QA1eqn} \\
    \frac{dQ_A^{K}}{dt} &= \underbrace{\lambda_{P_DA_1}P_D^{K}A_1}_{\text{formation of $Q_A^{K}$}} - \underbrace{\lambda_{Q_A}Q_A^{K}}_{\text{dissociation of $Q_A^{K}$}} - \underbrace{d_{Q_A}Q_A^{K}}_{\text{internalisation}}. \label{QAKeqn}
\end{align}
\subsubsection{Equation for Pembrolizumab in the TS ($A_1$)}
We assume that pembrolizumab is administered intravenously at times $t_1$, $t_2$, \dots, $t_n$ with doses $\xi_{1}$, $\xi_{2}$, \dots, $\xi_{n}$ respectively, assuming that the duration of infusion is negligible in comparison to the time period of interest. We also account for pembrolizumab depletion due to binding to unbound PD-1, replenishment due to PD-1/pembrolizumab complex dissociation, and elimination of pembrolizumab. It is important to note that the administered dose is not equal to the corresponding change in concentration in the TS. For simplicity, we assume linear pharmacokinetics so that, for some scaling factor $f_\mathrm{pembro}$, we have that
\begin{equation}
        \frac{dA_{1}}{dt} =\underbrace{\sum_{j=1}^{n} \xi_{j}f_\mathrm{pembro}\delta\left(t-t_j \right)}_{\text{infusion}} + \underbrace{\lambda_{Q_A}\left(Q_A^{T_8}+ Q_A^{T_1} + Q_A^{K}\right)}_{\text{dissociation of $Q_A^{T_8}$, $Q_A^{T_1}$, and $Q_A^{K}$}} - \underbrace{\lambda_{P_DA_1}\left(P_D^{T_8} + P_D^{T_1} + P_D^{K}\right)A_1}_{\text{formation of $Q_A^{T_8}$, $Q_A^{T_1}$, and $Q_A^{K}$}} - \underbrace{d_{A_1}A_{1}}_{\text{elimination}}. \label{A1eqn}
\end{equation}
\subsubsection{Equation for Unbound PD-L1 in the TS ($P_L$)}
We also know that PD-L1 is expressed on the surface of viable cancer cells \citep{Zheng2019}, mature DCs \citep{Oh2020}, effector CD8+ T cells \citep{Zheng2022, Kowanetz2018}, effector Th1 cells \citep{Chen2004}, effector Tregs \citep{Gianchecchi2018}, and M2 macrophages \citep{Zhu2020}. For brevity, we denote $\mathcal{X}$ as the set of PD-L1-expressing cells in the TS, so that $\mathcal{X}:=\set{C, D, T_8, T_1, T_r, M_2}$. Furthermore, $\lambda_{P_LX}$ denotes the synthesis rate of unbound PD-L1 on the surface of $X \in \mathcal{X}$. We must take into account the synthesis of PD-L1, its depletion due to binding to unbound PD-1, replenishment due to PD-1/PD-L1 complex dissociation, and the degradation of PD-L1. Hence,
\begin{equation}
    \frac{dP_{L}}{dt} = \underbrace{\sum_{X \in \mathcal{X}}\lambda_{P_LX}X}_{\text{synthesis}} + \underbrace{\lambda_{Q}\left(Q^{T_8} + Q^{T_1} + Q^{K}\right)}_{\text{dissociation of $Q^{T_8}$, $Q^{T_1}$ and $Q^{K}$}} - \underbrace{\lambda_{P_DP_L}\left(P_D^{T_8} + P_D^{T_1} + P_D^{K}\right)P_L}_{\text{formation of $Q^{T_8}$, $Q^{T_1}$ and $Q^{K}$}} - \underbrace{d_{P_L}P_L}_{\text{degradation}}. \label{PLeqntemp}
\end{equation}
\subsubsection{Equations for the PD-1/PD-L1 Complex in the TS ($Q^{T_8}$, $Q^{T_1}$, and $Q^{K}$)}
\begin{figure}[ht]
    \centering
    \fbox{\includegraphics[width=0.75\textwidth]{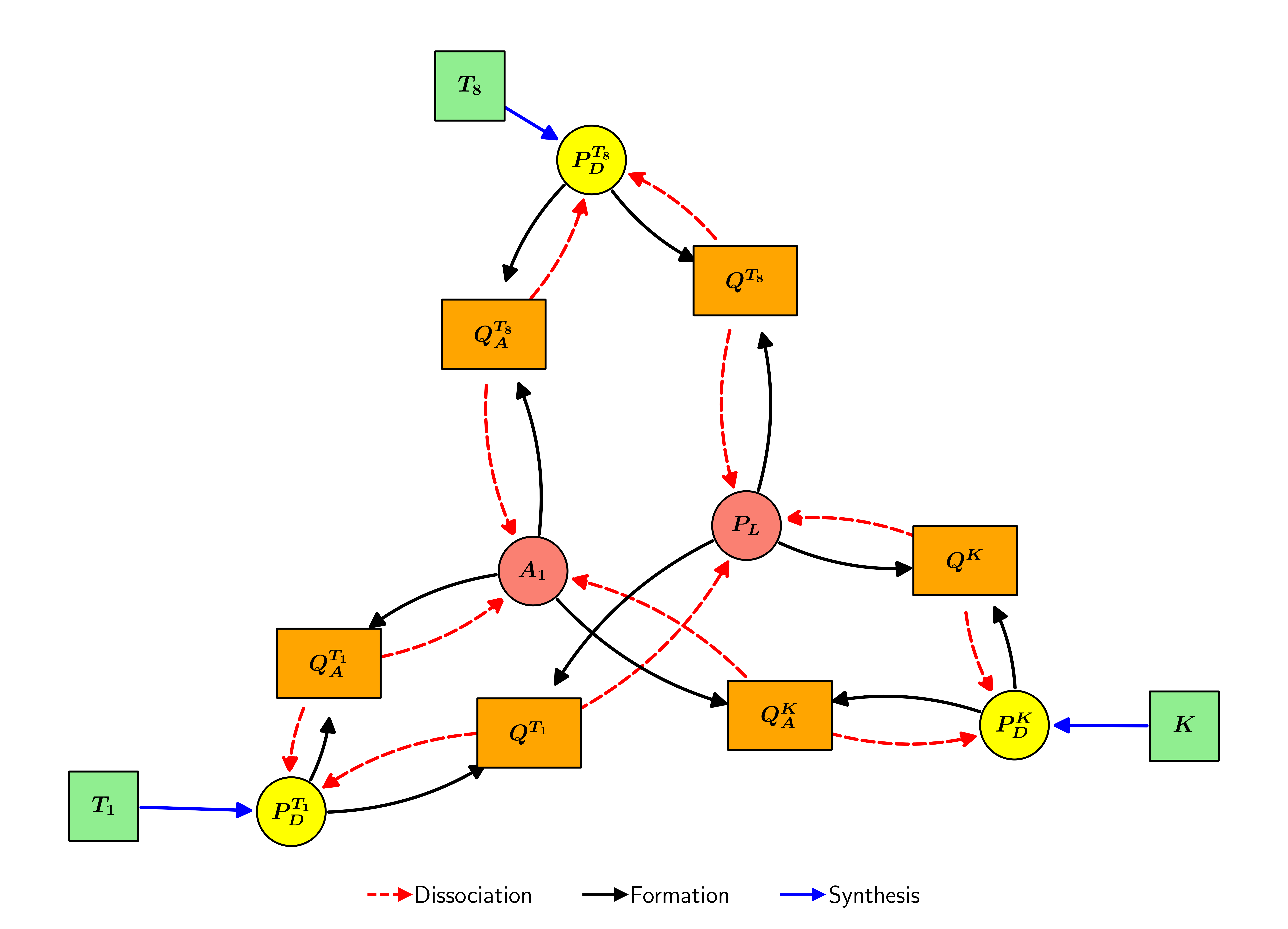}}
    \caption{\label{modeldiagramICITS}Schematic diagram of the interactions of immune checkpoint-associated components in the TS in the model.}
\end{figure}

PD-L1 binds to unbound PD-1 receptors on the surfaces of PD-1-expressing cells in a 1:1 ratio \citep{Cheng2013}, forming the PD-1/PD-L1 complex in a reversible chemical process. Considering $Q^{T_8}$ as an example, we can express its formation and dissociation via the reaction $P_{D}^{T_8}+P_{L} \underset{\lambda_{Q}}{\stackrel{\lambda_{P_{D}P_{L}}}{\rightleftharpoons}} Q^{T_8}$. We assume that the degradation is negligible relative to the dissociation, so that
\begin{equation}
\frac{dQ^{T_8}}{dt} =\underbrace{\lambda_{P_{D} P_{L}}P_{D}^{T_8}P_{L}}_{\text{formation}} - \underbrace{\lambda_{Q}Q^{T_8}}_{\text{dissociation}}.
\end{equation}
However, the dissociation rate constant of the PD-1/PD-L1 complex is $1.44 \mathrm{~s^{-1}}$, corresponding to a mean lifetime of less than 1 second \citep{Cheng2013}. As such, we employ a quasi-steady-state approximation (QSSA) for $Q^{T_8}$, so that $\frac{dQ^{T_8}}{dt}=0$, so that
\begin{align}
Q^{T_8} &= \frac{\lambda_{P_{D}P_{L}}}{\lambda_{Q}} P_{D}^{T_8}P_{L}. \label{Q8eqn}
\intertext{Similarly,}
Q^{T_1} &= \frac{\lambda_{P_{D}P_{L}}}{\lambda_{Q}} P_{D}^{T_1}P_{L}, \label{QT1eqn} \\ 
Q^{K} &= \frac{\lambda_{P_{D}P_{L}}}{\lambda_{Q}} P_{D}^{K}P_{L}. \label{QKeqn}
\end{align}
Furthermore, we can simplify \eqref{PD8eqntemp}~--~\eqref{PDKeqntemp} and \eqref{PLeqntemp} by substituting in \eqref{Q8eqn}~--~\eqref{QKeqn} so that
\begin{align}
    \frac{dP_D^{T_8}}{dt} &= \underbrace{\lambda_{P_D^{T_8}}T_8}_{\text{synthesis}} + \underbrace{\lambda_{Q_A}Q_A^{T_8}}_{\substack{\text{dissociation} \\ \text{of $Q_A^{T_8}$}}} - \underbrace{\lambda_{P_DA_1}P_D^{T_8}A_1}_{\text{binding to $A_1$}} - \underbrace{d_{P_D}P_D^{T_8}}_{\text{degradation}}, \label{PD8eqn} \\
    \frac{dP_D^{T_1}}{dt} &= \underbrace{\lambda_{P_D^{T_1}}T_1}_{\text{synthesis}} + \underbrace{\lambda_{Q_A}Q_A^{T_1}}_{\substack{\text{dissociation} \\ \text{of $Q_A^{T_1}$}}} - \underbrace{\lambda_{P_DA_1}P_D^{T_1}A_1}_{\text{binding to $A_1$}} - \underbrace{d_{P_D}P_D^{T_1}}_{\text{degradation}}, \label{PD1eqn} \\
    \frac{dP_D^{K}}{dt} &= \underbrace{\lambda_{P_D^{K}}K}_{\text{synthesis}} + \underbrace{\lambda_{Q_A}Q_A^{K}}_{\substack{\text{dissociation} \\ \text{of $Q_A^{K}$}}} - \underbrace{\lambda_{P_DA_1}P_D^{K}A_1}_{\text{binding to $A_1$}} - \underbrace{d_{P_D}P_D^{K}}_{\text{degradation}}, \label{PDKeqn} \\
    \frac{dP_{L}}{dt} &= \underbrace{\sum_{X \in \mathcal{X}}\lambda_{P_L X}X}_{\text{synthesis}} - \underbrace{d_{P_L}P_L}_{\text{degradation}}. \label{PLeqn}
\end{align}
A diagram encompassing the interactions of immune checkpoint-associated components in the TS is shown in \autoref{modeldiagramICITS}.
\subsubsection{Equations for Unbound PD-1 receptors on Cells in the TDLN ($P_D^{8\mathrm{LN}}$ and $P_D^{1\mathrm{LN}}$)}
The equations for $P_D^{8\mathrm{LN}}$ and $P_D^{1\mathrm{LN}}$ follow identically to that of \eqref{PD8eqn}~--~\eqref{PD1eqn}. For simplicity, we assume that the formation and dissociation rates of the PD-1/pembrolizumab complex are identical in the TDLN and the TS so that
\begin{align}
    \frac{dP_D^{8\mathrm{LN}}}{dt} &= \underbrace{\lambda_{P_D^{8\mathrm{LN}}}T_A^8}_{\text{synthesis}} + \underbrace{\lambda_{Q_A}Q_A^{8\mathrm{LN}}}_{\substack{\text{dissociation} \\ \text{of $Q_A^{8\mathrm{LN}}$}}} - \underbrace{\lambda_{P_DA_1}P_D^{8\mathrm{LN}}A_1^\mathrm{LN}}_{\text{binding to $A_1^\mathrm{LN}$}} - \underbrace{d_{P_D}P_D^{8\mathrm{LN}}}_{\text{degradation}}, \label{PD8LNeqn} \\
    \frac{dP_D^{1\mathrm{LN}}}{dt} &= \underbrace{\lambda_{P_D^{1\mathrm{LN}}}T_A^1}_{\text{synthesis}} + \underbrace{\lambda_{Q_A}Q_A^{1\mathrm{LN}}}_{\substack{\text{dissociation} \\ \text{of $Q_A^{1\mathrm{LN}}$}}} - \underbrace{\lambda_{P_DA_1}P_D^{1\mathrm{LN}}A_1^\mathrm{LN}}_{\text{binding to $A_1^\mathrm{LN}$}} - \underbrace{d_{P_D}P_D^{1\mathrm{LN}}}_{\text{degradation}}. \label{PD1LNeqn}
\end{align}
\subsubsection{Equations for the PD-1/pembrolizumab Complex on Cells in the TDLN ($Q_A^{8\mathrm{LN}}$ and $Q_A^{1\mathrm{LN}}$)}
The equations for $Q_A^{8\mathrm{LN}}$ and $Q_A^{1\mathrm{LN}}$ follow identically to that of \eqref{QA8eqn}~--~\eqref{QA1eqn}. For simplicity, we assume that the rates of PD-1 receptor internalisation are identical in the TDLN and the TS, so that
\begin{align}
    \frac{dQ_A^{8\mathrm{LN}}}{dt} = \underbrace{\lambda_{P_DA_1}P_D^{8\mathrm{LN}}A_1^\mathrm{LN}}_{\text{formation of $Q_A^{8\mathrm{LN}}$}} - \underbrace{\lambda_{Q_A}Q_A^{8\mathrm{LN}}}_{\text{dissociation of $Q_A^{8\mathrm{LN}}$}} - \underbrace{d_{Q_A}Q_A^{8\mathrm{LN}}}_{\text{internalisation}}, \label{QA8LNeqn} \\
    \frac{dQ_A^{1\mathrm{LN}}}{dt} = \underbrace{\lambda_{P_DA_1}P_D^{1\mathrm{LN}}A_1^\mathrm{LN}}_{\text{formation of $Q_A^{1\mathrm{LN}}$}} - \underbrace{\lambda_{Q_A}Q_A^{1\mathrm{LN}}}_{\text{dissociation of $Q_A^{1\mathrm{LN}}$}} - \underbrace{d_{Q_A}Q_A^{1\mathrm{LN}}}_{\text{internalisation}}. \label{QA1LNeqn}
\end{align}
\subsubsection{Equation for Pembrolizumab in the TDLN ($A_1^\mathrm{LN}$)}
The equation for $A_1^\mathrm{LN}$ follows identically to that of \eqref{A1eqn} so that
\begin{equation}
        \frac{dA_{1}^\mathrm{LN}}{dt} =\underbrace{\sum_{j=1}^{n} \xi_{j}f_\mathrm{pembro}\delta\left(t-t_j \right)}_{\text{infusion}} + \underbrace{\lambda_{Q_A}\left(Q_A^{8\mathrm{LN}} + Q_A^{1\mathrm{LN}}\right)}_{\text{dissociation of $Q_A^{8\mathrm{LN}}$ and $Q_A^{1\mathrm{LN}}$}} - \underbrace{\lambda_{P_DA_1}\left(P_D^{8\mathrm{LN}} + P_D^{1\mathrm{LN}}\right)A_1^\mathrm{LN}}_{\text{formation of $Q_A^{8\mathrm{LN}}$ and $Q_A^{1\mathrm{LN}}$}} - \underbrace{d_{A_1}A_{1}^\mathrm{LN}}_{\text{elimination}}. \label{A1LNeqn}
\end{equation}
\subsubsection{Equation for Unbound PD-L1 in the TDLN ($P_L^\mathrm{LN}$)}
We recall that PD-L1 is expressed on the surface of mature DCs, effector CD8+ T cells, effector Th1 cells, and effector Tregs. We denote $\mathcal{Y}$ as the set of PD-L1-expressing cells in the TDLN, so that $\mathcal{Y}:=\set{D^\mathrm{LN}, T_A^8, T_A^1, T_A^r}$, with $\lambda_{P_L^\mathrm{LN}Y}$ denoting the synthesis rate of unbound PD-L1 on the surface of $Y \in \mathcal{Y}$. The equation for $P_L^\mathrm{LN}$ follows identically to \eqref{PLeqn} so that
\begin{equation}
\begin{split}
    \frac{dP_{L}^\mathrm{LN}}{dt} &= \underbrace{\sum_{Y \in \mathcal{Y}} \lambda_{P_L^\mathrm{LN}Y}Y}_{\text{synthesis}} - \underbrace{d_{P_L}P_L^\mathrm{LN}}_{\text{degradation}}. \label{PLLNeqn}
\end{split}
\end{equation}
\subsubsection{Equations for the PD-1/PD-L1 Complex in the TDLN ($Q^{8\mathrm{LN}}$ and $Q^{1\mathrm{LN}}$)}
\begin{figure}[H]
    \centering
    \fbox{\includegraphics[width=0.75\textwidth]{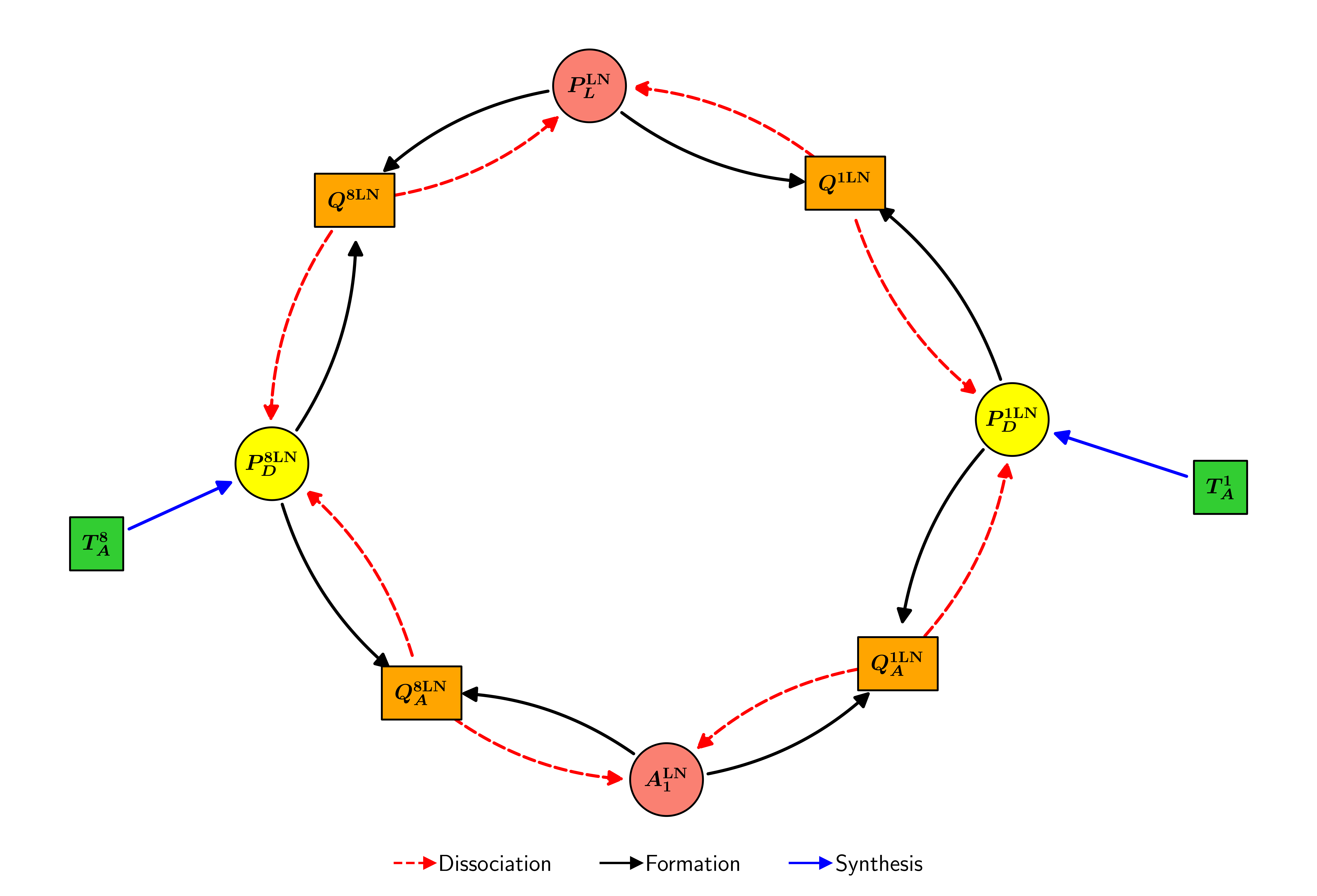}}
    \caption{\label{modeldiagramICITDLN}Schematic diagram of the interactions of immune checkpoint-associated components in the TDLN in the model.}
\end{figure}

For simplicity, we assume that the formation and dissociation rates of the PD-1/PD-L1 complex are identical in the TDLN and the TS. The equations for $Q^{8\mathrm{LN}}$ and $Q^{1\mathrm{LN}}$ follow identically from \eqref{Q8eqn}~--~\eqref{QT1eqn} so that
\begin{align}
Q^{8\mathrm{LN}} &= \frac{\lambda_{P_{D}P_{L}}}{\lambda_{Q}} P_{D}^{8\mathrm{LN}}P_{L}^\mathrm{LN}, \label{Q8LNeqn} \\
Q^{1\mathrm{LN}} &= \frac{\lambda_{P_{D}P_{L}}}{\lambda_{Q}} P_{D}^{1\mathrm{LN}}P_{L}^\mathrm{LN}. \label{Q1LNeqn}
\end{align}
A diagram encompassing the interactions of immune checkpoint-associated components in the TDLN is shown in \autoref{modeldiagramICITDLN}.
We note that throughout the model, the PD-1/PD-L1 complex appears only within an inhibition constant, making its absolute magnitude less important since it always appears as a ratio. One thing to note is that activated CD8+ T cells and Th1 cells also express PD-1 receptors and PD-L1 ligands, and we assume that effector and activated cells express these in equal amounts. However, as discussed in \Cref{TDLNssinit}, the ratio between effector and activated T cells can be assumed to remain roughly constant. Since the PD-1/PD-L1 complex concentration is linearly proportional to the product of PD-L1 concentration and unbound PD-1 receptor concentration, and PD-1/PD-L1-mediated inhibition of T cell proliferation in the TDLN appears only as a ratio, it is sufficient to consider only PD-1, PD-L1, and PD-1/PD-L1 concentrations on effector cells, as this will be appropriately scaled by the corresponding inhibition constants. Furthermore, this also justifies using the PD-1/PD-L1 complex concentration on effector T cells as a proxy for its concentration on activated T cells that have not yet undergone division, given that their ratio to effector cells remains roughly constant and that PD-1/PD-L1-mediated inhibition of T cell activation in the TDLN appears only as a ratio.
\subsection{Model Parameters}
The model parameter values are estimated in Appendix B and are listed in \autoref{tableofparams}.
\begin{center}
\begin{longtable}{|lp{140pt}lp{100pt}p{100pt}|}
\caption{\label{tableofparams}Parameter values for the model. TDLN denotes the tumour-draining lymph node, whilst TS denotes the tumour site. est. denotes estimated parameters.}\\
\hline \textbf{Parameter} & \textbf{Description} & \textbf{Value} & \textbf{Unit} & \textbf{References} \\
\hline
$f_\mathrm{pembro}$ & $A_1$/$A_1^\mathrm{LN}$ dose scaling factor & $1.17 \times 10^{12}$ & $\left(\mathrm{molec/cm^3}\right)/\mathrm{mg}$ & est.\\
\hline
$\mathcal{A}_{D_0}$ & Source of $D_0$ & $9.89 \times 10^{5}$ & $\left(\mathrm{cell/cm^{3}}\right)\mathrm{day^{-1}}$ & est. \\
$\mathcal{A}_{T_0^8}$ & Source of $T_0^8$ & $3.88 \times 10^{5}$ & $\left(\mathrm{cell/cm^{3}}\right)\mathrm{day^{-1}}$ & est. \\
$\mathcal{A}_{T_0^4}$ & Source of $T_0^4$ & $1.77 \times 10^{5}$ & $\left(\mathrm{cell/cm^{3}}\right)\mathrm{day^{-1}}$ & est. \\
$\mathcal{A}_{T_0^r}$ & Source of $T_0^r$ & $4.43 \times 10^{4}$ & $\left(\mathrm{cell/cm^{3}}\right)\mathrm{day^{-1}}$ & est. \\
$\mathcal{A}_{M_0}$ & Source of $M_0$ & $1.00 \times 10^{6}$ & $\left(\mathrm{cell/cm^{3}}\right)\mathrm{day^{-1}}$ & est. \\
$\mathcal{A}_{K_0}$ & Source of $K_0$ & $3.67 \times 10^{5}$ & $\left(\mathrm{cell/cm^{3}}\right)\mathrm{day^{-1}}$ & est. \\
\hline
$\lambda_{C}$ & Growth rate of $C$ & $1.77 \times 10^{-1}$ & $\mathrm{day^{-1}}$ & fitted \\
$\lambda_{CT_8}$ & Elimination rate of $C$ by $T_8$ & $2.58 \times 10^{-8}$ & $\left(\mathrm{cell/cm^{3}}\right)^{-1}\mathrm{day^{-1}}$ & fitted \\
$\lambda_{CK}$ & Elimination rate of $C$ by $K$ & $2.58 \times 10^{-8}$ & $\left(\mathrm{cell/cm^{3}}\right)^{-1}\mathrm{day^{-1}}$ & est. \\
$\lambda_{CI_{\upalpha}}$ & Necrosis rate of $C$ by $I_{\upalpha}$ & $1.21 \times 10^{-1}$ & $\mathrm{day^{-1}}$ & est. \\
$\lambda_{CI_{\upgamma}}$ & Necrosis rate of $C$ by $I_{\upgamma}$ & $2.43 \times 10^{-2}$ & $\mathrm{day^{-1}}$ & est. \\
$\lambda_{HN_c}$ & Production rate of $H$ by $N_c$ & $1.52 \times 10^{-14}$ & $\left(\mathrm{g/cell}\right)\mathrm{day^{-1}}$ & est. \\
$\lambda_{SN_c}$ & Production rate of $S$ by $N_c$ & $1.23 \times 10^{-14}$ & $\left(\mathrm{g/cell}\right)\mathrm{day^{-1}}$ & est. \\
$\lambda_{DH}$ & Maturation rate of $D_0$ by $H$ & $1.98 \times 10^{-1}$ & $\mathrm{day^{-1}}$ & est. \\
$\lambda_{DS}$ & Maturation rate of $D_0$ by $S$ & $1.98 \times 10^{-2}$ & $\mathrm{day^{-1}}$ & est. \\
$\lambda_{D_0K}$ & Killing rate of $D_0$ by $K$ & $2.21 \times 10^{-7}$ & $\left(\mathrm{cell/cm^3}\right)^{-1}\mathrm{day^{-1}}$ & est. \\
$\lambda_{DD^\mathrm{LN}}$ & Migration rate of $D$ to TDLN & $1.68 \times 10^{-2}$ & $\mathrm{day^{-1}}$ & est. \\
$\lambda_{T_0^8 T_A^8}$ & Kinetic rate constant for $T_0^8$ activation & $1.12 \times 10^{-10}$ & $\left(\mathrm{cell/cm^3}\right)^{-1}\mathrm{day^{-1}}$ & est. \\
$\lambda_{T_A^8T_8}$ & Kinetic rate constant for $T_A^8$ migration to the TS & $4.75 \times 10^{-1}$ & $\mathrm{day}^{-1}$ & est. \\
$\lambda_{T_8 I_2}$ & Growth rate of $T_8$ by $I_2$ & $1.61 \times 10^{-3}$ & $\mathrm{day}^{-1}$ & est. \\
$\lambda_{T_8 C}$ & Exhaustion rate of $T_8$ due to $C$ exposure & $7.08 \times 10^{-3}$ & $\mathrm{day}^{-1}$ & est. \\
$\lambda_{T_\mathrm{ex} A_1}$ & Reinvigoration rate of $T_\mathrm{ex}$ by $A_1$ & $2.25 \times 10^{-3}$ & $\mathrm{day}^{-1}$ & est. \\
$\lambda_{T_0^4 T_A^1}$ & Kinetic rate constant for $T_0^4$ activation into $T_A^1$ & $4.05 \times 10^{-10}$ & $\left(\mathrm{cell/cm^3}\right)^{-1}\mathrm{day^{-1}}$ & est. \\
$\lambda_{T_A^1 T_1}$ & Kinetic rate constant for $T_A^1$ migration to the TS & $2.66 \times 10^{-2}$ & $\mathrm{day}^{-1}$ & est. \\
$\lambda_{T_1 I_2}$ & Growth rate of $T_1$ by $I_2$ & $2.00 \times 10^{-3}$ & $\mathrm{day}^{-1}$ & est. \\
$\lambda_{T_1T_r}$ & Conversion rate of $T_1$ to $T_r$ by $Q^{T_1}$ & $4.00 \times 10^{-3}$ & $\mathrm{day}^{-1}$ & est. \\
$\lambda_{T_0^r T_A^r}$ & Kinetic rate constant for $T_0^r$ activation into $T_A^r$ & $4.24 \times 10^{-8}$ & $\left(\mathrm{cell/cm^3}\right)^{-1}\mathrm{day^{-1}}$ & est. \\
$\lambda_{T_A^r T_r}$ & Kinetic rate constant for $T_A^r$ migration to the TS & $3.51$ & $\mathrm{day}^{-1}$ & est. \\
$\lambda_{M_1I_\upalpha}$ & Polarisation rate of $M_0$ to $M_1$ by $I_{\upalpha}$ & $6.92 \times 10^{-1}$ & $\mathrm{day^{-1}}$ & est. \\
$\lambda_{M_1I_\upgamma}$ & Polarisation rate of $M_0$ to $M_1$ by $I_\upgamma$ & $8.06 \times 10^{-1}$ & $\mathrm{day^{-1}}$ & est. \\
$\lambda_{M_2I_{10}}$ & Polarisation rate of $M_0$ to $M_2$ by $I_{10}$ & $4.38 \times 10^{-1}$ & $\mathrm{day^{-1}}$ & est. \\
$\lambda_{M_2I_\upbeta}$ & Polarisation rate of $M_0$ to $M_2$ by $I_\upbeta$ & $4.90 \times 10^{-1}$ & $\mathrm{day^{-1}}$ & est. \\
$\lambda_{MI_{\upgamma}}$ & Polarisation rate of $M_2$ to $M_1$ by $I_{\upgamma}$ & $1.71 \times 10^{-2}$ & $\mathrm{day^{-1}}$ & est. \\
$\lambda_{MI_{\upalpha}}$ & Polarisation rate of $M_2$ to $M_1$ by $I_{\upalpha}$ & $1.43 \times 10^{-2}$ & $\mathrm{day^{-1}}$ & est. \\
$\lambda_{MI_{\upbeta}}$ & Polarisation rate of $M_1$ to $M_2$ by $I_{\upbeta}$ & $8.11 \times 10^{-3}$ & $\mathrm{day^{-1}}$ & est. \\
$\lambda_{KI_2}$ & Maturation rate of $K_0$ by $I_2$ & $9.24 \times 10^{-1}$ & $\mathrm{day^{-1}}$ & est. \\
$\lambda_{KD_0}$ & Maturation rate of $K_0$ by $D_0$ & $3.08 \times 10^{-1}$ & $\mathrm{day^{-1}}$ & est. \\
$\lambda_{KD}$ & Maturation rate of $K_0$ by $D$ & $1.54$ & $\mathrm{day^{-1}}$ & est. \\
$\lambda_{I_2 T_8}$ & Production rate of $I_2$ by $T_8$ & $7.44 \times 10^{-16}$ & $\left(\mathrm{g/cell}\right)\mathrm{day^{-1}}$ & est. \\
$\lambda_{I_2 T_1}$ & Production rate of $I_2$ by $T_1$ & $2.18 \times 10^{-15}$ & $\left(\mathrm{g/cell}\right)\mathrm{day^{-1}}$ & est. \\
$\lambda_{I_{\upgamma} T_8}$ & Production rate of $I_{\upgamma}$ by $T_8$ & $1.26 \times 10^{-17}$ & $\left(\mathrm{g/cell}\right)\mathrm{day^{-1}}$ & est. \\
$\lambda_{I_{\upgamma}T_1}$ & Production rate of $I_{\upgamma}$ by $T_1$ & $4.40 \times 10^{-18}$ & $\left(\mathrm{g/cell}\right)\mathrm{day^{-1}}$ & est. \\
$\lambda_{I_{\upgamma} K}$ & Production rate of $I_{\upgamma}$ by $K$ & $1.16 \times 10^{-16}$ & $\left(\mathrm{g/cell}\right)\mathrm{day^{-1}}$ & est. \\
$\lambda_{I_{\upalpha}T_8}$ & Production rate of $I_{\upalpha}$ by $T_8$ & $3.24 \times 10^{-16}$ & $\left(\mathrm{g/cell}\right)\mathrm{day^{-1}}$ & est. \\
$\lambda_{I_{\upalpha}T_1}$ & Production rate of $I_{\upalpha}$ by $T_1$ & $5.36 \times 10^{-16}$ & $\left(\mathrm{g/cell}\right)\mathrm{day^{-1}}$ & est. \\
$\lambda_{I_{\upalpha}M_1}$ & Production rate of $I_{\upalpha}$ by $M_1$ & $1.97 \times 10^{-16}$ & $\left(\mathrm{g/cell}\right)\mathrm{day^{-1}}$ & est. \\
$\lambda_{I_{\upalpha}K}$ & Production rate of $I_{\upalpha}$ by $K$ & $5.66 \times 10^{-16}$ & $\left(\mathrm{g/cell}\right)\mathrm{day^{-1}}$ & est. \\
$\lambda_{I_{\upbeta}C}$ & Production rate of $I_{\upbeta}$ by $C$ & $1.33 \times 10^{-11}$ & $\left(\mathrm{g/cell}\right)\mathrm{day^{-1}}$ & est. \\
$\lambda_{I_{\upbeta}T_r}$ & Production rate of $I_{\upbeta}$ by $T_r$ & $7.68 \times 10^{-11}$ & $\left(\mathrm{g/cell}\right)\mathrm{day^{-1}}$ & est. \\
$\lambda_{I_{\upbeta}M_2}$ & Production rate of $I_{\upbeta}$ by $M_2$ & $9.54 \times 10^{-11}$ & $\left(\mathrm{g/cell}\right)\mathrm{day^{-1}}$ & est. \\
$\lambda_{I_{10}C}$ & Production rate of $I_{10}$ by $C$ & $1.94 \times 10^{-17}$ & $\left(\mathrm{g/cell}\right)\mathrm{day^{-1}}$ & est. \\
$\lambda_{I_{10}M_2}$ & Production rate of $I_{10}$ by $M_2$ & $3.89 \times 10^{-17}$ & $\left(\mathrm{g/cell}\right)\mathrm{day^{-1}}$ & est. \\
$\lambda_{I_{10}T_r}$ & Production rate of $I_{10}$ by $T_r$ & $7.34 \times 10^{-18}$ & $\left(\mathrm{g/cell}\right)\mathrm{day^{-1}}$ & est. \\
$\lambda_{I_{10}I_{2}}$ & Production ratio of $I_{10}$ by $I_{2}$ & $3$ & dimensionless & \citep{Lo2016} est. \\
$\lambda_{P_D^{T_8}}$ & Synthesis rate of $P_D^{T_8}$ & $9.26 \times 10^{2}$ & $\left(\mathrm{molec/cell}\right)\mathrm{day^{-1}}$ & est. \\
$\lambda_{Q_A}$ & Dissociation rate of the PD-1/pembrolizumab complex & $2.6$ & $\mathrm{day}^{-1}$ & \citep{Li2021} \\
$\lambda_{Q}$ & Dissociation rate of the PD-1/PD-L1 complex & $1.24 \times 10^{5}$ & $\mathrm{day}^{-1}$ & \citep{Cheng2013} \\
$\lambda_{P_DA_1}$ & Formation rate of the PD-1/pembrolizumab complex & $4.69 \times 10^{-13}$ & $\left(\mathrm{molec/cm^3}\right)^{-1}\mathrm{day}^{-1}$ & fitted \\
$\lambda_{P_DP_L}$ & Formation rate of the PD-1/PD-L1 complex & $2.64 \times 10^{-11}$ & $\left(\mathrm{molec/cm^3}\right)^{-1}\mathrm{day}^{-1}$ & \citep{Cheng2013} \\
$\lambda_{P_D^{T_1}}$ & Synthesis rate of $P_D^{T_1}$ & $6.88 \times 10^{2}$ & $\left(\mathrm{molec/cell}\right)\mathrm{day^{-1}}$ & est. \\
$\lambda_{P_D^{K}}$ & Synthesis rate of $P_D^{K}$ & $1.85 \times 10^{2}$ & $\left(\mathrm{molec/cell}\right)\mathrm{day^{-1}}$ & est. \\
$\lambda_{P_LC}$ & Synthesis rate of $P_L$ by $C$ & $2.50 \times 10^{5}$ & $\left(\mathrm{molec/cell}\right)\mathrm{day^{-1}}$ & est. \\
$\lambda_{P_LD}$ & Synthesis rate of $P_L$ by $D$ & $2.46 \times 10^{4}$ & $\left(\mathrm{molec/cell}\right)\mathrm{day^{-1}}$ & est. \\
$\lambda_{P_LT_8}$ & Synthesis rate of $P_L$ by $T_8$ & $2.07 \times 10^{3}$ & $\left(\mathrm{molec/cell}\right)\mathrm{day^{-1}}$ & est. \\
$\lambda_{P_LT_1}$ & Synthesis rate of $P_L$ by $T_1$ & $2.89 \times 10^{3}$ & $\left(\mathrm{molec/cell}\right)\mathrm{day^{-1}}$ & est. \\
$\lambda_{P_LT_r}$ & Synthesis rate of $P_L$ by $T_r$ & $2.89 \times 10^{3}$ & $\left(\mathrm{molec/cell}\right)\mathrm{day^{-1}}$ & est. \\
$\lambda_{P_LM_2}$ & Synthesis rate of $P_L$ by $M_2$ & $3.71 \times 10^{5}$ & $\left(\mathrm{molec/cell}\right)\mathrm{day^{-1}}$ & est. \\
$\lambda_{P_D^{8\mathrm{LN}}}$ & Synthesis rate of $P_D^{8\mathrm{LN}}$ & $9.27 \times 10^{2}$ & $\left(\mathrm{molec/cell}\right)\mathrm{day^{-1}}$ & est. \\
$\lambda_{P_D^{1\mathrm{LN}}}$ & Synthesis rate of $P_D^{1\mathrm{LN}}$ & $6.89 \times 10^{2}$ & $\left(\mathrm{molec/cell}\right)\mathrm{day^{-1}}$ & est. \\
$\lambda_{P_L^\mathrm{LN}D^\mathrm{LN}}$ & Synthesis rate of $P_L^\mathrm{LN}$ by $D^\mathrm{LN}$ & $2.46 \times 10^{4}$ & $\left(\mathrm{molec/cell}\right)\mathrm{day^{-1}}$ & est. \\
$\lambda_{P_L^\mathrm{LN}T_A^8}$ & Synthesis rate of $P_L^\mathrm{LN}$ by $T_A^8$ & $2.07 \times 10^{3}$ & $\left(\mathrm{molec/cell}\right)\mathrm{day^{-1}}$ & est. \\
$\lambda_{P_L^\mathrm{LN}T_A^1}$ & Synthesis rate of $P_L^\mathrm{LN}$ by $T_A^1$ & $2.89 \times 10^{3}$ & $\left(\mathrm{molec/cell}\right)\mathrm{day^{-1}}$ & est. \\
$\lambda_{P_L^\mathrm{LN}T_A^r}$ & Synthesis rate of $P_L^\mathrm{LN}$ by $T_A^r$ & $2.89 \times 10^{3}$ & $\left(\mathrm{molec/cell}\right)\mathrm{day^{-1}}$ & est. \\
\hline
$K_{CI_{\upalpha}}$ & Half-saturation constant of $I_\upalpha$ for $C$ & $5.30 \times 10^{-11}$ & $\mathrm{g/cm^{3}}$ & est. \\
$K_{CI_{\upgamma}}$ & Half-saturation constant of $I_\upgamma$ for $C$ & $1.69 \times 10^{-11}$ & $\mathrm{g/cm^{3}}$ & est. \\
$K_{DH}$ & Half-saturation constant of $H$ for $D$ & $1.01 \times 10^{-8}$ & $\mathrm{g/cm^{3}}$ & est. \\
$K_{DS}$ & Half-saturation constant of $S$ for $D$ & $3.25 \times 10^{-8}$ & $\mathrm{g/cm^{3}}$ & est.\\
$K_{T_8I_2}$ & Half-saturation constant of $I_2$ for $T_8$ & $2.00 \times 10^{-12}$ & $\mathrm{g/cm^{3}}$ & est. \\
$K_{T_8 C}$ & Half-saturation constant $T_8$ exhaustion due to $C$ exposure & $3.31 \times 10^{8}$ & $\left(\mathrm{cell}/\mathrm{cm^3}\right)\ \mathrm{day}$ & est. \\
$K_{T_\mathrm{ex} A_1}$ & Half-saturation constant of $T_\mathrm{ex}$ reinvigoration by $A_1$ & $2.05 \times 10^{14}$ & $\mathrm{molec/cm^3}$ & est. \\
$K_{T_1I_2}$ & Half-saturation constant of $I_2$ for $T_1$ & $2.00 \times 10^{-12}$ & $\mathrm{g/cm^{3}}$ & est. \\
$K_{T_1Q^{T_1}}$ & Half-saturation constant of $T_1$ conversion to $T_r$ by $Q^{T_1}$ & $2.02 \times 10^{5}$ & $\mathrm{molec/cm^{3}}$ & est. \\
$K_{M_1 I_\upalpha}$ & Half-saturation constant of $I_\upalpha$ for $M_1$ & $5.30 \times 10^{-11}$ & $\mathrm{g/cm^{3}}$ & est. \\
$K_{M_1 I_\upgamma}$ & Half-saturation constant of $I_\upgamma$ for $M_1$ & $1.69 \times 10^{-11}$ & $\mathrm{g/cm^{3}}$ & est. \\
$K_{M_2 I_{10}}$ & Half-saturation constant of $I_{10}$ for $M_{2}$ & $1.15 \times 10^{-10}$ & $\mathrm{g/cm^{3}}$ & est. \\
$K_{M_2 I_{\upbeta}}$ & Half-saturation constant of $I_{\upbeta}$ for $M_{2}$ & $1.51 \times 10^{-6}$ & $\mathrm{g/cm^{3}}$ & est. \\
$K_{MI_{\upgamma}}$ & Half-saturation constant of $I_{\upgamma}$ for $M_{1}/M_{2}$ & $1.69 \times 10^{-11}$ & $\mathrm{g/cm^{3}}$ & est.\\
$K_{MI_{\upalpha}}$ & Half-saturation constant of $I_{\upalpha}$ for $M_{1}/M_{2}$ & $5.30 \times 10^{-11}$ & $\mathrm{g/cm^{3}}$ & est. \\
$K_{MI_{\upbeta}}$ & Half-saturation constant of $I_{\upbeta}$ for $M_{1}/M_{2}$ & $1.51 \times 10^{-6}$ & $\mathrm{g/cm^{3}}$ & est. \\
$K_{KI_2}$ & Half-saturation constant of $I_2$ for $K$ & $2.00 \times 10^{-12}$ & $\mathrm{g/cm^{3}}$ & est. \\
$K_{KD_0}$ & Half-saturation constant of $D_0$ for $K$ & $1.46 \times 10^{6}$ & $\mathrm{cell/cm^{3}}$ & est. \\
$K_{KD}$ & Half-saturation constant of $D$ for $K$ & $4.78 \times 10^{5}$ & $\mathrm{cell/cm^{3}}$ & est. \\
$K_{I_{10}I_{2}}$ & Half-saturation constant of $I_{2}$ for $I_{10}$ & $2.00 \times 10^{-12}$ & $\mathrm{g/cm^{3}}$ & est. \\
\hline
$C_0$ & Carrying capacity of $C$ & $8.15 \times 10^{7}$ & $\mathrm{cell/cm^3}$ & fitted \\
$K_{CI_{\upbeta}}$ & Inhibition constant of $T_8$ and $K$ elimination of $C$ by $I_{\upbeta}$ & $1.51 \times 10^{-6}$ & $\mathrm{g/cm^{3}}$ & est. \\
$K_{CQ^{T_8}}$ & Inhibition constant of $T_8$ elimination of $C$ by $Q^{T_8}$ & $6.68 \times 10^{5}$ & $\mathrm{molec/cm^{3}}$ & est. \\
$K_{CQ^K}$ & Inhibition constant of $K$ elimination of $C$ by $Q^K$ & $3.62 \times 10^{6}$ & $\mathrm{molec/cm^3}$ & est. \\
$K_{D_0I_\upbeta}$ & Inhibition constant of $K$ elimination of $D_0$ by $I_\upbeta$ & $1.51 \times 10^{-6}$ & $\mathrm{g/cm^{3}}$ & est. \\
$V_\mathrm{TS}$ & Volume of the TS & $2.76 \times 10^{1}$ & $\mathrm{cm^3}$ & \citep{Park2017} est. \\
$V_\mathrm{LN}$ & Volume of the TDLN & $9.20 \times 10^{-2}$ & $\mathrm{cm^3}$ & \citep{Rssler2017} est. \\
$K_{T_0^8T_A^r}$ & Inhibition constant of $T_0^8$ activation by $T_A^r$ & $1.56 \times 10^{6}$ & $(\mathrm{cell/cm^3})\mathrm{~day}$ & est. \\
$K_{T_0^8 Q^{8\mathrm{LN}}}$ & Inhibition constant of $T_0^8$ activation by $Q^{8\mathrm{LN}}$ & $1.27 \times 10^{5}$ & $(\mathrm{molec/cm^3})\mathrm{~day}$ & est. \\
$K_{T_A^8T_A^r}$ & Inhibition constant of $T_A^8$ proliferation by $T_A^r$ & $3.80 \times 10^{6}$ & $(\mathrm{cell/cm^3})\mathrm{~day}$ & est. \\
$K_{T_A^8 Q^{8\mathrm{LN}}}$ & Inhibition constant of $T_A^8$ proliferation by $Q^{8\mathrm{LN}}$ & $3.10 \times 10^{5}$ & $(\mathrm{molec/cm^3})\mathrm{~day}$ & est. \\
$K_{T_8T_r}$ & Inhibition constant of $I_2$-mediated growth of $T_8$ by $T_r$ & $1.45 \times 10^{5}$ & $\mathrm{cell/cm^{3}}$ & est. \\
$K_{T_8 I_{10}}$ & Inhibition constant of $T_8$ death by $I_{10}$ & $1.15 \times 10^{-10}$ & $\mathrm{g/cm^{3}}$ & est. \\
$K_{T_\mathrm{ex} I_{10}}$ & Inhibition constant of $T_\mathrm{ex}$ death by $I_{10}$ & $1.15 \times 10^{-10}$ & $\mathrm{g/cm^{3}}$ & est. \\
$K_{T_0^4 T_A^r}$ & Inhibition constant of $T_0^4$ activation by $T_A^r$ & $1.17 \times 10^{6}$ & $(\mathrm{cell/cm^3})\mathrm{~day}$ & est. \\
$K_{T_0^4 Q^{1\mathrm{LN}}}$ & Inhibition constant of $T_0^4$ activation by $Q^{1\mathrm{LN}}$ & $6.41 \times 10^{5}$ & $(\mathrm{molec/cm^3})\mathrm{~day}$ & est. \\
$K_{T_A^1 T_A^r}$ & Inhibition constant of $T_A^1$ proliferation by $T_A^r$ & $3.23 \times 10^{6}$ & $(\mathrm{cell/cm^3})\mathrm{~day}$ & est. \\
$K_{T_A^1 Q^{1\mathrm{LN}}}$ & Inhibition constant of $T_A^1$ proliferation by $Q^{1\mathrm{LN}}$ & $1.76 \times 10^{6}$ & $(\mathrm{molec/cm^3})\mathrm{~day}$ & est. \\
$K_{T_1T_r}$ & Inhibition constant of $I_2$-mediated growth of $T_1$ by $T_r$ & $1.45 \times 10^{5}$ & $\mathrm{cell/cm^{3}}$ & est. \\
$K_{KI_\upbeta}$ & Inhibition constant of NK cell activation by $I_\upbeta$ & $1.51 \times 10^{-6}$ & $\mathrm{g/cm^{3}}$ & est. \\
$K_{I_\upgamma T_r}$ & Inhibition constant of T cell production of $I_\upgamma$ by $T_r$ & $1.45 \times 10^{5}$ & $\mathrm{cell/cm^{3}}$ & est. \\
\hline
$d_{N_c}$ & Removal rate of $N_c$ & $6.55 \times 10^{-1}$ & $\mathrm{day}^{-1}$ & est. \\
$d_{H}$ & Degradation rate of $H$ & $5.55$ & $\mathrm{day}^{-1}$ & \citep{Zandarashvili2013} est. \\
$d_{S}$ & Degradation rate of $S$ & $1.39$ & $\mathrm{day}^{-1}$ & \citep{Goicoechea2003, Zhang2021cal} est. \\
$d_{D_0}$ & Death rate of $D_0$ & $3.57 \times 10^{-2}$ & $\mathrm{day}^{-1}$ & \citep{Ruedl2000} est. \\
$d_{D}$ & Death rate of $D$ & $3.15 \times 10^{-1}$ & $\mathrm{day}^{-1}$ & \citep{Kamath2002} est. \\
$d_{T_0^8}$ & Death rate of $T_0^8$ & $3.22 \times 10^{-2}$ & $\mathrm{day}^{-1}$ & \citep{Takada2009} est. \\
$d_{T_8}$ & Death rate of $T_8$ & $9 \times 10^{-3}$ & $\mathrm{day}^{-1}$ & \citep{Hellerstein1999} \\
$d_{T_\mathrm{ex}}$ & Death rate of $T_\mathrm{ex}$ & $9 \times 10^{-3}$ & $\mathrm{day}^{-1}$ & \citep{Hellerstein1999} \\
$d_{T_0^4}$ & Death rate of $T_0^4$ & $4.03 \times 10^{-2}$ & $\mathrm{day}^{-1}$ & \citep{Takada2009} est. \\
$d_{T_1}$ & Death rate of $T_1$ & $8 \times 10^{-3}$ & $\mathrm{day}^{-1}$ & \citep{Hellerstein1999}\\
$d_{T_0^r}$ & Death rate of $T_0^r$ & $2.2 \times 10^{-3}$ & $\mathrm{day}^{-1}$ & \citep{Kumbhari2020n2} \\
$d_{T_r}$ & Death rate of $T_r$ & $6.30 \times 10^{-2}$ & $\mathrm{day}^{-1}$ & \citep{VukmanovicStejic2006} est. \\
$d_{M_0}$ & Death rate of $M_0$ & $0.73$ & $\mathrm{day}^{-1}$ & \citep{Patel2017} \\
$d_{M_1}$ & Death rate of $M_1$ & $0.99$ & $\mathrm{day}^{-1}$ & \citep{Patel2017} \\
$d_{M_2}$ & Death rate of $M_2$ & $1.35 \times 10^{-1}$ & $\mathrm{day}^{-1}$ & \citep{Patel2017} \\
$d_{K_0}$ & Death rate of $K_0$ & $6.93 \times 10^{-2}$ & $\mathrm{day}^{-1}$ & \citep{Wu2020, Vivier2008, Lowry2017} est. \\
$d_{K}$ & Death rate of $K$ & $6.93 \times 10^{-2}$ & $\mathrm{day}^{-1}$ & \citep{Wu2020, Vivier2008, Lowry2017} est. \\
$d_{I_2}$ & Degradation rate of $I_2$ & $1.45 \times 10^{2}$ & $\mathrm{day}^{-1}$ & \citep{Lotze1985n1} est. \\
$d_{I_\upgamma}$ & Degradation rate of $I_\upgamma$ & $3.33 \times 10^{1}$ & $\mathrm{day}^{-1}$ & \citep{Balachandran2013} est. \\
$d_{I_{\upalpha}}$ & Degradation rate of $I_{\upalpha}$ & $5.48 \times 10^{1}$ & $\mathrm{day}^{-1}$ & \citep{Ma2015, Oliver1993-vl} est. \\
$d_{I_{\upbeta}}$ & Degradation rate of $I_{\upbeta}$ & $3.99 \times 10^{2}$ & $\mathrm{day}^{-1}$ & \citep{TiradoRodriguez2014} est. \\
$d_{I_{10}}$ & Degradation rate of $I_{10}$ & $6.16$ & $\mathrm{day}^{-1}$ & \citep{Huhn1997} est. \\
$d_{P_D}$ & Degradation rate of unbound PD-1 receptors & $3.36 \times 10^{-1}$ & $\mathrm{day}^{-1}$ & \citep{Lassman2021} \\
$d_{Q_A}$ & Internalisation rate of the PD-1/pembrolizumab complex & $0.43$ & $\mathrm{day}^{-1}$ & \citep{Li2021} \\
$d_{A_1}$ & Elimination rate of $A_1/A_1^\mathrm{LN}$ & $2.92 \times 10^{-2}$ & $\mathrm{day}^{-1}$ & \citep{Li2017, Li2019pembro, Ahamadi2016} est. \\
$d_{P_L}$ & Degradation rate of unbound PD-L1 & $1.39$ & $\mathrm{day}^{-1}$ & \citep{Cha2019} \\
\hline
$\tau_m$ & DC migration time from the TS to the TDLN & $0.75$ & $\mathrm{day}$ & \citep{Catron2004} est. \\
$\tau_8^\mathrm{act}$ & CD8+ T cell activation time & $2$ & $\mathrm{day}$ & \citep{Kinjyo2015} \\
$\Delta_8^0$ & Time taken for first CTL division & $1.63$ & $\mathrm{day}$ & \citep{Plambeck2022} \\
$n^8_\mathrm{max}$ & Maximal number of CTL divisions in the TDLN & $10$ & dimensionless & \citep{Kaech2001, Masopust2007} est. \\
$\Delta_8$ & Time taken for successive CTL divisions & $0.36$ & $\mathrm{day}$ & \citep{Kaech2001} \\
$\tau_{T_A^8}$ & Time taken for CTL division program & $4.87$ & $\mathrm{day}$ & est. \\
$\tau_a$ & T cell migration time from the TDLN to the TS & $0.27$ & $\mathrm{day}$ & est. \\
$\tau_l$ & Time for CTL to become exhausted in TS & $10$ & $\mathrm{day}$ & \citep{Blank2019, McLane2019} est. \\
$\tau_4^\mathrm{act}$ & CD4+ T cell activation time & $1.5$& $\mathrm{day}$ & \citep{JelleyGibbs2000} est. \\
$\Delta_1^0$ & Time taken for first Th1 cell division & $0.77$ & $\mathrm{day}$ & \citep{Kaech2002} est. \\
$n^1_\mathrm{max}$ & Maximal number of Th1 cell divisions in the TDLN & $9$ & dimensionless & \citep{Homann2001} est. \\
$\Delta_1$ & Time taken for successive Th1 cell divisions & $0.42$ & $\mathrm{day}$ & \citep{Kaech2002} est. \\
$\tau_{T_A^1}$ & Time taken for Th1 cell division program & $4.13$ & $\mathrm{day}$ & est. \\
$\tau_r^\mathrm{act}$ & Treg activation time & $1.5$ & $\mathrm{day}$ & \citep{JelleyGibbs2000} est. \\
$\Delta_r^0$ & Time taken for first Treg division & $0.77$ & $\mathrm{day}$ & \citep{Kaech2002} est. \\
$n^r_\mathrm{max}$ & Maximal number of Treg divisions in the TDLN & $6$ & dimensionless & \citep{DarrasseJze2009} est. \\
$\Delta_r$ & Time taken for successive Treg divisions & $0.42$ & $\mathrm{day}$ & \citep{Kaech2002} est. \\
$\tau_{T_A^r}$ & Time taken for Treg division program & $2.87$ & $\mathrm{day}$ & est. \\
\hline
\end{longtable}
\end{center}
\subsection{Model Reduction via QSSA}
We note that the degradation rates of cytokines and DAMPs are, in general, orders of magnitude larger than those of immune and cancer cells. In particular, IL-2, IFN-$\upgamma$, TNF, and TGF-$\upbeta$ evolve on a very fast timescale, with degradation rates significantly higher than all other species in the model, causing them to equilibrate much more rapidly. As such, we perform a quasi-steady-state approximation (QSSA) and reduce the model by setting \eqref{il2eqn}~--~\eqref{tgfbetaeqn} to $0$ and solving for $I_2$, $I_\upgamma$, $I_\upalpha$, and $I_\upbeta$ in terms of the other parameters and variables in the model. This minimally affects the system's evolution after a very short period of transient behaviour \citep{Snowden2017}, and we justify this by observing that, empirically, the deviation in system trajectories remains negligible for nearby parameter choices. Performing the QSSA leads to
\begin{align}
    \frac{dI_2}{dt} &= 0 \implies I_2 = \frac{1}{d_{I_2}}\left(\lambda_{I_2 T_8}T_8 + \lambda_{I_2 T_1}T_1\right), \label{il2qssa}\\
    \frac{dI_\upgamma}{dt} &= 0 \implies I_\upgamma = \frac{1}{d_{I_\upgamma}}\left[\left(\lambda_{I_{\upgamma} T_8}T_8 + \lambda_{I_{\upgamma} T_1}T_1\right)\frac{1}{1+T_r/K_{I_{\upgamma}T_r}} + \lambda_{I_{\upgamma} K}K\right], \label{ifngammaqssa}\\ 
    \frac{dI_\upalpha}{dt} &= 0 \implies I_\upalpha = \frac{1}{d_{I_{\upalpha}}}\left(\lambda_{I_{\upalpha}T_8}T_8 + \lambda_{I_{\upalpha}T_1}T_1 + \lambda_{I_{\upalpha}M_1}M_1 + \lambda_{I_{\upalpha}K}K\right), \label{tnfqssa}\\
    \frac{dI_\upbeta}{dt} &= 0 \implies I_\upbeta = \frac{1}{d_{I_{\upbeta}}}\left(\lambda_{I_{\upbeta}C}C + \lambda_{I_{\upbeta}T_r}T_r + \lambda_{I_{\upbeta}M_2}M_2\right).\label{ibetaqssa}
\end{align}
We note that this reduction is valid since the timescale of IFN-$\upgamma$, the slowest of the ``fast'' species, is significantly shorter than the timescales of all ``slow'' species in the model.
\section{Initial Conditions \label{initcond section}}
For all species in the model, we assume a constant solution history, where the history for each species is set to its respective initial condition.
\subsection{Initial Conditions for Cells in the TS}
We chose the initial conditions for cells in the TS to be as in \autoref{firstmodelinitcondmainbody}, with justification for the choice of these values in Appendix A.1.
\begin{table}[H]
\centering
\begin{tabular}{|c|c|c|c|c|c|c|}
\hline
$C$ & $N_c$ & $D_0$ & $D$ & $T_8$ & $T_\mathrm{ex}$ & $T_1$ \\ 
\hline
$1.79 \times 10^{7}$ & $1.99 \times 10^{6}$ & $1.63 \times 10^{6}$ & $8.29 \times 10^{5}$ & $2.43 \times 10^{5}$ & $2.09 \times 10^{5}$ & $1.04 \times 10^{5}$ \\ 
\hline
$T_r$ & $M_0$ & $M_1$ & $M_2$ & $K_0$ & $K$ & \\ 
\hline
$2.12 \times 10^{5}$ & $6.67 \times 10^{5}$ & $6.61 \times 10^{5}$ & $1.23 \times 10^{6}$ & $3.06 \times 10^{5}$ & $5.20 \times 10^{6}$ &\\ 
\hline
\end{tabular}
\caption{\label{firstmodelinitcondmainbody}TS initial condition cell densities for the model. All values are in $\mathrm{cell/cm^3}$.}
\end{table}
\subsection{Initial Conditions for Cells in the TDLN}
We estimated the initial conditions for T cells in the TDLN to be as in \Cref{TDLNfirstmodelinitcondmainbody}, with justification for the choice of these values in Appendix B.12.
\begin{table}[H]
\centering
\begin{tabular}{|c|c|c|c|c|c|}
\hline
$T_0^8$ & $T_A^8$ & $T_0^4$ & $T_A^1$ & $T_0^r$ & $T_A^r$ \\
\hline
$1.20 \times 10^{7}$ & $1.11 \times 10^{6}$ & $4.40 \times 10^{6}$ & $1.01 \times 10^{7}$ & $9.95 \times 10^{4}$ & $7.84 \times 10^{5}$ \\
\hline
\end{tabular}
\caption{\label{TDLNfirstmodelinitcondmainbody}TDLN initial condition cell densities for the model. All values are in $\mathrm{cell/cm^3}$.}
\end{table}
We also estimated the initial condition for $D^\mathrm{LN}$ to be $1.05 \times 10^{7} ~\mathrm{cell/cm^3}$, with justification for this in Appendix B.7.1.
\subsection{Initial Conditions for DAMPs\label{DAMPssinitsec}}
We chose the DAMP initial conditions to be as in \autoref{DAMPtable}, with justification for the choice of these values in Appendix A.3.
\begin{table}[H]
    \centering
    \begin{tabular}{|c|c|}
    \hline
    \textbf{DAMP} & \textbf{Initial Condition} \\
    \hline 
    $H$ & $5.76 \times 10^{-9}$\\
    $S$ & $2.00 \times 10^{-8}$ \\
    \hline
    \end{tabular}
    \caption{\label{DAMPtable}DAMP initial conditions for the model. All values are in units of $\mathrm{g/cm^3}$.}
\end{table}
\subsection{Initial Conditions for Cytokines}
We chose the cytokine initial conditions to be as in \autoref{cytokinetable}, with justification for the choice of these values in Appendix A.4 and Appendix B.6.
\begin{table}[H]
    \centering
    \begin{tabular}{|c|c|}
    \hline
    \textbf{Cytokine} & \textbf{Initial Condition} \\
    \hline 
    $I_2$ & $2.81 \times 10^{-12}$ \\
    $I_\upgamma$ & $1.82 \times 10^{-11}$ \\
    $I_\upalpha$ & $5.85 \times 10^{-11}$\\
    $I_\upbeta$ & $9.32 \times 10^{-7}$ \\
    $I_{10}$ & $4.60 \times 10^{-11}$ \\
    \hline 
    \end{tabular}
    \caption{\label{cytokinetable}Cytokine initial conditions for the model. All values are in units of $\mathrm{g/cm^3}$.}
\end{table}
\subsection{Initial Conditions for Immune Checkpoint-Associated Components in the TS}
We chose the TS immune checkpoint-associated component initial conditions to be as in \autoref{TSICItable}, with justification for the choice of these values given in Appendix B.10.
\begin{table}[H]
    \centering
    \begin{tabular}{|c|c|}
    \hline
    \textbf{Protein} & \textbf{Initial Condition} \\
    \hline 
    $P_D^{T_8}$ & $6.70 \times 10^{8}$ \\
    $P_D^{T_1}$ & $2.13 \times 10^{8}$ \\
    $P_D^{K}$ & $2.87 \times 10^{9}$ \\
    $P_L$ & $3.57 \times 10^{12}$ \\
    $Q^{T_8}$ & $5.09 \times 10^{5}$ \\
    $Q^{T_1}$ & $1.62 \times 10^{5}$ \\
    $Q^{K}$ & $2.18 \times 10^{6}$ \\
    \hline
    \end{tabular}
    \caption{\label{TSICItable}TS immune checkpoint-associated component initial conditions for the model. All values are in units of $\mathrm{molec/cm^3}$.}
\end{table}
We also set the initial condition for all pembrolizumab-associated components in the TS to be 0, as shown in \autoref{TSpembrotable}.
\begin{table}[H]
    \centering
    \begin{tabular}{|c|c|}
    \hline
    \textbf{Protein} & \textbf{Initial Condition} \\
    \hline 
    $Q_A^{T_8}$ & $0$ \\
    $Q_A^{T_1}$ & $0$ \\
    $Q_A^{K}$ & $0$ \\
    $A_1$ & $0$ \\
    \hline
    \end{tabular}
    \caption{\label{TSpembrotable}Initial conditions for pembrolizumab-associated components in the TS in the model. All values are in units of $\mathrm{molec/cm^3}$.}
\end{table}
\subsection{Initial Conditions for Immune Checkpoint-Associated Components in the TDLN}
We chose the TDLN immune checkpoint-associated component initial conditions to be as in \autoref{TDLNICItable}, with justification for the choice of these values in Appendix B.12.
\begin{table}[H]
    \centering
    \begin{tabular}{|c|c|}
    \hline
    \textbf{Protein} & \textbf{Initial Condition} \\
    \hline 
    $P_{D}^{8\mathrm{LN}}$ & $3.06 \times 10^{9}$ \\
    $P_{D}^{1\mathrm{LN}}$ & $2.07 \times 10^{10}$ \\
    $P_{L}^{\mathrm{LN}}$ & $2.10 \times 10^{11}$ \\
    $Q^{8\mathrm{LN}}$ & $1.37 \times 10^{5}$ \\
    $Q^{1\mathrm{LN}}$ & $9.23 \times 10^{5}$ \\
    \hline
    \end{tabular}
    \caption{\label{TDLNICItable}TDLN immune checkpoint-associated component initial conditions for the model. All values are in units of $\mathrm{molec/cm^3}$.}
\end{table}
We also set the initial condition for all pembrolizumab-related quantities in the TDLN to be 0, as shown in \autoref{TDLNpembrotable}.
\begin{table}[H]
    \centering
    \begin{tabular}{|c|c|}
    \hline
    \textbf{Protein} & \textbf{Initial Condition} \\
    \hline 
    $Q_A^{8\mathrm{LN}}$ & $0$ \\
    $Q_A^{1\mathrm{LN}}$ & $0$ \\
    $A_1^\mathrm{LN}$ & $0$ \\
    \hline
    \end{tabular}
    \caption{\label{TDLNpembrotable}Initial conditions for pembrolizumab-associated components in the TDLN in the model. All values are in units of $\mathrm{molec/cm^3}$.}
\end{table}
\section{Results}
\begin{figure}[H]
    \centering
    \includegraphics[width=0.95\textwidth]{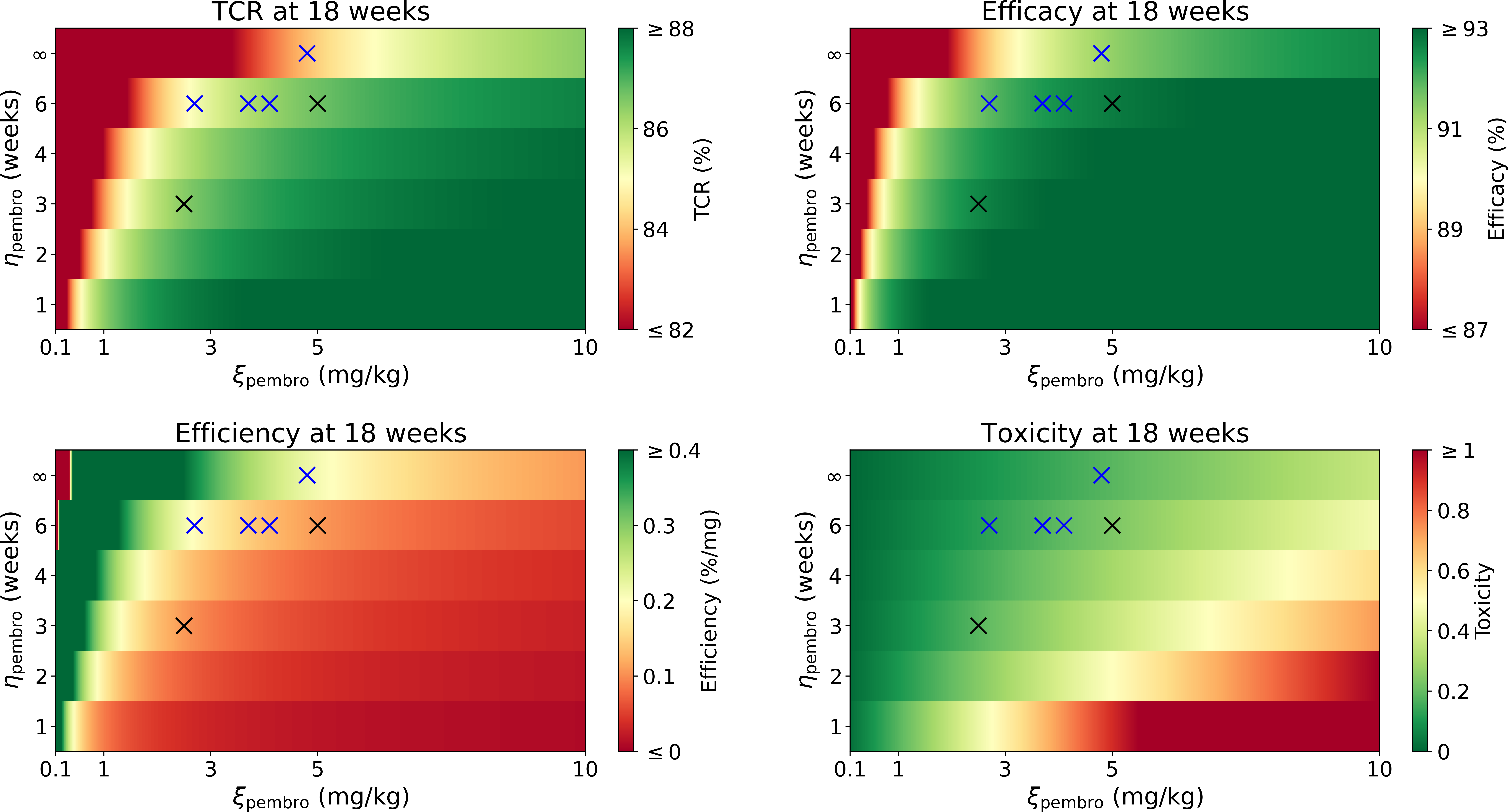}
    \caption{\label{pembroplots}TCR (top left), efficacy (top right), efficiency (bottom left), and toxicity (bottom right) at $18$ weeks for $\eta_\mathrm{pembro} \in \set{1,2,3,4,6, \infty}$ weeks. We sweep across $\xi_\mathrm{pembro} \in [0.1,10]$ mg/kg with an increment of $0.0125$ mg/kg. The FDA-approved regimens (Treatments 1 and 2) for mMCRC are shown in black, and the optimal regimens (Treatments 3--6) are shown in blue.}
\end{figure}

\begin{figure}[H]
    \centering
\includegraphics[width=0.65\textwidth]{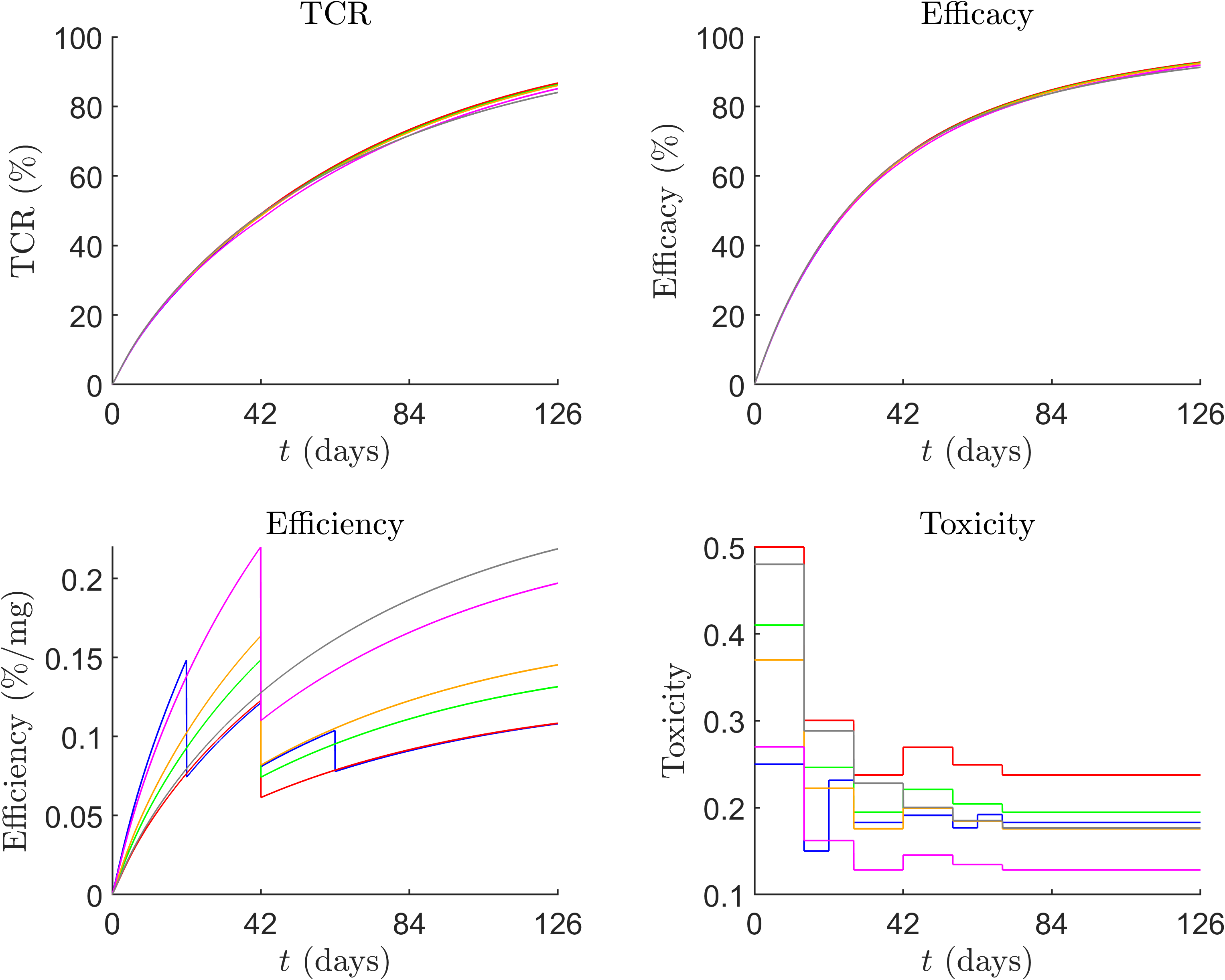}
    \caption{\label{allmetricsplot}Time traces of TCR (top left), efficacy (top right), efficiency (bottom left), and toxicity (bottom right) for Treatments 1--6 in blue, red, green, orange, magenta, and grey, respectively.}
\end{figure}

\begin{figure}[H]
    \centering
\includegraphics[width=0.65\textwidth]{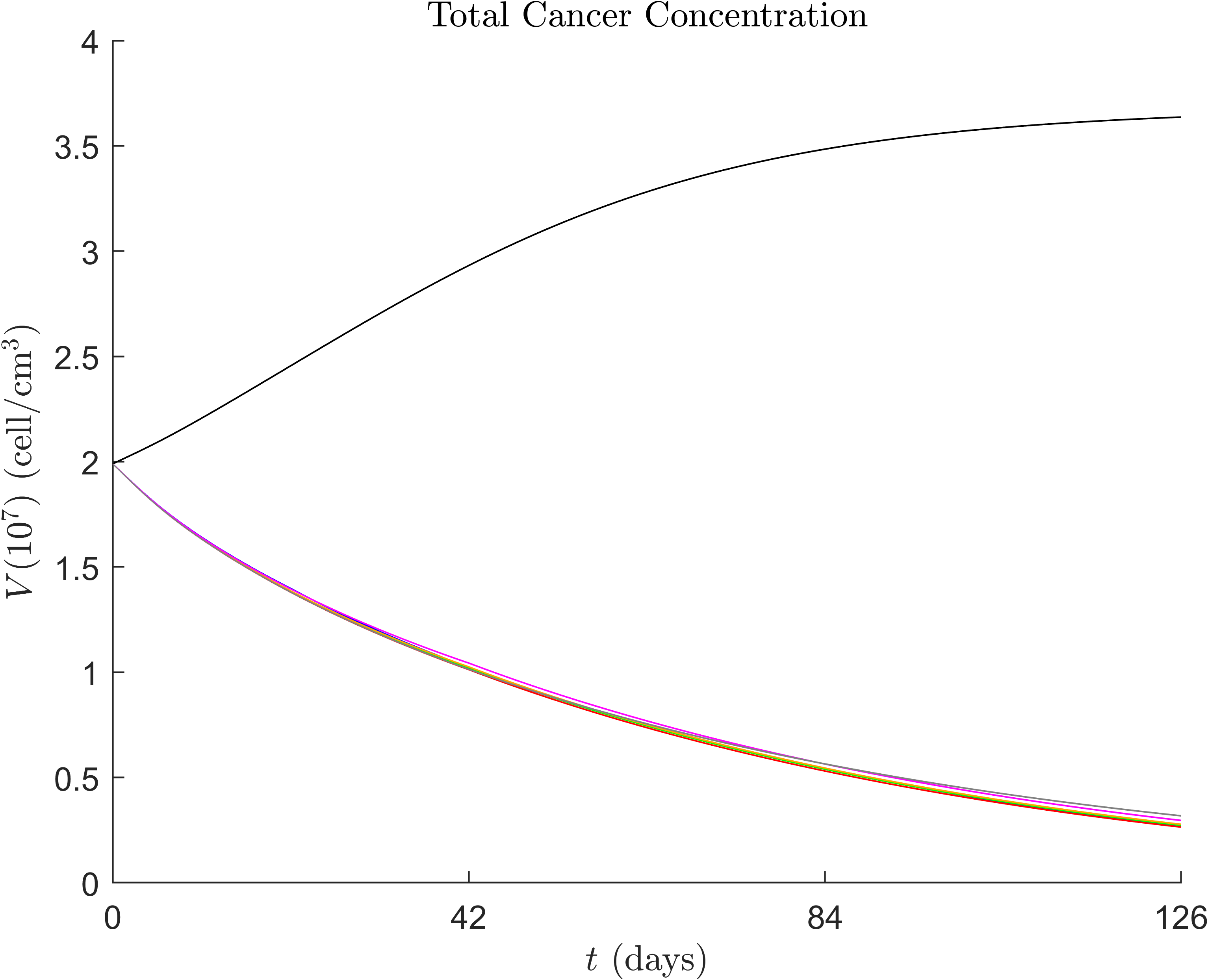}
    \caption{\label{pembrooptplot}Time traces of $V$ up to $18$ weeks from commencement, with no treatment in black, and Treatments 1--6 in blue, red, green, orange, magenta, and grey, respectively.}
\end{figure}

\begin{figure}[H]
\centering
\includegraphics[width=\textwidth]{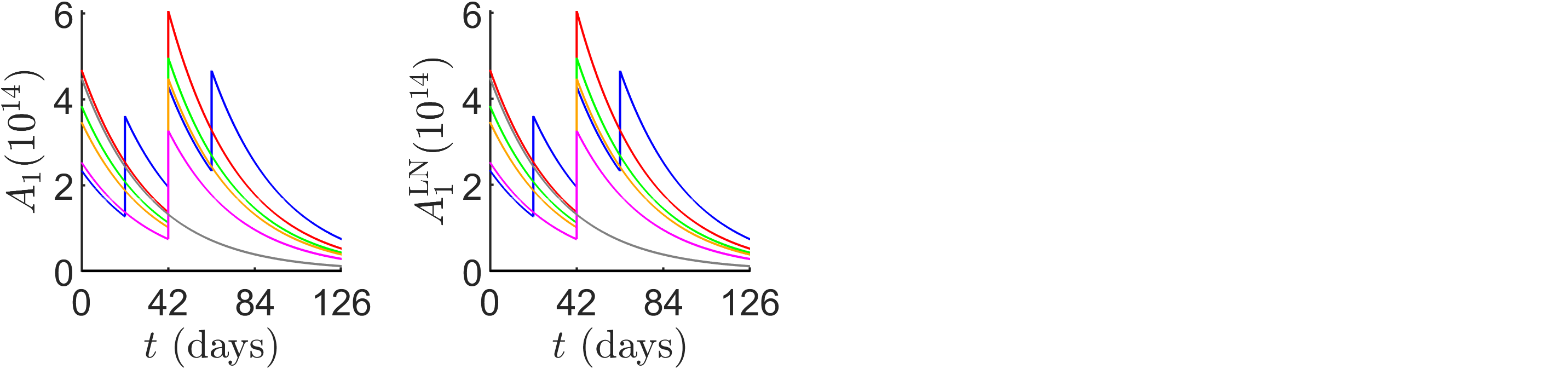}
    \caption{\label{comparisonoftreatment}Time traces of variables in the model, with the units of the variables as in \autoref{modelvars}. Time traces with no treatment are in black, and Treatments 1--6 in blue, red, green, orange, magenta, and grey, respectively.}
\end{figure}

We now aim to optimise neoadjuvant pembrolizumab therapy for laMCRC. For simplicity, we assume that pembrolizumab is given at a constant dosage, and the spacing between consecutive pembrolizumab infusions is constant. We also assume that the patient receives pembrolizumab at $t=0 \mathrm{~days}$, and we consider a treatment regimen lasting for at most $12$ weeks so that the latest allowed infusion occurs before $t=84 \mathrm{~days}$, and we simulate to $18 \mathrm{~weeks} = 126 \mathrm{~days}$. Furthermore, we assume that the patient has a mass of $m = 80$ kg. In our optimisation of pembrolizumab therapy, we consider the following four objectives: tumour concentration reduction (TCR), efficacy, efficiency, and toxicity. \\~\\
We define $V(\xi_\mathrm{pembro};\eta_\mathrm{pembro},t) = C(\xi_\mathrm{pembro};\eta_\mathrm{pembro},t) + N_c(\xi_\mathrm{pembro};\eta_\mathrm{pembro},t)$ as the total cancer concentration at time $t$ under treatment with pembrolizumab doses of $\xi_\mathrm{pembro}$ (in mg/kg) at a dosing interval of $\eta_\mathrm{pembro}$ (in weeks), omitting the $\xi_\mathrm{pembro}$ and $\eta_\mathrm{pembro}$ arguments in the case that no treatment is given. In particular, $\eta_\mathrm{pembro} = \infty$ weeks denotes a single dose of treatment, given at $t=0 \mathrm{~days}$. We define the TCR from this regimen to be
\begin{align}
    \operatorname{TCR}(\xi_\mathrm{pembro};\eta_\mathrm{pembro},t) := \frac{V(0) - V(\xi_\mathrm{pembro};\eta_\mathrm{pembro},t)}{V(0)} \times 100\%.
\intertext{We also define the efficacy similarly as}
    \operatorname{efficacy}(\xi_\mathrm{pembro};\eta_\mathrm{pembro},t) := \frac{V(t) - V(\xi_\mathrm{pembro};\eta_\mathrm{pembro},t)}{V(t)} \times 100\%.
\end{align}
In particular, the efficacy represents the extent of tumour density shrinkage throughout its growth course in comparison to no treatment, whereas the TCR reveals how much the tumour density has reduced since the commencement of treatment. We see that the TCR and efficacy are linearly related so that an increase in treatment efficacy results in increased TCR, and vice versa, via the formula
\begin{equation}
    \operatorname{efficacy}(\xi_\mathrm{pembro};\eta_\mathrm{pembro},t) = \left(1 - \frac{V(0)}{V(t)}\right) \times 100\% + \frac{V(0)}{V(t)} \times \operatorname{TCR}\left(\xi_\mathrm{pembro};\eta_\mathrm{pembro},t\right). \label{tcreff}
\end{equation}
We can also consider the efficiency of the treatment regimen, with a dosing interval of $\eta_\mathrm{pembro}$ weeks and dosage $\xi_\mathrm{pembro}$ mg/kg given by 
\begin{equation}
    \operatorname{efficiency}(\xi_\mathrm{pembro};\eta_\mathrm{pembro},t) := \frac{\operatorname{TCR}(\xi_\mathrm{pembro};\eta_\mathrm{pembro},t)}{\xi_\mathrm{pembro}m\left(\lfloor \min\left(t,84\right)/7\eta_\mathrm{pembro}\rfloor + \theta(84-t)\right)}, 
\end{equation}
where $\theta(t)$ is the Heaviside function which equals $1$ if $t\geq 0$, and $0$ otherwise. In particular, \\ $\xi_\mathrm{pembro}m\left(\lfloor \min\left(t,84\right)/7\eta_\mathrm{pembro}\rfloor + \theta(84-t)\right)$ is the total dose of pembrolizumab administered by time $t$, recalling that no treatment is given for $t \geq 84$ days. This corresponds to the ratio between the TCR percentage and the total dose of pembrolizumab administered.\\~\\
Finally, we can define the toxicity of the treatment regimen, noting that large enough pembrolizumab concentrations can potentially cause hepatotoxicity and ocular toxicity \citep{Wang2018tox, Martins2019}, as well as increase the probability of serious infections and malignancies. Experiments show that dosages of pembrolizumab between $0.1\mathrm{~mg/kg}$ and $10\mathrm{~mg/kg}$, given every 2 weeks, are safe and tolerable \citep{Robert2015, Chatterjee2016}. We thus assume that the threshold for pembrolizumab toxicity is $10\mathrm{~mg/kg}$ every 2 weeks, with higher doses being deemed toxic. To rigorise this notion, we define the toxicity of the treatment regimen, with a dosing interval of $\eta_\mathrm{pembro}$ weeks and dosage $\xi_\mathrm{pembro}$ mg/kg, as 
\begin{equation}
\operatorname{toxicity}(\xi_\mathrm{pembro};\eta_\mathrm{pembro},t) := \max\left( \frac{\underset{s \in [0,t]}{\max} \, A_1(\xi_\mathrm{pembro};\eta_\mathrm{pembro},s)}{\underset{s \in [0,t]}{\max} \, A_1(10;2,s)}, \frac{\underset{s \in [0,t]}{\max} \, A_1^\mathrm{LN}(\xi_\mathrm{pembro};\eta_\mathrm{pembro},s)}{\underset{s \in [0,t]}{\max} \, A_1^\mathrm{LN}(10;2,s)} \right).
\end{equation}
In particular, $A_1(\xi_\mathrm{pembro};\eta_\mathrm{pembro},s)$ and $A_1^\mathrm{LN}(\xi_\mathrm{pembro};\eta_\mathrm{pembro},s)$ denote the concentrations of $A_1$ and $A_1^\mathrm{LN}$ at time $s$, with pembrolizumab doses of $\xi_\mathrm{pembro}$ at a dosing interval of $\eta_\mathrm{pembro}$, respectively. In particular, the toxicity quantifies the ratio of the maximum pembrolizumab concentrations from the regimen to those of a $10\mathrm{~mg/kg}$ dose given every 2 weeks, taking the highest value of this ratio between the TDLN and TS. A toxicity greater than 1 indicates a toxic and unsafe regimen, whereas a toxicity of 1 or less signifies a non-toxic and safe regimen, with lower toxicity values corresponding to safer treatments.\\~\\
Furthermore, it is beneficial for us to use the two FDA-approved pembrolizumab regimens for the first-line treatment of mMCRC in adults as a benchmark for comparison \citep{fda_pembrolizumab}:
\begin{itemize}
    \item Treatment 1: 200 mg of pembrolizumab administered by intravenous infusion over a duration of $30$ minutes every 3 weeks until disease progression or unacceptable toxicity.
    \item Treatment 2: 400 mg of pembrolizumab administered by intravenous infusion over a duration of $30$ minutes every 6 weeks until disease progression or unacceptable toxicity.
\end{itemize}
These correspond to the following parameter values in the model:
\begin{itemize}
    \item Treatment 1: $\xi_{j} = 200 \mathrm{~mg}$, $t_j = 21(j-1)$, $n = 4$, $\xi_\mathrm{pembro} = 2.5 ~\mathrm{mg/kg}$, $\eta_\mathrm{pembro} = 3$ weeks,
    \item Treatment 2: $\xi_{j} = 400 \mathrm{~mg}$, $t_j = 42(j-1)$, $n = 2$, $\xi_\mathrm{pembro} = 5 ~\mathrm{mg/kg}$, $\eta_\mathrm{pembro} = 6$ weeks.
\end{itemize}
Denoting the dosing interval of pembrolizumab as $\eta_\mathrm{pembro}$, we perform a sweep across the space $\eta_\mathrm{pembro} \in \set{1,2,3,4,6, \infty}$ weeks. The finite values are integer factors of $12$ weeks, and each $\eta_\mathrm{pembro}$ corresponds to a distinct number of doses administered. This approach ensures practicality whilst preventing any artefacts that could occur from selecting a treatment regimen that ends at a fixed time of 18 weeks. Taking practicality constraints into account, we consider linearly spaced dosages in the domain $\xi_\mathrm{pembro} \in [0.1, 10]$ mg/kg, with a spacing of $0.0125$ mg/kg. This corresponds to $\xi_j \in [0.1m, 10m] \mathrm{~mg} = [8, 800] \mathrm{~mg}$ with an increment of $1 \mathrm{~mg}$.\\~\\
We can determine the optimal pembrolizumab therapy by considering the regimen that achieves an acceptable TCR at $18$ weeks whilst maximising treatment efficiency as much as possible and ensuring a toxicity of less than 1. The TCRs of Treatment 1 and Treatment 2 at $18$ weeks were calculated to be $86.47\%$ and $86.71\%$, respectively. As such, to ensure that the TCR of the optimal treatment is comparable to current FDA-approved pembrolizumab regimens, we consider threshold TCRs of $86.25\%$, $86\%$, and $85\%$. We also consider constraints due to practicality, so that $\xi_\mathrm{pembro}$ is an integer multiple of $0.1$ mg/kg, corresponding to an integer multiple of $8\mathrm{~mg}$, leaving the domain for $\eta_\mathrm{pembro}$ unchanged. Denoting the space of $(\xi_\mathrm{pembro},\eta_\mathrm{pembro})$ pairs that satisfy these criteria as $\mathcal{S}^\mathrm{prac}$, the optimal pembrolizumab dosing and spacing, denoted $\xi_\mathrm{pembro}^\mathrm{opt}$ and $\eta_\mathrm{pembro}^\mathrm{opt}$, respectively, for a given threshold TCR $\mathcal{T}_\mathrm{thresh}$, satisfy
\begin{equation}
\left(\xi_\mathrm{pembro}^\mathrm{opt},\eta_\mathrm{pembro}^\mathrm{opt} \right) = \underset{\substack{\operatorname{TCR}(\xi_\mathrm{pembro};\eta_\mathrm{pembro},126) \geq \mathcal{T}_\mathrm{thresh}\\ (\xi_\mathrm{pembro},\eta_\mathrm{pembro}) \in \mathcal{S}^\mathrm{prac} \\ \operatorname{toxicity}(\xi_\mathrm{pembro};\eta_\mathrm{pembro},126)\leq 1}}{\operatorname{argmax}} \ \operatorname{efficiency}\left(\xi_\mathrm{pembro}; \eta_\mathrm{pembro}, 126\right). \label{pembroopteqn}
\end{equation}
Solutions of \eqref{pembroopteqn} with the previously given threshold TCRs compared to Treatments 1 and 2 are shown in \autoref{opttable}. Noting that neoadjuvant treatment consisting of just a single dose of pembrolizumab is significantly more convenient and cost-efficient in a treatment setting, we also consider the optimal single-dose treatment regimen using a threshold TCR of $\mathcal{T}_\mathrm{thresh} = 84\%$, denoting this as Treatment 6.
\begin{table}[H]
    \centering
    \resizebox{\columnwidth}{!}{%
    \begin{tabular}{|c|c|c|c|c|c|c|c|c|c|}
    \hline
    & \textbf{Tx} & $\mathcal{T}_\mathrm{thresh}$& $\xi_\mathrm{pembro}^\mathrm{opt}$ & \textbf{Dosage} & \textbf{Spacing} & \textbf{TCR} & \textbf{Efficacy} & \textbf{Efficiency} & \textbf{Toxicity} \\
    & \textbf{Num.} & \textbf{(\%)} & \textbf{(mg/kg)} & \textbf{(mg)} & \textbf{(weeks)} & \textbf{(\%)} & \textbf{(\%)} & \textbf{(\%/mg)} &\\
    \hline 
    \multirow{2}{*}{\rotatebox[origin=c]{90}{\textbf{FDA}}} & 1 & --- & --- & 200 & 3 & 86.47 & 92.60 & $1.08 \times 10^{-1}$ & $1.83 \times 10^{-1}$ \\
    & 2 & --- & --- & 400 & 6 & 86.71 & 92.73 & $1.08 \times 10^{-1}$ & $2.37 \times 10^{-1}$ \\
    \hline
    \multirow{4}{*}{\rotatebox[origin=c]{90}{\textbf{Optimal}}} & 3 & 86.25 & 4.1 & 328 & 6 & 86.30 & 92.51 & $1.32 \times 10^{-1}$ & $1.95 \times 10^{-1}$ \\
    & 4 & 86 & 3.7 & 296 & 6 & 86.05 & 92.37 & $1.45 \times 10^{-1}$ & $1.76 \times 10^{-1}$ \\
    & 5 & 85 & 2.7 & 216 & 6 & 85.12 & 91.86 & $1.97 \times 10^{-1}$ & $1.28 \times 10^{-1}$ \\
    & 6 & 84 & 4.8 & 384 & $\infty$ & 84.02 & 91.26 & $2.19 \times 10^{-1}$ & $1.76 \times 10^{-1}$ \\
    \hline
    \end{tabular}
    \caption{\label{opttable}Comparison of $\xi_\mathrm{pembro}^\mathrm{opt}$, dosage, spacing ($\eta_\mathrm{pembro}^\mathrm{opt}$), TCR, efficacy, efficiency, and toxicity between FDA-approved regimens for mMCRC and optimal treatment regimens for various $\mathcal{T}_\mathrm{thresh}$, assuming a patient mass of 80 kg. Tx No.\ denotes the treatment number, with FDA-approved therapies for mMCRC labelled as Treatments 1 and 2, and optimal regimens labelled as Treatments 3--6.} %
    }
\end{table}
Heatmaps of TCR, efficacy, efficiency, and toxicity at $t = 18$ weeks for various $\eta_\mathrm{pembro}$ and $\xi_\mathrm{pembro}$ values are shown in \autoref{pembroplots}. All simulations were done in MATLAB using the dde23 solver with the initial conditions stated in \Cref{initcond section}.
Time traces of TCR, efficacy, efficiency, and toxicity for Treatments 1--6 are shown in \autoref{allmetricsplot}. 
Time traces for the total cancer concentration, $V$, with Treatments 1--6 compared to no treatment are shown in \autoref{pembrooptplot}. 
We can also compare the effects of optimal pembrolizumab therapies and FDA-approved regimens to those of no treatment on the TME, with time traces of model variables shown in \autoref{comparisonoftreatment} and immune cell and cytokine concentrations at 18 weeks shown in \autoref{finalconcentrationstable}.
\begin{figure}[H]
    \centering
    \includegraphics[width=\textwidth]{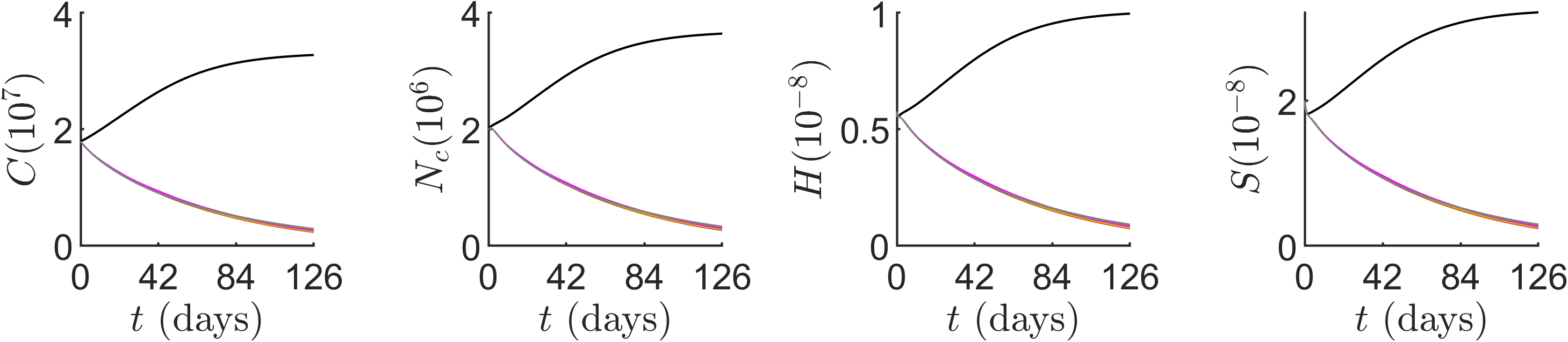}
\end{figure}
\begin{figure}[H]
    \centering
    \includegraphics[width=\textwidth]{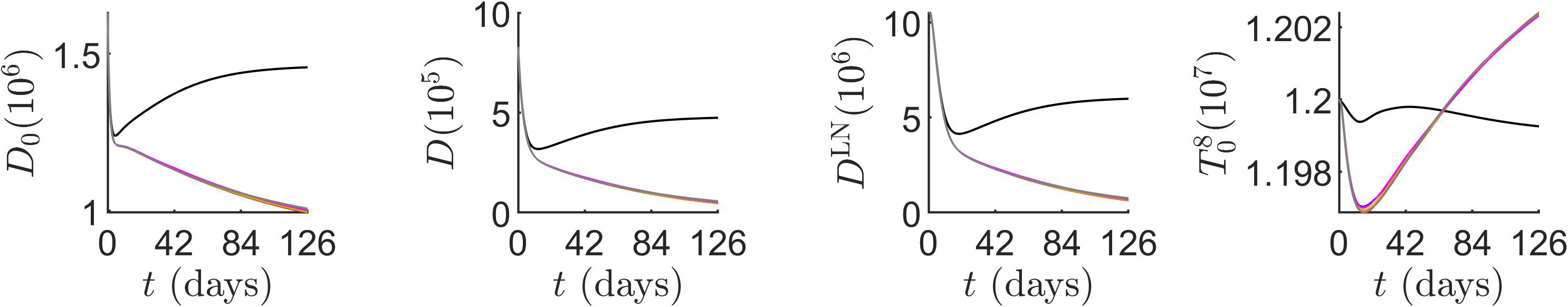}
\end{figure}
\begin{figure}[H]
    \centering
    \includegraphics[width=\textwidth]{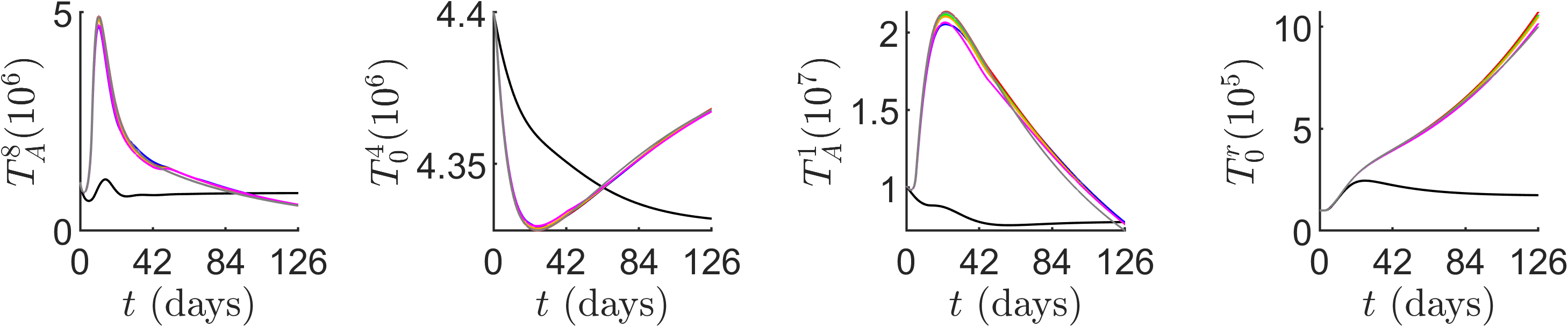}
\end{figure}
\begin{figure}[H]
    \centering
    \includegraphics[width=\textwidth]{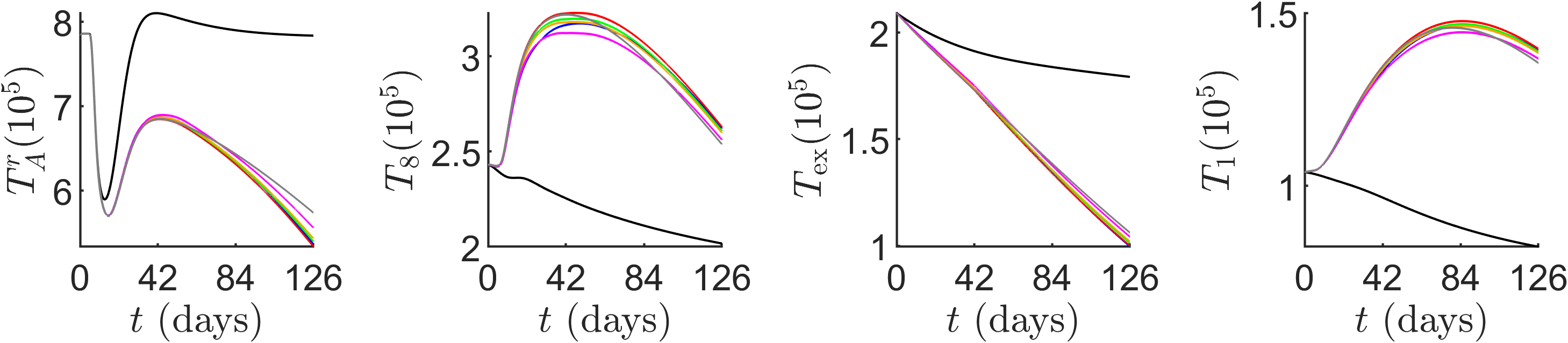}
\end{figure}
\begin{figure}[H]
    \centering
    \includegraphics[width=\textwidth]{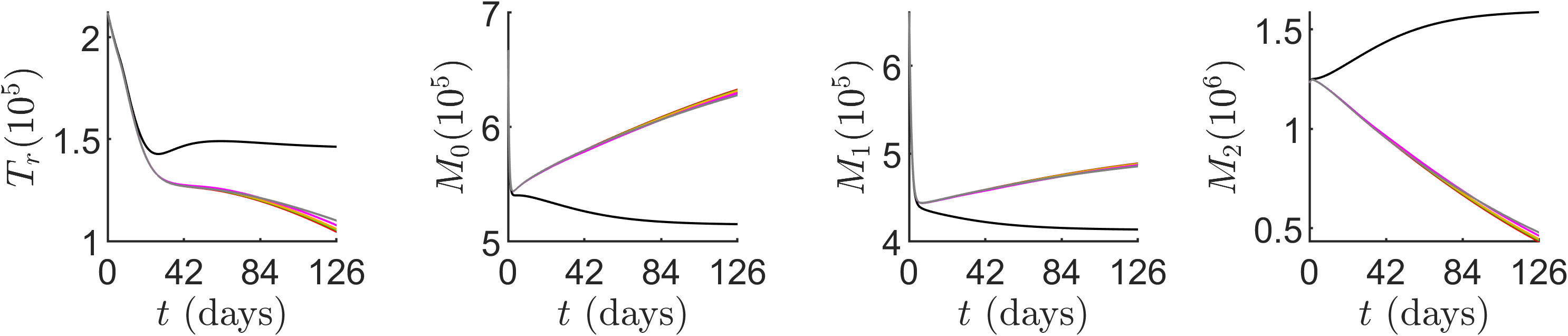}
\end{figure}
\begin{figure}[H]
    \centering
    \includegraphics[width=\textwidth]{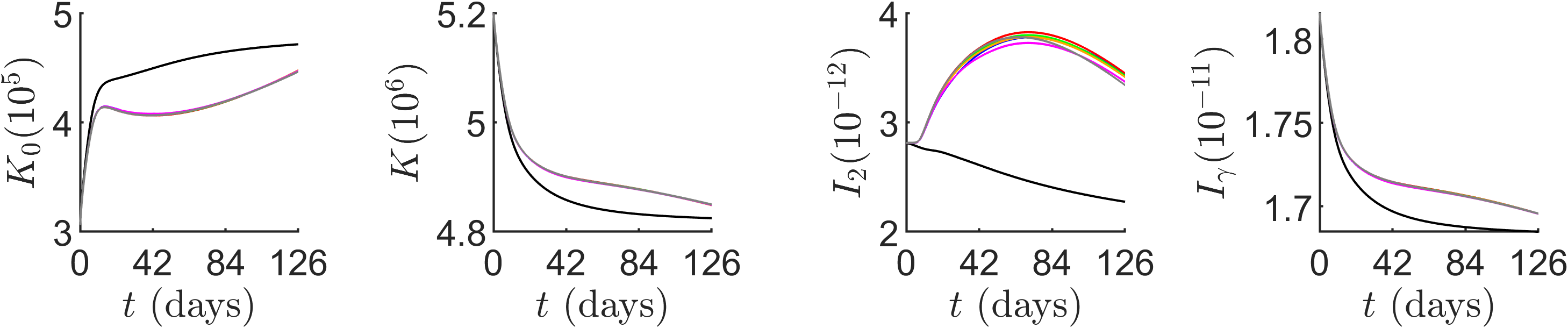}
\end{figure}
\begin{figure}[H]
    \centering
    \includegraphics[width=\textwidth]{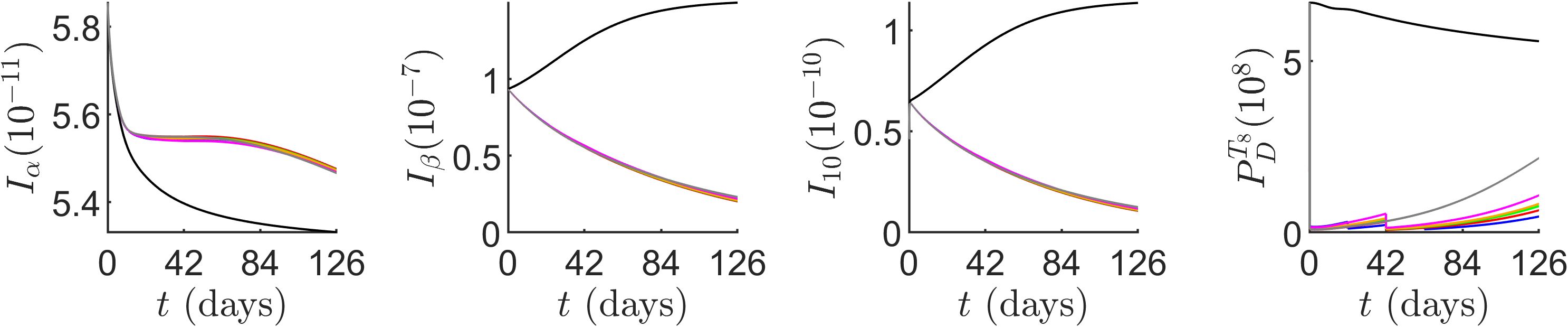}
\end{figure}
\begin{figure}[H]
    \centering
    \includegraphics[width=\textwidth]{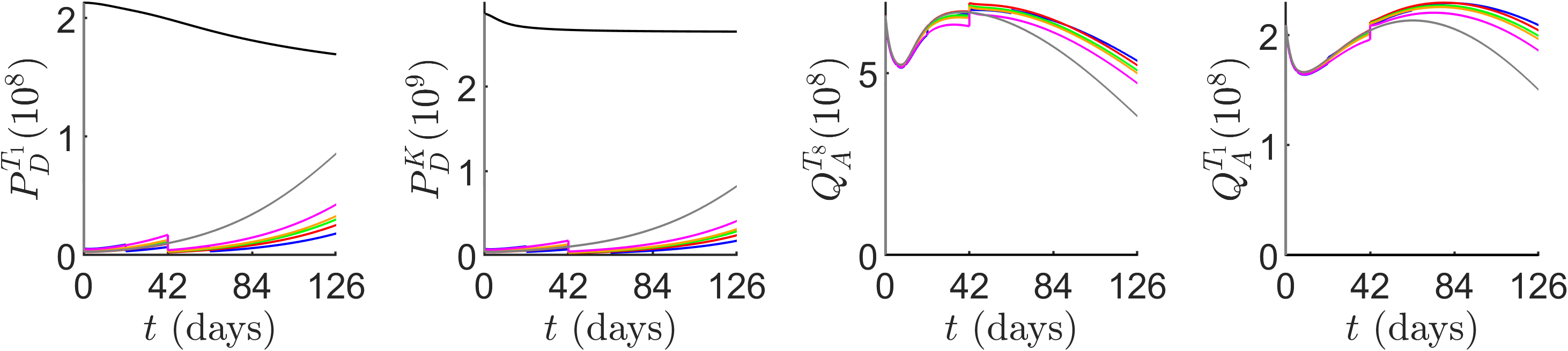}
\end{figure}
\begin{figure}[H]
    \centering
    \includegraphics[width=\textwidth]{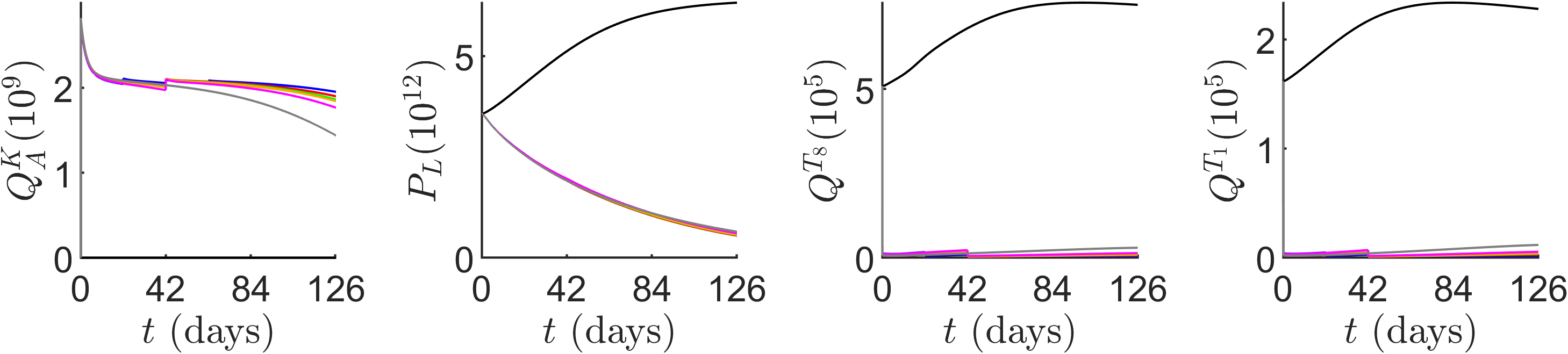}
\end{figure}
\begin{figure}[H]
    \centering
    \includegraphics[width=\textwidth]{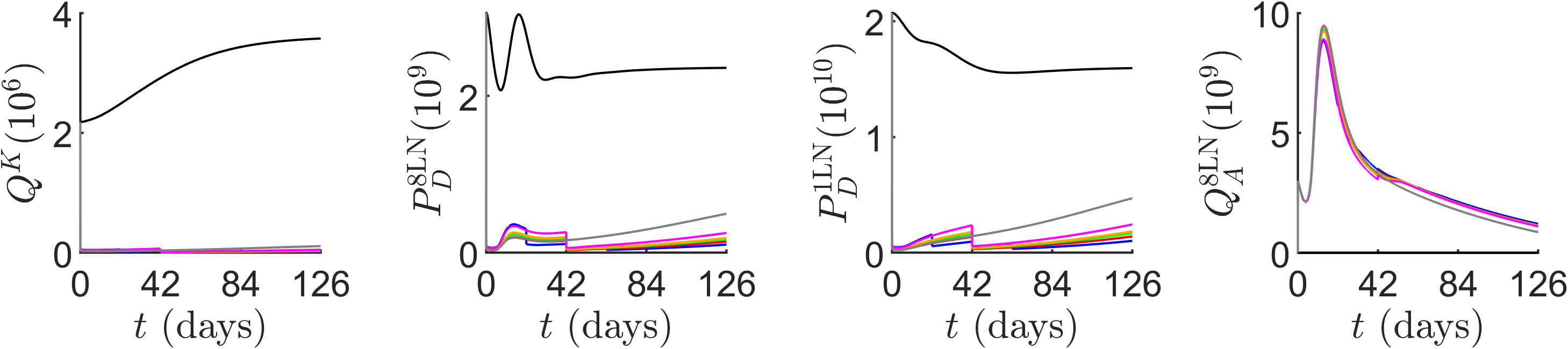}
\end{figure}
\begin{figure}[H]
    \centering
    \includegraphics[width=\textwidth]{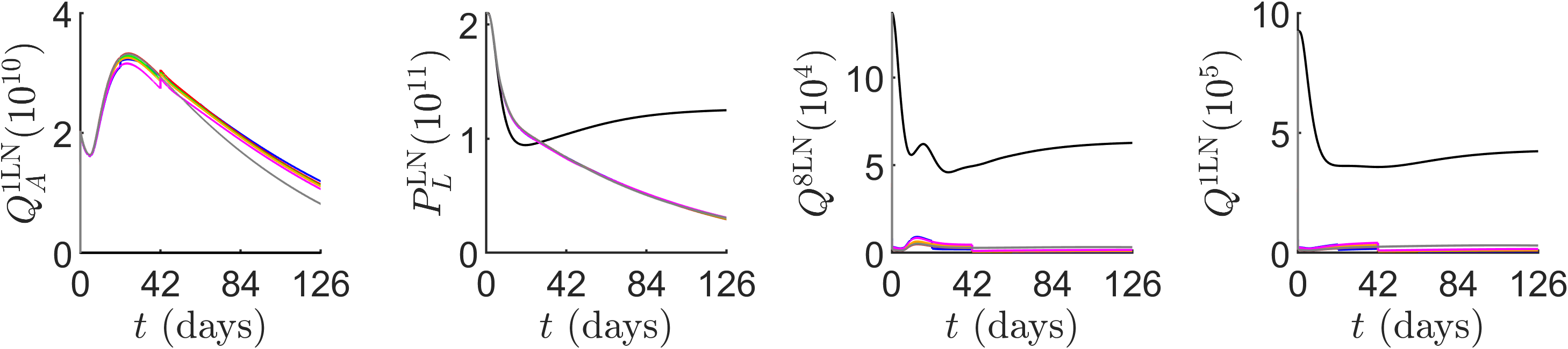}
\end{figure}
\begin{table}[ht]
    \centering
    \resizebox{\columnwidth}{!}{%
    \begin{tabular}{|c|c|c|c|c|c|c|c|}
    \hline
    & \textbf{No Tx} & \textbf{Treatment} & \textbf{Treatment} & \textbf{Treatment} & \textbf{Treatment} & \textbf{Treatment} & \textbf{Treatment} \\
    & & \textbf{1} & \textbf{2} & \textbf{3} & \textbf{4} & \textbf{5} & \textbf{6} \\
    \hline
$\bm{C}$ & $3.27 \times 10^{7}$ & $2.41 \times 10^{6}$ & $2.37 \times 10^{6}$ & $2.44 \times 10^{6}$ & $2.49 \times 10^{6}$ & $2.65 \times 10^{6}$ & $2.85 \times 10^{6}$ \\
$\bm{N_c}$ & $3.64 \times 10^{6}$ & $2.79 \times 10^{5}$ & $2.74 \times 10^{5}$ & $2.82 \times 10^{5}$ & $2.87 \times 10^{5}$ & $3.06 \times 10^{5}$ & $3.27 \times 10^{5}$ \\
$\bm{H}$ & $9.96 \times 10^{-9}$ & $7.66 \times 10^{-10}$ & $7.52 \times 10^{-10}$ & $7.75 \times 10^{-10}$ & $7.88 \times 10^{-10}$ & $8.39 \times 10^{-10}$ & $8.98 \times 10^{-10}$ \\
$\bm{S}$ & $3.22 \times 10^{-8}$ & $2.50 \times 10^{-9}$ & $2.45 \times 10^{-9}$ & $2.53 \times 10^{-9}$ & $2.57 \times 10^{-9}$ & $2.73 \times 10^{-9}$ & $2.92 \times 10^{-9}$ \\
$\bm{D_0}$ & $1.46 \times 10^{6}$ & $9.96 \times 10^{5}$ & $9.95 \times 10^{5}$ & $9.97 \times 10^{5}$ & $9.98 \times 10^{5}$ & $1.00 \times 10^{6}$ & $1.01 \times 10^{6}$ \\
$\bm{D}$ & $4.74 \times 10^{5}$ & $4.86 \times 10^{4}$ & $4.77 \times 10^{4}$ & $4.91 \times 10^{4}$ & $4.99 \times 10^{4}$ & $5.30 \times 10^{4}$ & $5.63 \times 10^{4}$ \\
$\bm{D^\mathrm{LN}}$ & $5.98 \times 10^{6}$ & $6.58 \times 10^{5}$ & $6.45 \times 10^{5}$ & $6.63 \times 10^{5}$ & $6.73 \times 10^{5}$ & $7.12 \times 10^{5}$ & $7.51 \times 10^{5}$ \\
$\bm{T_0^8}$ & $1.20 \times 10^{7}$ & $1.20 \times 10^{7}$ & $1.20 \times 10^{7}$ & $1.20 \times 10^{7}$ & $1.20 \times 10^{7}$ & $1.20 \times 10^{7}$ & $1.20 \times 10^{7}$ \\
$\bm{T_A^8}$ & $8.55 \times 10^{5}$ & $5.82 \times 10^{5}$ & $5.69 \times 10^{5}$ & $5.74 \times 10^{5}$ & $5.77 \times 10^{5}$ & $5.88 \times 10^{5}$ & $5.68 \times 10^{5}$ \\
$\bm{T_8}$ & $2.02 \times 10^{5}$ & $2.62 \times 10^{5}$ & $2.63 \times 10^{5}$ & $2.61 \times 10^{5}$ & $2.60 \times 10^{5}$ & $2.56 \times 10^{5}$ & $2.54 \times 10^{5}$ \\
$\bm{T_{\mathrm{ex}}}$ & $1.79 \times 10^{5}$ & $9.97 \times 10^{4}$ & $1.00 \times 10^{5}$ & $1.01 \times 10^{5}$ & $1.02 \times 10^{5}$ & $1.04 \times 10^{5}$ & $1.06 \times 10^{5}$ \\
$\bm{T_0^4}$ & $4.33 \times 10^{6}$ & $4.37 \times 10^{6}$ & $4.37 \times 10^{6}$ & $4.37 \times 10^{6}$ & $4.37 \times 10^{6}$ & $4.37 \times 10^{6}$ & $4.37 \times 10^{6}$ \\
$\bm{T_A^1}$ & $7.77 \times 10^{6}$ & $7.75 \times 10^{6}$ & $7.62 \times 10^{6}$ & $7.62 \times 10^{6}$ & $7.63 \times 10^{6}$ & $7.64 \times 10^{6}$ & $7.21 \times 10^{6}$ \\
$\bm{T_1}$ & $8.24 \times 10^{4}$ & $1.39 \times 10^{5}$ & $1.40 \times 10^{5}$ & $1.39 \times 10^{5}$ & $1.38 \times 10^{5}$ & $1.37 \times 10^{5}$ & $1.36 \times 10^{5}$ \\
$\bm{T_0^r}$ & $1.74 \times 10^{5}$ & $1.06 \times 10^{6}$ & $1.07 \times 10^{6}$ & $1.06 \times 10^{6}$ & $1.05 \times 10^{6}$ & $1.02 \times 10^{6}$ & $1.00 \times 10^{6}$ \\
$\bm{T_A^r}$ & $7.84 \times 10^{5}$ & $5.36 \times 10^{5}$ & $5.34 \times 10^{5}$ & $5.40 \times 10^{5}$ & $5.43 \times 10^{5}$ & $5.56 \times 10^{5}$ & $5.74 \times 10^{5}$ \\
$\bm{T_r}$ & $1.46 \times 10^{5}$ & $1.05 \times 10^{5}$ & $1.05 \times 10^{5}$ & $1.06 \times 10^{5}$ & $1.06 \times 10^{5}$ & $1.08 \times 10^{5}$ & $1.10 \times 10^{5}$ \\
$\bm{M_0}$ & $5.15 \times 10^{5}$ & $6.32 \times 10^{5}$ & $6.32 \times 10^{5}$ & $6.32 \times 10^{5}$ & $6.31 \times 10^{5}$ & $6.30 \times 10^{5}$ & $6.28 \times 10^{5}$ \\
$\bm{M_1}$ & $4.14 \times 10^{5}$ & $4.88 \times 10^{5}$ & $4.89 \times 10^{5}$ & $4.88 \times 10^{5}$ & $4.88 \times 10^{5}$ & $4.87 \times 10^{5}$ & $4.86 \times 10^{5}$ \\
$\bm{M_2}$ & $1.59 \times 10^{6}$ & $4.39 \times 10^{5}$ & $4.33 \times 10^{5}$ & $4.41 \times 10^{5}$ & $4.46 \times 10^{5}$ & $4.63 \times 10^{5}$ & $4.79 \times 10^{5}$ \\
$\bm{K_0}$ & $4.72 \times 10^{5}$ & $4.47 \times 10^{5}$ & $4.48 \times 10^{5}$ & $4.48 \times 10^{5}$ & $4.47 \times 10^{5}$ & $4.47 \times 10^{5}$ & $4.46 \times 10^{5}$ \\
$\bm{K}$ & $4.82 \times 10^{6}$ & $4.85 \times 10^{6}$ & $4.85 \times 10^{6}$ & $4.85 \times 10^{6}$ & $4.85 \times 10^{6}$ & $4.85 \times 10^{6}$ & $4.85 \times 10^{6}$ \\
$\bm{I_2}$ & $2.27 \times 10^{-12}$ & $3.44 \times 10^{-12}$ & $3.45 \times 10^{-12}$ & $3.43 \times 10^{-12}$ & $3.41 \times 10^{-12}$ & $3.37 \times 10^{-12}$ & $3.34 \times 10^{-12}$ \\
$\bm{I_\upgamma}$ & $1.68 \times 10^{-11}$ & $1.70 \times 10^{-11}$ & $1.70 \times 10^{-11}$ & $1.70 \times 10^{-11}$ & $1.70 \times 10^{-11}$ & $1.70 \times 10^{-11}$ & $1.70 \times 10^{-11}$ \\
$\bm{I_\upalpha}$ & $5.33 \times 10^{-11}$ & $5.47 \times 10^{-11}$ & $5.48 \times 10^{-11}$ & $5.47 \times 10^{-11}$ & $5.47 \times 10^{-11}$ & $5.47 \times 10^{-11}$ & $5.47 \times 10^{-11}$ \\
$\bm{I_\upbeta}$ & $1.50 \times 10^{-6}$ & $2.06 \times 10^{-7}$ & $2.03 \times 10^{-7}$ & $2.07 \times 10^{-7}$ & $2.10 \times 10^{-7}$ & $2.20 \times 10^{-7}$ & $2.31 \times 10^{-7}$ \\
$\bm{I_{10}}$ & $1.14 \times 10^{-10}$ & $1.08 \times 10^{-11}$ & $1.06 \times 10^{-11}$ & $1.09 \times 10^{-11}$ & $1.10 \times 10^{-11}$ & $1.17 \times 10^{-11}$ & $1.24 \times 10^{-11}$ \\
\hline
\end{tabular}
\caption{\label{finalconcentrationstable}Comparison of final immune cell and cytokine concentrations at 18 weeks between no treatment and Treatments 1--6. Units of variables are as in \autoref{modelvars}.}%
}
\end{table}
The results from \autoref{pembroplots}, \autoref{allmetricsplot}, \autoref{pembrooptplot}, \autoref{comparisonoftreatment}, and \autoref{finalconcentrationstable} will be discussed in detail in \Cref{discussionsection}.
\section{Discussion\label{discussionsection}}
We can see from \autoref{allmetricsplot}, \autoref{pembrooptplot}, and \autoref{comparisonoftreatment} that Treatments 1--6 are highly effective in eradicating cancer cells, with TCRs of approximately 84.0\%--86.7\% at 18 weeks. However, we must note that we take into account that laMCRC patients will not have been treated with chemotherapy/other therapies and that these treatments are given as first-line treatments. We observe that higher doses at larger intervals are comparable to smaller doses at shorter intervals, which is consistent with clinical and experimental observations for other cancers \citep{Simeone2020, DubPelletier2023, Lala2020}. It is difficult to compare our results to those of pre-existing clinical trials for laMCRC due to the lack of time-series data, widely varying treatment regimens tested, and the broad range of outcomes found. Focusing on Treatment 2, which appears to be the primary focus of ongoing clinical trials, a TCR of $86.71\%$ at 18 weeks, following the cessation of treatment at 12 weeks, is consistent with the extent of response observed. Therefore, we consider the model to be accurate; however, additional experimental data is needed for further verification. We also observed that slight variations in the initial conditions had minimal impact on model trajectories after a few days and that reasonable choices of initial conditions did not lead to negativity for any model variables (not shown). \\~\\
Furthermore, analysing immune cell trajectories from \autoref{comparisonoftreatment} offers potential explanations for behaviour in the TME and identifies key factors that contribute to maximising cancer reduction. One of the most important observations with pembrolizumab therapy is that the concentration of activated and effector pro-inflammatory immune cells significantly increases. Notably, the concentrations of effector CD8+ T cells, effector Th1 cells, and M1 macrophages increase by approximately 28.5\%, 68\%, and 18\% by 18 weeks compared to no treatment, respectively. In particular, this leads to enhanced tumour cell lysis and increased production of pro-inflammatory cytokines, which drives macrophage polarisation into the pro-inflammatory M1 phenotype, resulting in a positive feedback loop. \\~\\
Of particular note is the increase in TNF and IFN-$\upgamma$ concentrations, since they directly induce necroptosis of cancer cells, causing the release of DAMPs, which in turn induce DC maturation and T cell activation. However, we observe that as treatment progresses, the tumour burden decreases, leading to a decrease in the magnitude of necrotic cancer cells and decreased DAMP release and DC maturation. Consequently, T cell activation decreases, explaining the gradual decrease in the concentrations of effector T cells in the TS and TDLN after a couple of months. Nonetheless, the concentration of effector and activated pro-inflammatory T cells remains significantly higher than without treatment. \\~\\
Similarly, there is a significant decrease in the concentration of activated and effector anti-inflammatory cells, including Tregs and M2 macrophages, which decrease by approximately 27\% and 72\%, respectively, by 18 weeks. Decreased effector Treg concentration leads to decreased inhibition of pro-inflammatory T cell activation and proliferation, reduced suppression of IFN-$\upgamma$ production, and decreased inhibition of IL-2-mediated pro-inflammatory T cell growth in the TS. This also results in decreased concentrations of anti-inflammatory cytokines, including IL-10 and TGF-$\upbeta$, which are reduced by 18 weeks by approximately 90\% and 86\%, respectively, compared to no treatment. As a result, there is decreased polarisation of macrophages to the M2 phenotype, further decreasing anti-inflammatory cytokine production and reinforcing a positive feedback cycle. \\~\\
A particularly potent positive feedback cycle occurs with respect to TGF-$\upbeta$. As cancer cells are eliminated, and the concentrations of M2 macrophages and effector Tregs decrease, TGF-$\upbeta$ concentrations decrease, leading to reduced inhibition of cancer cell lysis by NK cells and effector CD8+ T cells, diminished suppression of NK cell activation, and reduced M2 macrophage polarisation. This further lowers the number of viable cancer cells, Tregs, and M2 macrophages, perpetuating the decline in TGF-$\upbeta$ concentration and amplifying the anti-tumour response.\\~\\
Likewise, the rise in activated pro-inflammatory immune cells leads to an increased concentration of IL-2, a key growth factor for effector CD8+ and Th1 cells and an activator of resting NK cells. With pembrolizumab therapy, the concentration of IL-2 increases by approximately 50\% by 18 weeks, further promoting the expansion of activated NK cells, as well as effector Th1 and CD8+ T cells. This, in turn, enhances IL-2 production, driving further Th1 and CD8+ T cell proliferation, creating yet another positive feedback cycle.\\~\\
Furthermore, pembrolizumab therapy leads to an approximately 31.5\% decrease in immature dendritic cells (DCs) and an 89\% decrease in mature DCs by 18 weeks, compared to no treatment. This reduction in mature DCs is driven by a lower tumour burden, resulting in decreased cancer cell necrosis and DAMP release. These findings align with clinical evidence linking an increased presence of immature DCs to a higher risk of metastasis and poorer prognosis in colorectal cancer (CRC) \citep{Gulubova2011}, and make sense since decreased DAMP release in the presence of high TNF and IFN-$\upgamma$ concentrations implies a low concentration of necrotic cancer cells and thus cancer cells overall. Additionally, pembrolizumab treatment significantly increases the M1/M2 macrophage ratio, reaching approximately 1.06 by 18 weeks compared to 0.26 without treatment. This is consistent with clinical findings, which show that higher M1/M2 macrophage ratios are associated with improved survival in CRC \citep{Vyrynen2021}, as expected.\\~\\
Another key observation is the substantial decrease in exhausted CD8+ T cells with pembrolizumab treatment, declining by approximately 43\% by 18 weeks compared to no treatment. This reduction, in conjunction with the increased concentration of effector CD8+ T cells, showcases two important findings: (a) pembrolizumab increases the concentration of cytotoxic CD8+ T cells through reinvigorating exhausted CD8+ T cells, and (b) the concentration of exhausted CD8+ T cells plays a major role in treatment efficacy. \\~\\
We can also analyse the impact of pembrolizumab therapy on the concentrations of PD-1, the PD-1/pembrolizumab complex, and the PD-1/PD-L1 complex in the TS and the TDLN. As expected, pembrolizumab therapy significantly reduces the concentration of unbound PD-1 receptors on PD-1-expressing cells in both the TS and TDLN, decreasing by approximately 95\% at trough and 98\% at peak. The concentration of the PD-1/PD-L1 complex on cells in the TS and TDLN also decreases by approximately 99\% throughout treatment, as nearly all PD-1 receptors are bound to pembrolizumab as part of the PD-1/pembrolizumab complex. This is also due, in part, to a reduction in the concentration of M2 macrophages and cancer cells during treatment, leading to a significant reduction in PD-L1 concentration, which decreases by approximately 91\% by 18 weeks compared to no treatment. Consequently, there is enhanced lysis of cancer cells by effector CD8+ T cells and activated NK cells, reduced inhibition of pro-inflammatory T cell proliferation and activation, and a decreased number of activated and effector Tregs. Thus, we see that treatment efficacy and success are directly correlated with the extent of PD-1 receptor engagement and reduction in PD-1/PD-L1 complex concentration. \\~\\
However, we note that PD-1 receptor engagement by pembrolizumab saturates at low doses, with the KEYNOTE-001 study finding that 2 mg/kg of pembrolizumab is sufficient to saturate unbound PD-1 receptors and achieve maximum anti-tumour activity \citep{Patnaik2015}. As a result, the optimal dosing regimens are more efficient and exhibit lower overall dosing than the FDA-approved regimens for metastatic MSI-H/dMMR CRC while still achieving comparable efficacy and TCR.\\~\\
It is also beneficial for us to compare and analyse the time traces of TCR, efficacy, efficiency, and toxicity of Treatments 1 and 2, and the optimal therapies as shown in \autoref{allmetricsplot}. As expected, the TCRs and efficacies of Treatments 1 and 2 are similar to those of Treatments 3--6 throughout the treatment period, with TCR and efficacy being monotonically increasing functions of time. Due to the lower dosages and larger intervals of the optimal regimen, the optimal treatments are significantly more efficient than Treatments 1 and 2. In particular, Treatment 6, which consists of a single dose of pembrolizumab, becomes the most efficient after $t = 42$ days, as no further pembrolizumab is administered beyond this point. However, before the administration of a second dose, Treatments 3--5 exhibit greater efficiency due to their smaller yet still efficacious doses, while Treatment 6 becomes increasingly efficient as the treatment progresses. Finally, as expected, the toxicity of all treatments is generally a non-increasing function of time, but if the pembrolizumab concentration is sufficiently high, small spikes in toxicity may occur following dose administration.\\~\\
We now shift our focus to \autoref{pembroplots}. We see that TCR increases as the dosing increases and spacing decreases, though with diminishing returns at higher doses or shorter intervals. In particular, the TCRs and efficacies of all optimal regimens are high, with minimal deviations near these regions.\\~\\
In the spirit of completeness, we verify that Treatments 1 and 2 are non-toxic and compare their toxicity to that of the optimal regimens found. As expected, Treatments 1 and 2 are non-toxic, with toxicities of $1.83 \times 10^{-1}$ and $2.37 \times 10^{-1}$, respectively, whilst the optimal regimens have lower or comparable toxicity. Treatment 5, consisting of 216 mg of pembrolizumab administered every 6 weeks, is of interest as it achieves comparable TCR and efficacy to that of Treatments 1 and 2 while being significantly less toxic, making it a potentially better option for individuals with impaired renal function or other vulnerable populations. \\~\\
Unsurprisingly, the regimens of FDA-approved treatments for mMCRC are quite efficient, with the efficiency of Treatments 1 and 2 being approximately $1.08 \times 10^{-1} \mathrm{\%/mg}$ by 18 weeks. However, these pale in comparison to the other optimal regimens, particularly Treatment 6, which has an efficiency of $2.19 \times 10^{-1} \mathrm{\%/mg}$ --- more than twice that of Treatments 1 and 2. There is also a clear transition between efficient and inefficient treatments, marked by the rapid shift in efficiency as one deviates from local optima. A treatment is inefficient if its TCR is low, regardless of the dosing and spacing (corresponding to the top left inefficient region in \autoref{pembroplots}), or if an excessive amount of pembrolizumab is administered, regardless of the TCR (corresponding to the bottom right inefficient region in \autoref{pembroplots}). Of note is that administering only a single dose of pembrolizumab before surgery offers significant convenience for patients and cost-effectiveness for hospitals. As such, Treatment 6 is of particular value since it achieves a TCR comparable to that of the other optimal regimens, including Treatments 1 and 2, whilst maintaining comparable toxicity despite only consisting of a single dose of $384$ mg of neoadjuvant pembrolizumab. \\~\\
Striking a balance between TCR, efficiency, and toxicity is difficult, and the current FDA-approved regimens for mMCRC do this quite well in the case of laMCRC. Nonetheless, the optimal regimens defined by Treatments 3--6 in \autoref{opttable} are more efficient, lead to comparable TCRs, and are more cost-effective and convenient than current regimens, all while maintaining practicality and safety. Treatment 5 is potentially ideal for vulnerable populations due to its lower toxicity, while Treatment 6 offers greater convenience and cost-effectiveness, and maintains efficacy. Moreover, lower single doses of pembrolizumab are still effective, with a single dose of $320\mathrm{~mg}$ (equivalent to $4\mathrm{~mg/kg}$) achieving a TCR of $83.06\%$, and a single dose of $200 \mathrm{~mg}$ (equivalent to $2.5 \mathrm{~mg/kg}$) achieving a TCR of $79.42\%$. The associated toxicities are $1.47 \times 10^{-1}$ and $9.18 \times 10^{-2}$, respectively, which are significantly lower than those of Treatments 1 and 2. Administering a single 200 mg dose of pembrolizumab before surgery has proven highly effective for achieving long-term tumour eradication in a phase 1b clinical trial involving resectable stage III/IV melanoma \citep{Huang2019}. In particular, 30\% of patients experienced $>90\%$ tumour eradication, and all of these patients remained disease-free at a median follow-up of 25 months. A single medium-to-high dose of pembrolizumab shows promising potential for successful and cost-effective treatment, with the IMHOTEP and RESET-C trials highlighting its possible efficacy and safety in laMCRC.\\~\\
In the construction of our model of neoadjuvant pembrolizumab therapy in laMCRC, we combined experimental data with mathematical estimation methods to determine the model parameters and initial conditions. These values are meant to represent the ``typical'' parameters belonging to a person with laMCRC; however, they can vary significantly between individuals. In particular, the flexible and modular structure of our model provides a gateway for incorporating a patient's individual immune profile and disease characteristics into the modelling of their disease. This capability serves as an important avenue for further investigation by enabling the simulation of the temporal evolution of relevant model species on an individual basis, which can assist in personalised treatment optimisation. This is particularly important since traditional regimens, despite being easier to administer on a large scale, may be far less effective and beneficial compared to personalised therapeutic approaches.\\~\\
It should be noted that the model has several limitations, many of which exist for simplicity, but addressing these issues offers exciting avenues for future research.
\begin{itemize}
    \item We ignored spatial effects in the model; however, their resolution can provide information about the distribution and clustering of different immune cell types in the TME and their clinical implications \citep{Barua2018, Maley2015}. 
    \item We assumed that the death rates were constant throughout the T cell proliferation program; however, linear death rates were shown to markedly improve the quality of fit of Deenick et al.'s model \citep{Deenick2003} to experimental data \citep{DeBoer2006}. 
    \item We considered only the M1/M2 macrophage dichotomy; however, their plasticity motivates the description of their phenotypes as a continuum, giving them the ability to adapt their functions to achieve mixtures of M1/M2 responses and functions \citep{Mills2012}.
    \item In the optimisation of neoadjuvant pembrolizumab therapy, we restricted ourselves to treatments with constant dosing and spacing as is common in the literature; however, varying dosages and dosing frequencies may result in improved regimens.
    \item We did not consider T cell avidity, the overall strength of a TCR-pMHC interaction, which governs whether a cancer cell will be successfully killed \citep{Kumbhari2020n2}. In particular, high-avidity T cells are necessary for lysing cancer cells and durable tumour eradication, while low-avidity T cells are ineffective and may inhibit high-avidity T cells \citep{Kumbhari2021, Chung2014}. 
    \item We also did not consider the influence of cytokines in the TDLN for T cell activation and proliferation, which are important in influencing effector T cell differentiation \citep{Curtsinger1999, Raphael2015}.
    \item The definition of toxicity does not account for its potential origins in autoimmunity, which is a crucial component of certain adverse effects \citep{Wang2018tox}.
\end{itemize}
In this work, we have provided a framework for mathematically modelling many immune cell types in the TME, using experimental data to govern parameter estimation, and finally analysing and optimising neoadjuvant pembrolizumab therapy in laMCRC for TCR, efficiency, and toxicity. We conclude that a single medium-to-high dose of pembrolizumab is more efficient and demonstrates comparable or greater efficacy and TCR than current FDA-approved regimens for mMCRC, whilst maintaining practicality and safety. In addition, the versatility and power of the methods and equations herein can be easily adapted to attain a more comprehensive understanding of other cancers and improve healthcare as a result.
\section{CRediT Authorship Contribution Statement}
\textbf{Georgio Hawi}: conceptualisation, data curation, formal analysis, funding acquisition, investigation, methodology, project administration, resources, software, validation, visualisation, writing --- original draft, writing --- review \& editing. \\
\textbf{Peter S. Kim}: conceptualisation, formal analysis, funding acquisition, investigation, methodology, project administration, resources, supervision, validation, visualisation, writing --- original draft, writing --- review \& editing. \\
\textbf{Peter P. Lee}: conceptualisation, formal analysis, investigation, methodology, project administration, resources, supervision, validation, visualisation, writing --- original draft, writing --- review \& editing.
\section{Declaration of Competing Interests}
The authors declare that they have no known competing financial interests or personal relationships that could have appeared to influence the work reported in this paper.
\section{Data Availability}
All data, code, and procedures are available within the manuscript and its Supporting Information files.
\section{Acknowledgements}
This work was supported by an Australian Government Research Training Program Scholarship. PSK gratefully acknowledges support from the Australian Research Council Discovery Project (DP230100485).
\putbib[References.bib]
\end{bibunit}
\appendix
\counterwithin{figure}{section}
\counterwithin{table}{section}
\resetlinenumber
\nolinenumbers
\title{Appendix A: Optimisation of neoadjuvant pembrolizumab therapy for locally advanced MSI-H/dMMR colorectal cancer using data-driven delay integro-differential equations}
\maketitle
\setcounter{page}{1}
\begin{bibunit}[vancouver]
\section{Steady States and Initial Conditions}
We estimate all steady states and initial conditions under the assumption that pembrolizumab has not been and will not be administered.
\subsection{Steady States and Initial Conditions for Cells in the TS\label{tsitess}}
Digital cytometry has proved itself to be a powerful technique in characterising immune cell populations from individual patients' bulk tissue transcriptomes without requiring physical cell isolation \cite{Newman2019supp, Le2020supp, Gong2011supp, Liebner2013supp, Newman2015supp}. In particular, RNA-sequencing (RNA-seq) deconvolution of tumour gene expressions has been very useful in determining immune profiles and adjusting treatment accordingly. For all algorithms outlined in the sequel, we aggregate the estimates by taking the median of the relevant non-zero values elementwise and then normalising such that their sums become $1$. \\~\\
To estimate immune cell population proportions in locally advanced MSI-H/dMMR colorectal cancer (laMCRC), we applied multiple algorithms and then synthesised their results to obtain estimates for all cell types in the model. We first used the ImmuCellAI algorithm \cite{Miao2020supp}, which estimates the abundance of 24 immune cell types from gene expression data and has also been shown to be highly accurate in predicting immunotherapy response. These immune cell types include 18 T cell subsets, including CD4+ T cells which incorporate T helper cells (namely Th1 cells, Th2 cells, Th17 cells, and T follicular helper cells), regulatory T cells (including natural Tregs (nTregs), induced Tregs (iTregs), and type 1 regulatory T cells (Tr1s)), naive CD4+ T cells (CD4\_naive) and other CD4+ T cells (CD4\_T). In addition, they include naive CD8+ T cells (CD8\_T), cytotoxic T cells or CTLs (Tc), exhausted CD8+ T cells (Tex), central memory T cells (Tcm), effector memory T cells (Tem), natural killer T cells (NKT), $\gamma\delta$ T cells (Tgd), and mucosal-associated invariant T cells (MAIT). ImmuCellAI also estimates the abundance of DCs, B cells, monocytes, macrophages, and NK cells. Direct correspondences between state variables in the model and ImmuCellAI cell types are shown in \Cref{ImmuCellmap}.
\begin{table}[H]
    \centering
    \begin{tabular}{|c|c|}
    \hline
       \textbf{State Variable}  & \textbf{ImmuCellAI Cell Type} \\
       \hline
       $T_8$ & Tc \\
       $T_\mathrm{ex}$ & Tex\\
       $T_r$ & nTreg\\
       $D_0$, $D$ & DC\\
       $T_1$ & Th1\\
       $M_0$, $M_1$, $M_2$ & Macrophage\\
       $K_0$, $K$ & NK\\
       \hline
    \end{tabular}
    \caption{Mappings between state variables of the model and ImmuCellAI immune cell types.\label{ImmuCellmap}}
\end{table}
Using the UCSC Xena web portal \cite{Goldman2020supp}, RSEM normalised RNA-seq gene expression profiles of patients from the TCGA COAD and TCGA READ projects \cite{COADREAD2012supp} were acquired, featuring patients with colorectal adenocarcinomas. Corresponding clinical and biospecimen data were downloaded from the GDC portal \cite{Grossman2016supp} and included tumour dimensions, necrotic cell percentage, AJCC TNM stage, and MSIsensor and MANTIS MSI statuses. We filtered for samples from primary tumours and with non-empty necrosis percentage data from patients with AJCC stage III CRC and at least one of MANTIS score $>0.4$ or MSIsensor score $>3.5\%$, as these are the default thresholds for MSI-H \cite{Kautto2016supp}. We used the stage IIIC samples to infer steady states and the stage IIIA and stage IIIB samples to infer initial conditions. We also used the manually curated TIMEDB cell composition database \cite{Wang2022supp} to source tumour deconvolution estimates for each relevant individual sample.\\~\\
Additionally, the aggregated estimated cell proportions generated by ImmuCellAI for steady states and initial conditions, after normalisation, are shown in \Cref{ImmuCellAIsteady} and \Cref{ImmuCellAIinit}.
\begin{table}[H]
\centering
\begin{tabular}{|c|c|c|c|}
\hline
\textbf{Cell Type} & \textbf{Proportion} & \textbf{Cell Type} & \textbf{Proportion} \\
\hline
\textit{DC} & 0.055860 & \textit{nTreg} & 0.003787 \\
B\_cell & 0.107446 & iTreg & 0.003787 \\
Monocyte & 0.128762 & \textit{Th1} & 0.001895 \\
\textit{Macrophage} & 0.072902 & Th2 & 0.003705 \\
\textit{NK} & 0.138818 & Th17 & 0.002779 \\
Neutrophil & 0.150538 & Tfh & 0.003787 \\
CD4\_T & 0.040712 & CD8\_naive & 0.003677 \\
CD8\_T & 0.060652 & \textit{Tc} & 0.004668 \\
NKT & 0.137899 & \textit{Tex} & 0.003677 \\
Tgd & 0.062504 & MAIT & 0.004631 \\
CD4\_naive & 0.000937 & Tcm & 0.002786 \\
Tr1 & 0.003791 & Tem & 0.000000 \\
\hline
\end{tabular}
\caption{\label{ImmuCellAIsteady}TS steady-state cell proportions for the model, derived using RNA-sequencing deconvolution via ImmuCellAI. Values for italicised cell types are used in estimating TS cell populations in the model.}
\end{table}
\begin{table}[H]
\centering
\begin{tabular}{|c|c|c|c|}
\hline
\textbf{Cell Type} & \textbf{Proportion} & \textbf{Cell Type} & \textbf{Proportion} \\
\hline
\textit{DC} & 0.070785 & \textit{nTreg} & 0.005557 \\
B\_cell & 0.086875 & iTreg & 0.003705 \\
Monocyte & 0.059671 & \textit{Th1} & 0.002732 \\
\textit{Macrophage} & 0.073657 & Th2 & 0.002802 \\
\textit{NK} & 0.144167 & Th17 & 0.002758 \\
Neutrophil & 0.080897 & Tfh & 0.009193 \\
CD4\_T & 0.072488 & CD8\_naive & 0.005449 \\
CD8\_T & 0.091245 & \textit{Tc} & 0.006358 \\
NKT & 0.129328 & \textit{Tex} & 0.005475 \\
Tgd & 0.130304 & MAIT & 0.005475 \\
CD4\_naive & 0.000912 & Tcm & 0.002758 \\
Tr1 & 0.007410 & Tem & 0.000000 \\
\hline
\end{tabular}
\caption{\label{ImmuCellAIinit}Proportions of TS initial conditions for the model, derived using RNA-sequencing deconvolution via ImmuCellAI. Values for italicised cell types are used in estimating TS cell populations in the model.}
\end{table}
To determine the proportions of $D_0$ and $D$, $K_0$ and $K$, $M_0$ and $M_1$ and $M_2$, we used the CIBERSORTx algorithm \cite{Newman2019supp}, due to its high accuracy \cite{Sutton2022supp}. We followed a similar approach to \cite{Kirshtein2020supp} and \cite{Budithi2021supp} and applied CIBERSORTx B-mode on the refined gene expression data, using the validated LM22 signature matrix \cite{Newman2015supp}, which gave relative immune cell proportions of 22 immune cell types using 547 signature genes derived from microarray data. Direct correspondences between state variables in the model and keys of the LM22 signature matrix are shown in \Cref{LM22map}. 
\begin{table}[H]
    \centering
    \begin{tabular}{|c|c|}
    \hline
       \textbf{State Variable}  & \textbf{LM22 key} \\
       \hline
       $D_0$ & Dendritic cells resting\\
       $D$ & Dendritic cells activated\\
       $M_0$ & Macrophages M0 \\
       $M_1$ & Macrophages M1 \\
       $M_2$ & Macrophages M2 \\
       $K_0$ & NK cells resting \\
       $K$   & NK cells activated\\
       \hline
    \end{tabular}
    \caption{Mappings between state variables of the model and keys of the LM22 signature matrix.\label{LM22map}}
\end{table}
The aggregated estimated cell proportions generated by CIBERSORTx for steady states and initial conditions, after normalisation, are shown in \Cref{CIBERsteady} and \Cref{CIBERinitialcond}.
\begin{table}[H]
\centering
\begin{tabular}{|c|c|c|c|}
\hline
\textbf{Cell Type} & \textbf{Proportion} & \textbf{Cell Type} & \textbf{Proportion} \\
\hline
B cells naive & 0.059688 & NK cells activated & 0.083872 \\
B cells memory & 0.000000 & Monocytes & 0.018276 \\
Plasma cells & 0.005234 & \textit{Macrophages M0} & 0.069199 \\
T cells CD8 & 0.159313 & \textit{Macrophages M1} & 0.055534 \\
T cells CD4 naive & 0.000000 & \textit{Macrophages M2} & 0.214657 \\
T cells CD4 memory resting & 0.141747 & \textit{Dendritic cells resting} & 0.012246 \\
T cells CD4 memory activated & 0.026479 & \textit{Dendritic cells activated} & 0.004009 \\
T cells follicular helper & 0.004593 & Mast cells resting & 0.034480 \\
T cells regulatory (Tregs) & 0.013419 & Mast cells activated & 0.059208 \\
T cells gamma delta & 0.000000 & Eosinophils & 0.005205 \\
NK cells resting & 0.000000 & Neutrophils & 0.032841 \\
\hline
\end{tabular}
\caption{\label{CIBERsteady}TS steady-state cell proportions for the model, derived using RNA-sequencing deconvolution via CIBERSORTx. Values for italicised cell types are used in estimating TS cell populations in the model.}
\end{table}
\begin{table}[H]
\centering
\begin{tabular}{|c|c|c|c|}
\hline
\textbf{Cell Type} & \textbf{Proportion} & \textbf{Cell Type} & \textbf{Proportion} \\
\hline
B cells naive & 0.024411 & \textit{NK cells activated} & 0.085199 \\
B cells memory & 0.011051 & Monocytes & 0.019237 \\
Plasma cells & 0.002355 & \textit{Macrophages M0} & 0.051451 \\
T cells CD8 & 0.219758 & \textit{Macrophages M1} & 0.051039 \\
T cells CD4 naive & 0.000000 & \textit{Macrophages M2} & 0.094792 \\
T cells CD4 memory resting & 0.177310 & \textit{Dendritic cells resting} & 0.019598 \\
T cells CD4 memory activated & 0.062485 & \textit{Dendritic cells activated} & 0.009977 \\
T cells follicular helper & 0.007731 & Mast cells resting & 0.030947 \\
T cells regulatory (Tregs) & 0.032978 & Mast cells activated & 0.067909 \\
T cells gamma delta & 0.000000 & Eosinophils & 0.018542 \\
\textit{NK cells resting} & 0.005009 & Neutrophils & 0.008219 \\
\hline
\end{tabular}
\caption{\label{CIBERinitialcond}TS initial condition cell proportions for the model, derived using RNA-sequencing deconvolution via CIBERSORTx. Values for italicised cell types are used in estimating TS cell populations in the model.}
\end{table}
However, to determine the proportions of $K_0$ and $K$ at steady state, we could not use CIBERSORTx due to its nil results. Instead, we used a combination of biologically informed assumptions and data from physical experiments. It was determined in \cite{Mao2024supp} that the ratio of the proportions of cytotoxic, activated NK cells to resting NK cells decreases as CRC progresses. We thus assumed that $\overline{K} = 10\overline{K_0}$.\\~\\
We integrated the relative proportions within cell types for DCs, NK cells, and macrophages into the ImmuCellAI abundance estimates. We note that the density of immune cells in a healthy adult colon is approximately $3.37 \times 10^{7}\mathrm{~cell/g}$ \cite{Sender2023supp}, which, assuming a tissue density of $1.03 \mathrm{~g/cm^3}$, results in a total immune cell density of $3.47 \times 10^{7}\mathrm{~cell/cm^3}$. However, advanced cancer induces lymphadenopathy \cite{Lymphadenopathybooksupp, West2016supp}, which \cite{Sender2023supp} estimates results in an increase in the total number of lymphocytes of at most $10\%$. As such, we assume that there is a $10\%$ increase in lymphocyte concentration in laMCRC.\\~\\
Accounting for the high immunogenicity of tumours in laMCRC \cite{Lin2020supp}, we assumed at steady state that the density of cancer cells is equal to the total immune cell density. Taking into account lymphadenopathy and using data from \cite{Sender2023supp}, we assumed that the total immune cell density in laMCRC initially is approximately $3.72 \times 10^{7} \mathrm{~cell/cm^3}$ and at steady state is approximately $3.68 \times 10^{7} \mathrm{~cell/cm^3}$. From the TCGA biospecimen data, the median necrotic cancer cell percentage for stage IIIA/IIIB and stage IIIC MSI-H CRC samples is $10\%$. As such, denoting $\overline{\mathrm{TIC}}$ as the total immune cell density at steady state, and $N_p$ as the necrotic cell percentage, we have that
\begin{align}
    \overline{C} + \overline{N_c} &= \overline{\mathrm{TIC}},\\
    \frac{\overline{N_c}}{N_p} &=\frac{\overline{C}}{1-N_p}, \quad \overline{C} = \overline{\mathrm{TIC}}\times (1-N_p) \implies \overline{N_c} =\overline{\mathrm{TIC}} \times N_p.
\end{align}
Thus, at steady state, $C \approx 3.31 \times 10^{7} \mathrm{~cell/cm^3}$ and $N_c \approx 3.68 \times 10^{6} \mathrm{~cell/cm^3}$ in laMCRC. \\~\\
A retrospective cohort study by Burke et al.\ considered CRC patients at Leeds Teaching Hospitals NHS Trust over a 2-year interval who received no treatment and who underwent computed tomography (CT) scans twice, more than 5 weeks apart. It was found that in patients whose M category changed from M0 to M1, the median interval between CT scans was 155 days, and the median tumour doubling time was $172$ days \cite{Burke2020supp}. We assumed that these timeframes were similar to stage IIIA/IIIB MSI-H/dMMR CRC progression, and, as such, we assumed that it took $155$ days for $C$ and $N_c$ to reach their steady-state values. This corresponds to an initial condition for $C$ being $C(0) = 1.79 \times 10^{7} \mathrm{~cell/cm^3}$ and thus $N_c(0) \approx 1.99 \times 10^{6} \mathrm{~cell/cm^3}$.\\~\\
Combining everything, the resultant steady states and initial conditions for the model are shown in \Cref{firstmodelsteady} and \Cref{firstmodelinitcond}, respectively.
\begin{table}[H]
\centering
\begin{tabular}{|c|c|c|c|c|c|c|}
\hline
$C$ & $N_c$ & $D_0$ & $D$ & $T_8$ & $T_\mathrm{ex}$ & $T_1$ \\
\hline
$3.31 \times 10^{7}$ & $3.68 \times 10^{6}$ & $1.46 \times 10^{6}$ & $4.78 \times 10^{5}$ & $1.78 \times 10^{5}$ & $1.40 \times 10^{5}$ & $7.23 \times 10^{4}$ \\
\hline
$T_r$ & $M_0$ & $M_1$ & $M_2$ & $K_0$ & $K$ & \\
\hline
$1.45 \times 10^{5}$ & $5.16 \times 10^{5}$ & $4.14 \times 10^{5}$ & $1.60 \times 10^{6}$ & $4.82 \times 10^{5}$ & $4.82 \times 10^{6}$ & \\
\hline
\end{tabular}
\caption{\label{firstmodelsteady}TS steady-state cell densities for the model, combining estimates derived from ImmuCellAI and CIBERSORTx. All values are in $\mathrm{cell/cm^3}$.}     
\end{table}
\begin{table}[H]
\centering
\begin{tabular}{|c|c|c|c|c|c|c|}
\hline
$C$ & $N_c$ & $D_0$ & $D$ & $T_8$ & $T_\mathrm{ex}$ & $T_1$ \\
\hline
$1.79 \times 10^{7}$ & $1.99 \times 10^{6}$ & $1.63 \times 10^{6}$ & $8.29 \times 10^{5}$ & $2.43 \times 10^{5}$ & $2.09 \times 10^{5}$ & $1.04 \times 10^{5}$ \\
\hline
$T_r$ & $M_0$ & $M_1$ & $M_2$ & $K_0$ & $K$ & \\
\hline
$2.12 \times 10^{5}$ & $6.67 \times 10^{5}$ & $6.61 \times 10^{5}$ & $1.23 \times 10^{6}$ & $3.06 \times 10^{5}$ & $5.20 \times 10^{6}$ &\\
\hline
\end{tabular}
\caption{\label{firstmodelinitcond}TS initial condition cell densities for the model, combining estimates derived from ImmuCellAI and CIBERSORTx. All values are in $\mathrm{cell/cm^3}$.}
\end{table}
We note that, technically, ImmuCellAI is an enrichment-based method that does not provide absolute immune cell proportions but rather estimates abundances across various immune cell subtypes not reported by CIBERSORTx. However, normalising these abundances provides a good approximation of the true immune cell proportions, thereby allowing ImmuCellAI to be justifiably employed to estimate immune cell steady states and initial conditions.
\subsection{TDLN T Cell Steady States\label{TDLNssinit}}
To determine the steady-state values for $T_0^8$, $T_A^8$, $T_0^4$, $T_A^1$, and $T_A^r$, we used ImmuCellAI on the GSE26571 dataset from the NCBI Gene Expression Omnibus repository \cite{Clough2016supp, Barrett2012supp}, obtaining deconvolution results from TIMEDB. The dataset contains nine samples of lymph node metastases from 7 patients with colon adenocarcinoma, with data from \cite{Leydold2011supp}. However, the dataset's metadata does not contain AJCC TNM stages for patients. To estimate the TNM stages of the patients with lymph node metastases, we considered the samples of lymph node metastases for these patients and applied the ImmuCellAI algorithm to estimate their immune cell abundances, ignoring Tcm and Tem cell subtypes. Furthermore, we denote $T_A^{8\mathrm{LN}}$ and $T_A^{1\mathrm{LN}}$ as the total number of activated CD8+ T cells and activated Th1 cells in the TDLN, respectively, and let $T_r^{\mathrm{LN}}$ denote the total number of Tregs in the TDLN. Mappings between ImmuCellAI immune cell types and TDLN cell types in the model are shown in \Cref{ImmuCellmapTDLN}.
\begin{table}[H]
    \centering
    \begin{tabular}{|c|c|}
    \hline
       \textbf{State Variable}  & \textbf{ImmuCellAI Cell Type} \\
       \hline
       $T_0^8$ & CD8\_naive \\
       $T_A^{8\mathrm{LN}}$ & Tc\\
       $T_0^4$ & CD4\_naive\\
       $T_A^{1\mathrm{LN}}$ & Th1\\
       $T_0^r$, $T_r^{\mathrm{LN}}$ & nTreg \\
       \hline
    \end{tabular}
    \caption{Mappings between TDLN cell types in the model and ImmuCellAI immune cell types.\label{ImmuCellmapTDLN}}
\end{table}
We assumed that the lymph node metastases were from patients with a TNM stage of at least stage IIIC and performed 2-means clustering on the estimated cell proportions generated by ImmuCellAI to distinguish lymph node metastases as being from stage IIIC/IVA patients from those with more advanced disease. In particular, we considered the `nTreg', `Th1', `Th2', and `Cytotoxic' cell types as part of the clustering. We compared the individual coordinates of each cluster's centroid and note that lymph node metastases from stage IVB/IVC samples, which correspond to more advanced CRC progression, exhibit a higher proportion of Th2 cells and nTregs, alongside a lower proportion of Th1 cells and cytotoxic T cells compared to stage IIIC/IVA samples. Like before, we used lymph node metastases from stage IIIC/IVA patients to infer TDLN steady states. Aggregating the estimates, as before, and then normalising such that their sums become 1, results in the proportions as shown in \Cref{TDLNImmuCellAIsteady}.
\begin{table}[H]
\centering
\begin{tabular}{|c|c|c|c|}
\hline
\textbf{Cell Type} & \textbf{Proportion} & \textbf{Cell Type} & \textbf{Proportion} \\
\hline
DC & 0.176871 & Tr1 & 0.000930 \\
B\_cell & 0.116990 & \textit{nTreg} & 0.000929 \\
Monocyte & 0.056462 & iTreg & 0.003714 \\
Macrophage & 0.035633 & \textit{Th1} & 0.008357 \\
NK & 0.037605 & Th2 & 0.026991 \\
Neutrophil & 0.087545 & Th17 & 0.004162 \\
CD4\_T & 0.134744 & Tfh & 0.002777 \\
CD8\_T & 0.056912 & \textit{CD8\_naive} & 0.006467 \\
NKT & 0.130453 & \textit{Tc} & 0.000927 \\
Tgd & 0.105489 & Tex & 0.003719 \\
\textit{CD4\_naive} & 0.002324 & MAIT & 0.000000 \\
\hline
\end{tabular}
\caption{\label{TDLNImmuCellAIsteady}TDLN steady-state cell proportions for the model, derived using RNA-sequencing deconvolution via ImmuCellAI. Values for italicised cell types are used in estimating TDLN cell populations in the model.}
\end{table}
The density of immune cells in the lymph nodes of an adult is approximately $1.8 \times 10^{9}\mathrm{~cell/g}$ \cite{Sender2023supp}, which, assuming a tissue density of $1.03 \mathrm{~g/cm^3}$, results in a total immune cell density of $1.854 \times 10^{9}\mathrm{~cell/cm^3}$. We also assumed that in the TDLN, the number of activated CD8+ T cells having undergone $n^8_\mathrm{max}$ divisions is roughly half the number that has only undergone $n^8_\mathrm{max}-1$ divisions and so forth, and similarly for Th1 cells. Furthermore, we assumed that initially, and at steady state, $10\%$ of all Tregs are naive. Thus, we assumed that for $i=1,8$,
\begin{align*}
    T_A^i &= \frac{2^{n^i_\mathrm{max}}}{2^{n^i_\mathrm{max}+1}-1}T_A^{i\mathrm{LN}},
\intertext{and that}
    T_0^r &= \frac{T_r^{\mathrm{LN}}}{10}, \quad T_A^r = \frac{9}{10}\frac{2^{n^r_\mathrm{max}}}{2^{n^r_\mathrm{max}+1}-1}T_r^{\mathrm{LN}}.
\end{align*}
Combining everything, and incorporating T cell division numbers as justified in \Cref{TDLNsubsystemest}, the resultant steady states for the model are shown in \Cref{TDLNfirstmodelsteady}.
\begin{table}[H]
\centering
\begin{tabular}{|c|c|c|c|c|c|}
\hline
$T_0^8$ & $T_A^8$ & $T_0^4$ & $T_A^1$ & $T_0^r$ & $T_A^r$ \\
\hline
$1.20 \times 10^{7}$ & $8.60 \times 10^{5}$ & $4.31 \times 10^{6}$ & $7.76 \times 10^{6}$ & $1.72 \times 10^{5}$ & $7.81 \times 10^{5}$ \\
\hline
\end{tabular}
\caption{\label{TDLNfirstmodelsteady}TDLN steady-state cell densities for the model, using estimates derived from ImmuCellAI. All values are in $\mathrm{cell/cm^3}$.}
\end{table}
\subsection{Steady States and Initial Conditions for DAMPs\label{dampssinitappendix}}
We note that $1 \mathrm{~cm^3} = 1 \mathrm{~mL}$ for all DAMP measurements. To estimate DAMP steady states and initial conditions, we look at the respective experimental tissue concentration data, noting that this is more accurate than the more widely available serum/plasma concentration data. Nonetheless, we use serum/plasma concentration data, where relevant, to guide estimates if the corresponding tissue concentration data is limited. We note that DAMPs only appear in the model within an inhibition or half-saturation constant, making their absolute magnitude less important since they always appear as a ratio.
\subsubsection{Estimates for $H$}
In \cite{Zhang2019supp}, a study of blood samples from 144 patients with CRC was conducted, with the serum HMGB1 levels of patients with distant metastasis being $13.32\pm 6.12\mathrm{~\mu g/L}$, which was significantly higher than the concentration in patients with only lymphatic metastasis at $10.14 \pm 4.38\mathrm{~\mu g/L}$. We assumed that the serum concentrations and tissue concentrations of HMGB1 were similar, so we took the initial condition for $H$ to be $5.76 \times 10^{-9} \mathrm{~g/cm^3}$ and the steady state to be $1.01 \times 10^{-8} \mathrm{~g/cm^3}$.
\subsubsection{Estimates for $S$}
In epithelial ovarian cancer (EOC), calreticulin concentrations when no drugs are introduced were approximately $2 \times 10^{-2} \pm 2.5 \times 10^{-2} \mathrm{~\mu g}/\mathrm{mL}$ \cite{Abdullah2021supp}. Since surface calreticulin is produced by necrotic cancer cells, which have a larger population at steady state compared to initially, we assume that there is more surface calreticulin at steady state. We assume that calreticulin concentrations in EOC are similar to those in MSI-H/dMMR CRC, so we assume an initial condition for $S$ of $2.00 \times 10^{-8}\mathrm{~g/cm^3}$ and a steady state of $3.25 \times 10^{-8}\mathrm{~g/cm^3}$.
\subsection{Steady States and Initial Conditions for Cytokines\label{cytokinessinitappendix}}
To estimate cytokine steady states and initial conditions, we look at the respective experimental tissue concentration data, noting that $1 \mathrm{~cm^3} = 1 \mathrm{~mL}$ for all cytokine measurements. We note that cytokines only appear in the model within an inhibition or half-saturation constant, making their absolute magnitude less important since they always appear as a ratio.
\subsubsection{Estimates for $I_2$}
The tissue concentration of IL-2 in CRC is very low and was found to be below the lower limit of quantification in various experiments \cite{Calu2021supp, kim2014associationsupp}. In tumour supernatants of invasive ductal cancer, the median IL-2 concentration was found to be $2.1 \mathrm{~pg/mL}$ with the interquartile range being $2.0\mathrm{~pg/mL}\text{ -- }4.9\mathrm{~pg/mL}$ \cite{Autenshlyus2017supp}. We assume similar concentrations of IL-2 in the tissue of CRC patients. \\~\\
Taking into account the well-documented anti-tumour properties of IL-2 \cite{Zhao2021supp, West1989supp} and decreased IL-2 serum concentration in metastatic CRC patients compared to those without distant metastasis \cite{CzajkaFrancuz2020supp}, we assumed that $I_2$ has a steady-state value of $2.00 \times 10^{-12} \mathrm{~g/cm^3}$.
\subsubsection{Estimates for $I_\upgamma$}
It was found in \cite{Calu2021supp} that the median tissue concentration of IFN-$\upgamma$ in CRC patients was $15.2 \mathrm{~pg/mL}$, with the upper quartile concentration being approximately $16.9 \mathrm{~pg/mL}$. It was found in \cite{Sharp2017supp} that the serum concentration of IFN-$\upgamma$ in stage IV CRC patients (median $\approx 20.75\mathrm{~pg/mL}$) is significantly higher than that of stage I-III patients (median $\approx 1\mathrm{~pg/mL}$). We thus set the steady state of $I_\upgamma$ to $1.69 \times 10^{-11} \mathrm{~g/cm^3}$.
\subsubsection{Estimates for $I_\upalpha$}
It was found in \cite{Sharp2017supp} that in advanced CRC patients, i.e. those with stage III or stage IV disease, the mean TNF tissue concentration was $\approx 53 \mathrm{~pg/mL}$, with the concentration one standard deviation below the mean being approximately $16 \mathrm{~pg/mL}$. Furthermore, the serum TNF concentration in stage IV CRC patients (median $20.3\mathrm{~pg/mL}$) is significantly higher than in stage III CRC patients (median $16.0\mathrm{~pg/mL}$) \cite{KrzystekKorpacka2013supp}. We thus set the steady state of $I_\upalpha$ to $5.30 \times 10^{-11} \mathrm{~g/cm^3}$.
\subsubsection{Estimates for $I_\upbeta$}
It was found in \cite{Wang2012supp} that in CRC patients, the mean TGF-$\upbeta$ tissue concentration was $1311.5 \mathrm{~pg/mg}$, with the concentration one standard error above the mean being $1469.1 \mathrm{~pg/mg}$. Assuming a tissue density of $1.03 \mathrm{~g/mL}$, these correspond to tissue concentrations of $1.35 \times 10^6 \mathrm{~pg/mL}$ and $1.51 \times 10^6 \mathrm{~pg/mL}$, respectively. Furthermore, the serum TGF-$\upbeta$ concentration in stage IV CRC patients (mean $55\mathrm{~pg/mL}$) is significantly higher than in stage III CRC patients (mean $45\mathrm{~pg/mL}$) \cite{Shim1999supp}. We thus set the steady state of $I_\upbeta$ to $1.51 \times 10^{-6} \mathrm{~g/cm^3}$.
\subsubsection{Estimates for $I_{10}$}
It was found in \cite{Sharp2017supp} that in advanced CRC patients, i.e. those with stage III or stage IV disease, the mean IL-10 tissue concentration was $115 \mathrm{~pg/mL}$, with the concentration one standard deviation below the mean being approximately $46 \mathrm{~pg/mL}$. Furthermore, the serum IL-10 concentration in stage IV CRC patients (mean $36.02 \mathrm{~pg/mL}$) is significantly higher than in stage III CRC patients (mean $17.07\mathrm{~pg/mL}$) \cite{Stanilov2010supp}. We thus set the steady state of $I_{10}$ to $1.15 \times 10^{-10} \mathrm{~g/cm^3}$, with an initial condition of $4.60 \times 10^{-11} \mathrm{~g/cm^3}$.
\putbib[References.bib]
\end{bibunit}
\resetlinenumber
\nolinenumbers
\title{Appendix B: Optimisation of neoadjuvant pembrolizumab therapy for locally advanced MSI-H/dMMR colorectal cancer using data-driven delay integro-differential equations}
\maketitle
\setcounter{page}{1}
\begin{bibunit}[vancouver]
\section{Parameter Estimation\label{parameter estimation section 2}}
We estimate all parameters, where possible, under the assumption that no pembrolizumab has been or will be administered. The exceptions to this are the parameters directly related to pembrolizumab treatment, for which the assumptions are explicitly stated during estimation.

\subsection{TDLN Parameters \label{TDLNsubsystemest}}
\subsubsection{Estimate for $V_\mathrm{TS}$}
The mean tumour volume in CRC patients with a T stage of T4a or an N stage of N2 was found to be $27.56 \mathrm{~cm^3}$ and $27.57 \mathrm{~cm^3}$, respectively \cite{Park2017supp}. As such, we set $V_\mathrm{TS} = 2.76 \times 10^{1} \mathrm{~cm^3}$.
\subsubsection{Estimate for $V_\mathrm{LN}$}
The mean diameter of lymph nodes in CRC patients where cancer has metastasised was found to be $5.6 \mathrm{~mm}$ in \cite{Rssler2017supp}. Assuming a spherical lymph node, this corresponds to $V_\mathrm{LN} = \frac{4}{3} \times 2.8^3 \times \pi \mathrm{~mm^3} = 9.20 \times 10^{-2} \mathrm{~cm^3}$.
\subsubsection{Estimate for $\tau_m$}
In \cite{Catron2004supp}, it took $18$ hours for DCs, which acquired antigen from a site of subcutaneous injection, to arrive at the lymph node. We assume that this migration time is the same for DCs acquiring cancer antigens from the TS so that $\tau_m = 18 \mathrm{~hr} = 0.75 \mathrm{~day}$.
\subsubsection{Estimate for $\tau_a$}
To estimate $\tau_a$, we note that T cells in the TDLN travel at speeds of $11\text{ -- }14 \mathrm{~\mu m/min}$, in comparison to DCs which migrate at speeds of $3\text{ -- }6 \mathrm{~\mu m/min}$ \cite{anaya2013autoimmunitysupp}. We thus have that $\tau_a = \frac{4.5}{12.5} \tau_m \approx 0.27 \mathrm{~day}$.
\subsubsection{Estimates for CD8+ T cells}
Using data from \cite{Kinjyo2015supp}, we estimate that naive CD8+ T cells take 2 days to activate, and so set $\tau_8^\mathrm{act} = 2 \mathrm{~day}$. It was found in \cite{Plambeck2022supp} that activated CD8+ T cells required $39$ hours on average to complete their first cell division, and so we set $\Delta_8^0 = 39\mathrm{~hr} = 1.63 \mathrm{~day}$. Furthermore, the average division time for subsequent cell cycles is $8.6$ hours \cite{Plambeck2022supp}; however, it can vary between $5$ and $28$ hours. Thus, we set $\Delta_8 = 8.6\mathrm{~hr} = 0.36 \mathrm{~day}$. It was shown in \cite{Kaech2001supp} that fully activated CD8+ T cells divide between $7$ and $10$ times; however, they can divide more if persistent antigen exposure is present. Indeed, in Lymphocytic Choriomeningitis Virus (LCMV), CD8+ T cells can divide more than $15$ times \cite{Masopust2007supp}. We make a compromise and set $n^8_\mathrm{max} = 10$. We thus have that $\tau_{T_A^8} = 4.87 \mathrm{~day}$. Finally, it is widely accepted that T cell exhaustion can arise only days to weeks from the initial antigen exposure in the case of chronic antigen stimulation \cite{Blank2019supp, McLane2019supp}, so that we take $\tau_l = 10 \mathrm{~day}$.
\subsubsection{Estimates for Th1 cells}
We first note that Th1 cells are phenotypes of CD4+ T helper cells. It was found in \cite{JelleyGibbs2000supp} that CD4+ T cell priming takes between $1$ and $2$ days, and so we set $\tau_4^\mathrm{act} = 1.5\mathrm{~day}$. Compared to CD8+ T cells, CD4+ T cells appear to divide fewer times, with only approximately nine cell divisions, as in LCMV \cite{Homann2001supp}. We assume this is similar in MSI-H/dMMR CRC, and so set $n^1_\mathrm{max} = 9$. It takes between $12$ and $24$ hours for the first CD4+ T cell division to occur, with subsequent divisions occurring at a rate of approximately $10$ hours per cell division \cite{Kaech2002supp}. We thus set $\Delta_1^0 = 18.5 \mathrm{~hr} = 0.77 \mathrm{~day}$, and $\Delta_1 = 10 \mathrm{~hr} = 0.42 \mathrm{~day}$. This leads to $\tau_{T_A^1} = 4.13 \mathrm{~day}$.
\subsubsection{Estimates for Tregs}
We assume that the activation of Tregs takes the same amount of time as that of CD4+ T helper cells, so that $\tau_r^\mathrm{act} = 1.5\mathrm{~day}$. It was found in \cite{DarrasseJze2009supp} that in mice, 6 days after tumour implantation, 45$\%$ of Tregs in the TDLN had undergone at least 1 division, and 14$\%$ had undergone more than six divisions. We thus set $n^r_\mathrm{max} = 6$ and assume that the cell division rates of Tregs and CD4+ T helper cells are the same, so that $\Delta_r^0 = 0.77 \mathrm{~day}$ and $\Delta_r = 0.42 \mathrm{~day}$. We thus have that $\tau_{T_A^r} = 2.87 \mathrm{~day}$.
\subsection{Half-Saturation Constants}
We recall that for some species $X$, $K_X$ is denoted the half-saturation constant of $X$ in a term of the form
\begin{equation*}
    \frac{X}{K_X +X}.
\end{equation*}
For simplicity, we assume that if $\overline{X}$ denotes the steady-state value of $X$, then
\begin{equation}
    \frac{\overline{X}}{K_X + \overline{X}} = \frac{1}{2} \implies K_X = \overline{X}.
\end{equation}
This implies that
\begin{align*}
    K_{T_8C} = \overline{C}\tau_l &= 3.31 \times 10^{8} ~\left(\mathrm{cell}/\mathrm{cm^3}\right)\ \mathrm{day}, \\
    K_{KD_0} = \overline{D_0} &= 1.46 \times 10^{6} \mathrm{~cell/cm^3}, \\
    K_{KD} = \overline{D} &= 4.78 \times 10^{5} \mathrm{~cell/cm^3}, \\
    K_{DH} = \overline{H} &= 1.01 \times 10^{-8}\mathrm{~g/cm^3},\\
    K_{DS} = \overline{S} &= 3.25 \times 10^{-8} \mathrm{~g/cm^3}, \\
    K_{T_8I_2} = K_{T_1I_2} = K_{KI_2} = K_{I_{10} I_2} = \overline{I_2} &= 2.00 \times 10^{-12} \mathrm{~g/cm^3}, \\
    K_{C I_\upgamma} = K_{M_1 I_\upgamma} = K_{MI_\upgamma} = \overline{I_\upgamma} &= 1.69 \times 10^{-11} \mathrm{~g/cm^3}, \\
    K_{CI_\upalpha} = K_{M_1 I_\upalpha} = K_{MI_{\upalpha}} = \overline{I_\upalpha} &= 5.30 \times 10^{-11} \mathrm{~g/cm^3}, \\
    K_{M_2 I_\upbeta} = K_{MI_\upbeta} = \overline{I_\upbeta} &= 1.51 \times 10^{-6} \mathrm{~g/cm^3},\\
    K_{M_2 I_{10}} = \overline{I_{10}} &= 1.15 \times 10^{-10} \mathrm{~g/cm^3},\\
    K_{T_1Q^{T_1}} = \overline{Q^{T_1}} &= 2.02 \times 10^{5} \mathrm{~molec/cm^3}.
\end{align*}
To estimate $K_{T_\mathrm{ex} A_1}$, we note that the value of the geometric mean $C_\mathrm{avg}$ of pembrolizumab in serum at steady state varied minimally regardless of whether pembrolizumab was administered at $200\mathrm{~mg}$ every 3 weeks, or $400\mathrm{~mg}$ every 6 weeks \cite{Lala2020supp}. This was equal to approximately $50.8 \mathrm{~\mu g/mL}$, and we assume this to be the same in tissue, so we take $C_\mathrm{avg} = 5.08 \times 10^{-5} ~\mathrm{g/cm^3} = 2.05 \times 10^{14} ~ \mathrm{molec/cm^3}$, noting that the molecular mass of pembrolizumab is approximately $149,000 \mathrm{~g/mol}$ \cite{fda_pembrolizumabsupp}. Thus, we assume that $K_{T_\mathrm{ex} A_1} = 2.05 \times 10^{14} ~\mathrm{molec}/\mathrm{cm^3}$.
\subsection{Inhibition Constants}
We recall that for some species $X$, $K_X$ is denoted as the inhibition constant of $X$ in a term of the form
\begin{equation*}
    \frac{1}{1 + X/K_X}.
\end{equation*}
For simplicity, we assume that if $\overline{X}$ denotes the steady-state value of $X$, then
\begin{equation}
    \frac{1}{1 + \overline{X}/K_X} = \frac{1}{2} \implies K_X = \overline{X}.
\end{equation}
This implies that
\begin{align*}
K_{T_8T_r} = K_{T_1T_r} = K_{I_\upgamma T_r} = \overline{T_r} &= 1.45 \times 10^{5} \mathrm{~cell/cm^3}, \\
K_{CI_\upbeta} = K_{D_0I_\upbeta} = K_{KI_\upbeta} = \overline{I_\upbeta} &= 1.51 \times 10^{-6} \mathrm{~g/cm^3}, \\
K_{T_8I_{10}} = K_{T_\mathrm{ex}I_{10}} = \overline{I_{10}} &= 1.15 \times 10^{-10} \mathrm{~g/cm^3}, \\
K_{CQ^{T_8}} = \overline{Q^{T_8}} &= 6.68 \times 10^{5} ~\mathrm{molec/cm^3}, \\
K_{CQ^K} = \overline{Q^K} &= 3.62 \times 10^{6} ~\mathrm{molec/cm^3}, \\
K_{T_0^8T_A^r} = \tau_8^\mathrm{act}\overline{T_A^r} &= 1.56 \times 10^{6} ~(\mathrm{cell/cm^3})\mathrm{~day}, \\
K_{T_0^8 Q^{8\mathrm{LN}}} = \tau_8^\mathrm{act}\overline{Q^{8\mathrm{LN}}} &= 1.27 \times 10^{5} ~(\mathrm{molec/cm^3})\mathrm{~day}, \\
K_{T_A^8T_A^r} = \tau_{T_A^8}\overline{T_A^r} &= 3.80 \times 10^{6} ~(\mathrm{cell/cm^3})\mathrm{~day}, \\
K_{T_A^8 Q^{8\mathrm{LN}}} = \tau_{T_A^8}\overline{Q^{8\mathrm{LN}}} &= 3.10 \times 10^{5} ~(\mathrm{molec/cm^3})\mathrm{~day}, \\
K_{T_0^4 T_A^r} = \tau_4^\mathrm{act}\overline{T_A^r} &= 1.17 \times 10^{6} ~(\mathrm{cell/cm^3})\mathrm{~day}, \\
K_{T_0^4 Q^{1\mathrm{LN}}} = \tau_4^\mathrm{act}\overline{Q^{1\mathrm{LN}}} &= 6.41 \times 10^{5} ~(\mathrm{molec/cm^3})\mathrm{~day}, \\
K_{T_A^1 T_A^r} = \tau_{T_A^1}\overline{T_A^r} &= 3.23 \times 10^{6} ~(\mathrm{cell/cm^3})\mathrm{~day}, \\
K_{T_A^1 Q^{1\mathrm{LN}}} = \tau_{T_A^1}\overline{Q^{1\mathrm{LN}}} &= 1.76 \times 10^{6} ~(\mathrm{molec/cm^3})\mathrm{~day}.
\end{align*}
\subsection{\label{degratesection}Degradation Rates}
We recall the formula that the degradation rate of some species, $X$, is given by
\begin{equation}
    d_{X} = \frac{\ln 2}{t^X_{1/2}}
\end{equation}
where $t^X_{1/2}$ is the half-life of $X$.
\subsubsection{Estimate for $d_H$}
The half-life of HMGB1 was found to be approximately $3$ hours in the context of prostate cancer \cite{Zandarashvili2013supp}. We assume a similar value for MSI-H/dMMR CRC, and so
\begin{equation*}
    d_H = \frac{\ln 2}{3\mathrm{~hr}} = 5.55 \mathrm{~day^{-1}}.
\end{equation*}
\subsubsection{Estimate for $d_{S}$}
Surface calreticulin has a half-life of approximately $12$ hours \cite{Zhang2021calsupp, Goicoechea2003supp}. Thus, we have that
\begin{equation*}
    d_{S} = \frac{\ln 2}{12 \mathrm{~hr}} = 1.39 \mathrm{~day^{-1}}.
\end{equation*}
\subsubsection{Estimate for $d_{D_0}$}
The time taken for immature DCs to degrade is estimated to be 28 days in mice \cite{Ruedl2000supp}. We assume that this is similarly the case for humans, so that this corresponds to
\begin{equation*}
    d_{D_0} = \frac{1}{28 \mathrm{~day}} = 3.57 \times 10^{-2} \mathrm{~day^{-1}}.
\end{equation*}
\subsubsection{Estimate for $d_{D}$}
Mature DCs have a half-life of $1.5\text{ -- }2.9$ days in mice \cite{Kamath2002supp}. We assume that this is similarly the case for humans, and take $t_{1/2}^{D} = 2.2 \mathrm{~day}$ so that      
\begin{equation*}
    d_{D} = \frac{\ln 2}{2.2 \mathrm{~day}} = 3.15 \times 10^{-1} \mathrm{~day^{-1}}.
\end{equation*}
\subsubsection{Estimate for $d_{T_0^8}$}
The half-life of naive CD8+ T cells in the lymph node was estimated to be $21.5$ days in \cite{Takada2009supp} so that
\begin{equation*}
    d_{T_0^8} = \frac{\ln 2}{21.5 \mathrm{~day}} = 3.22 \times 10^{-2} \mathrm{~day^{-1}}.
\end{equation*}
\subsubsection{Estimate for $d_{T_8}$ and $d_{T_\mathrm{ex}}$}
It was measured in \cite{Hellerstein1999supp} that the mean death rate of circulating CD8+ T cells in HIV seronegative patients was $0.009 \mathrm{~day^{-1}}$. We assume that this is the case for MSI-H/dMMR CRC, and so we set $d_{T_8} = d_{T_\mathrm{ex}} = 0.009 \mathrm{~day^{-1}}$.
\subsubsection{Estimate for $d_{T_0^4}$}
The half-life of naive CD4+ T cells in the lymph node was estimated to be $17.2$ days in \cite{Takada2009supp} so that
\begin{equation*}
    d_{T_0^4} = \frac{\ln 2}{17.2 \mathrm{~day}} = 4.03 \times 10^{-2} \mathrm{~day^{-1}}.
\end{equation*}
\subsubsection{Estimate for $d_{T_1}$}
It was measured in \cite{Hellerstein1999supp} that the mean death rate of circulating CD4+ T cells in HIV seronegative patients was $0.008 \mathrm{~day^{-1}}$. We assume that this is the case for Th1 cells in MSI-H/dMMR CRC, and so we set $d_{T_1} = 0.008 \mathrm{~day^{-1}}$.
\subsubsection{Estimate for $d_{T_0^r}$}
The death rate of naive Tregs in the lymph node was estimated to be $2.2 \times 10^{-3} \mathrm{~day^{-1}}$ in \cite{Kumbhari2020n2supp}, and we assume that the death rate in MSI-H/dMMR CRC is similar, so that
\begin{equation*}
    d_{T_0^r} = 2.2 \times 10^{-3} \mathrm{~day^{-1}}.
\end{equation*}
\subsubsection{Estimate for $d_{T_r}$}
The mean half-life of Tregs in healthy adults was measured to be approximately $11$ days in \cite{VukmanovicStejic2006supp}. We assume that this is similarly the case for MSI-H/dMMR CRC and that this corresponds to
\begin{equation*}
    d_{T_{r}} = \frac{\ln 2}{11 \mathrm{~day}} = 6.30 \times 10^{-2} \mathrm{~day^{-1}}.
\end{equation*}
\subsubsection{Estimate for $d_{M_0}$}
The lifespan for naive macrophages was found in humans to be approximately $1.37$ days on average \cite{Patel2017supp}. This corresponds to
\begin{equation*}
    d_{M_0} = \frac{1}{1.37\mathrm{~day}} = 0.73 \mathrm{~day^{-1}}.
\end{equation*}
\subsubsection{Estimate for $d_{M_1}$}
The lifespan for M1 macrophages was found in humans to be approximately $1.01$ days on average \cite{Patel2017supp}. This corresponds to
\begin{equation*}
    d_{M_1} = \frac{1}{1.01\mathrm{~day}} = 0.99 \mathrm{~day^{-1}}.
\end{equation*}
\subsubsection{Estimate for $d_{M_2}$}
The lifespan for M2 macrophages was found in humans to be approximately $7.41$ days on average \cite{Patel2017supp}. This corresponds to
\begin{equation*}
    d_{M_2} = \frac{1}{7.41\mathrm{~day}} = 1.35 \times 10^{-1} \mathrm{~day^{-1}}.
\end{equation*}
\subsubsection{Estimate for $d_{K_0}$ and $d_{K}$}
The half-life of human NK cells varies between $1$ and $2$ weeks \cite{Wu2020supp, Vivier2008supp, Lowry2017supp}. We assume that the half-lives of resting and activated NK cells are both equal to $10$ days, so that
\begin{equation*}
    d_{K_0} = d_{K} = \frac{\ln 2}{10 \mathrm{~day}} = 6.93 \times 10^{-2} \mathrm{~day^{-1}}.
\end{equation*}
\subsubsection{Estimate for $d_{I_2}$}
The half-life of IL-2 varies between $5$ and $7$ minutes \cite{Lotze1985n1supp}. We take $t_{1/2}^{I_2} = 6.9 \mathrm{~min}$ so that
\begin{equation*}
    d_{I_2} = \frac{\ln 2}{6.9 \mathrm{~min}} = 1.45 \times 10^{2} \mathrm{~day^{-1}}.
\end{equation*}
\subsubsection{Estimate for $d_{I_\upgamma}$}
The half-life of IFN-$\upgamma$ varies between $25$ and $35$ minutes \cite{Balachandran2013supp}. We take $t_{1/2}^{I_\upgamma}$ to be $30$ minutes so that
\begin{equation*}
    d_{I_\upgamma} = \frac{\ln 2}{30 \mathrm{~min}} = 3.33 \times 10^{1} \mathrm{~day^{-1}}.
\end{equation*}
\subsubsection{Estimate for $d_{I_\upalpha}$}
The half-life of TNF varies between $15$ and $30$ minutes \cite{Ma2015supp, Oliver1993-vlsupp}. We take $t_{1/2}^{I_\upalpha}$ to be $18.2$ minutes, so that
\begin{equation*}
    d_{I_\upalpha} = \frac{\ln 2}{18.2 \mathrm{~min}} = 5.48 \times 10^{1} \mathrm{~day^{-1}}.
\end{equation*}
\subsubsection{Estimate for $d_{I_\upbeta}$}
The half-life of active TGF-$\upbeta$ is approximately $2\text{ -- }3$ minutes \cite{TiradoRodriguez2014supp}. We take $t_{1/2}^{I_\upbeta} = 2.5 \mathrm{~min}$, so that
\begin{equation*}
    d_{I_\upbeta} = \frac{\ln 2}{2.5 \mathrm{~min}} = 3.99 \times 10^{2} \mathrm{~day^{-1}}.
\end{equation*}
\subsubsection{Estimate for $d_{I_{10}}$}
The half-life of IL-10 varies between $2.7$ and $4.5$ hours \cite{Huhn1997supp}. We take $t_{1/2}^{I_{10}} = 2.7 \mathrm{~hr}$ so that
\begin{equation*}
    d_{I_{10}} = \frac{\ln 2}{2.7 \mathrm{~hr}} = 6.16 \mathrm{~day^{-1}}.
\end{equation*}
\subsubsection{Estimate for $d_{P_D}$}
The median lower bound on PD-1 half-life on human peripheral blood mononuclear cells was found to be $49.5$ hours based on leucine enrichment in \cite{Lassman2021supp}. Hence, we take $t_{1/2}^{P_D} = 49.5~\mathrm{hr}$ so that
\begin{equation*}
    d_{P_D} = \frac{\ln 2}{49.5 \mathrm{~hr}} = 3.36 \times 10^{-1} \mathrm{~day^{-1}}.
\end{equation*}
\subsubsection{Estimate for $d_{A_1}$}
The half-life of pembrolizumab varies between $22$ and $27$ days \cite{Yan2023supp, Longoria2016supp, Dang2015supp}. We take it to be $23.7$ days, consistent with models from Li et al.\ and Ahamadi et al.\ \cite{Li2017supp, Li2019pembrosupp, Ahamadi2016supp} so that
\begin{equation*}
    d_{A_1} = \frac{\ln 2}{23.7 \mathrm{~day}} = 2.92 \times 10^{-2} \mathrm{~day^{-1}}.
\end{equation*}
\subsubsection{Estimate for $d_{P_L}$}
The half-life of fully glycosylated PD-L1 is approximately 12 hours \cite{Cha2019supp}, with PD-L1 on immune cells being heavily glycosylated \cite{Li2016supp}. Thus, we take $t_{1/2}^{P_L} = 12 \mathrm{~hr}$ so that
\begin{equation*}
    d_{P_L} = \frac{\ln 2}{12 \mathrm{~hr}} = 1.39 \mathrm{~day^{-1}}.
\end{equation*}
\subsection{DAMP Parameters}
\subsubsection{Estimates for $H$}
Considering \eqref{Heqn} at steady state, we have that
\begin{equation*}
    \lambda_{HN_c}\overline{N_c} - d_{H}\overline{H}=0.
\end{equation*}
This leads to
\begin{align*}
    \lambda_{HN_c} &= 1.52 \times 10^{-14} ~\left(\mathrm{g/cell}\right)\mathrm{day^{-1}}.
\end{align*}
\subsubsection{Estimates for $S$}
Considering \eqref{Seqn} at steady state leads to the equation
\begin{equation*}
    \lambda_{SN_c}\overline{N_c} - d_{S}\overline{S} = 0.
\end{equation*}
This leads to
\begin{equation*}
    \lambda_{SN_c} = 1.23 \times 10^{-14} ~\left(\mathrm{g/cell}\right)\mathrm{day^{-1}}.
\end{equation*}
\subsection{Cytokine Production Parameters\label{cytokineparamestsection}}
To estimate many of the cytokine production parameters, we consider \eqref{il2eqn}~--~\eqref{il10eqn} at steady state and use the data from \cite{Cui2023supp}. For each immune cell, we assume that each cytokine's corresponding gene expression is proportional to its production rate by that cell.
\subsubsection{Estimates for $I_2$}
Using values from \cite{Cui2023supp} and considering \eqref{il2eqn} at steady state, or equivalently considering \eqref{il2qssa}, leads to the equations
\begin{equation*}
    \frac{\lambda_{I_2 T_8}}{0.114615876287774} = \frac{\lambda_{I_2 T_1}}{0.335763693785869},
\end{equation*}
and
\begin{equation*}
    \lambda_{I_2 T_8}\overline{T_8} + \lambda_{I_2 T_1}\overline{T_1} - d_{I_2}\overline{I_2} = 0.
\end{equation*}
Solving these simultaneously leads to
\begin{align*}
    \lambda_{I_2 T_8} &= 7.44 \times 10^{-16} ~\left(\mathrm{g/cell}\right)\mathrm{day^{-1}}, \\
    \lambda_{I_2 T_1} &= 2.18 \times 10^{-15} ~\left(\mathrm{g/cell}\right)\mathrm{day^{-1}}.
\end{align*}
Consequently, considering \eqref{il2qssa}, we have that
\begin{equation*}
    I_2(0) = \frac{1}{d_{I_2}}\left(\lambda_{I_2 T_8}T_8(0) + \lambda_{I_2 T_1}T_1(0)\right) = 2.81 \times 10^{-12} \mathrm{~g/cm^3}.
\end{equation*}
\subsubsection{Estimates for $I_\upgamma$}
Using values from \cite{Cui2023supp} and considering \eqref{ifngammaeqn} at steady state, or equivalently considering \eqref{ifngammaqssa}, leads to the equations
\begin{equation*}
\frac{\lambda_{I_{\upgamma} T_8}}{2 \times 0.0539973307184416} = \frac{\lambda_{I_{\upgamma} T_1}}{2 \times 0.0188926732394088} = \lambda_{I_{\upgamma} K},
\end{equation*}
and
\begin{equation*}
\left(\lambda_{I_{\upgamma} T_8}\overline{T_8} + \lambda_{I_{\upgamma} T_1}\overline{T_1}\right)\frac{1}{2} + \lambda_{I_{\upgamma} K}\overline{K} - d_{I_\upgamma}\overline{I_\upgamma}=0.
\end{equation*}
Solving these simultaneously leads to
\begin{align*}
    \lambda_{I_{\upgamma} T_8} &= 1.26 \times 10^{-17} ~\left(\mathrm{g/cell}\right)\mathrm{day^{-1}}, \\
    \lambda_{I_{\upgamma} T_1} &= 4.40 \times 10^{-18} ~\left(\mathrm{g/cell}\right)\mathrm{day^{-1}},\\
    \lambda_{I_{\upgamma} K} &= 1.16 \times 10^{-16} ~\left(\mathrm{g/cell}\right)\mathrm{day^{-1}}.
\end{align*}
Consequently, considering \eqref{ifngammaqssa}, we have that
\begin{equation*}
    I_\upgamma(0) = \frac{1}{d_{I_\upgamma}}\left[\left(\lambda_{I_{\upgamma} T_8}T_8(0) + \lambda_{I_{\upgamma} T_1}T_1(0)\right)\frac{1}{1+T_r(0)/K_{I_{\upgamma}T_r}} + \lambda_{I_{\upgamma} K}K(0)\right] = 1.82 \times 10^{-11} \mathrm{~g/cm^3}.
\end{equation*}
\subsubsection{Estimates for $I_\upalpha$}
Using values from \cite{Cui2023supp} and considering \eqref{tnfeqn} at steady state, or equivalently considering \eqref{tnfqssa}, leads to the equations
\begin{equation*}
\medmath{\frac{\lambda_{I_{\upalpha}T_8}}{0.0654443776961264} = \frac{\lambda_{I_{\upalpha}T_1}}{0.108187215112606} = \frac{\lambda_{I_{\upalpha}M_1}}{0.0396575742078822} = \frac{\lambda_{I_{\upalpha}K}}{0.114108294134927}},
\end{equation*}
and
\begin{equation*}
\lambda_{I_{\upalpha}T_8}\overline{T_8} + \lambda_{I_{\upalpha}T_1}\overline{T_1} + \lambda_{I_{\upalpha}M_1}\overline{M_1} + \lambda_{I_{\upalpha}K}\overline{K} - d_{I_{\upalpha}}\overline{I_{\upalpha}} = 0.
\end{equation*}
Solving these simultaneously leads to
\begin{align*}
    \lambda_{I_{\upalpha} T_8} &= 3.24 \times 10^{-16} ~\left(\mathrm{g/cell}\right)\mathrm{day^{-1}}, \\
    \lambda_{I_{\upalpha} T_1} &= 5.36 \times 10^{-16} ~\left(\mathrm{g/cell}\right)\mathrm{day^{-1}},\\
    \lambda_{I_{\upalpha} M_1} &= 1.97 \times 10^{-16} ~\left(\mathrm{g/cell}\right)\mathrm{day^{-1}},\\
    \lambda_{I_{\upalpha} K} &= 5.66 \times 10^{-16} ~\left(\mathrm{g/cell}\right)\mathrm{day^{-1}}.
\end{align*}
Consequently, considering \eqref{tnfqssa}, we have that
\begin{equation*}
    I_\upalpha(0) = \frac{1}{d_{I_{\upalpha}}}\left(\lambda_{I_{\upalpha}T_8}T_8(0) + \lambda_{I_{\upalpha}T_1}T_1(0) + \lambda_{I_{\upalpha}M_1}M_1(0) + \lambda_{I_{\upalpha}K}K(0)\right) = 5.85 \times 10^{-11} \mathrm{~g/cm^3}.
\end{equation*}
\subsubsection{Estimates for $I_\upbeta$}
Estimating the production constants for TGF-$\upbeta$ is slightly more complicated than it is for other cytokines. We assume that the results for fibroblastic reticular cells in \cite{Cui2023supp} translate directly to results for cancer-associated fibroblasts (CAFs), which are considered to be all fibroblasts found in the TME \cite{Zhang2023supp}. We assume that at steady state, CAFs produce twice as much TGF-$\upbeta$ as cancer cells in the TME, and denote the production rate of TGF-$\upbeta$ by CAFs as $\lambda_{I_\upbeta C_F}$. This, in conjunction with values from \cite{Cui2023supp}, and considering \eqref{tgfbetaeqn} at steady state, or equivalently considering \eqref{ibetaqssa}, leads to the equations
\begin{equation*}
    \frac{\lambda_{I_{\upbeta}C_F}}{2} = \frac{\lambda_{I_{\upbeta}C}}{1},
\end{equation*}
and
\begin{equation*}
\frac{\lambda_{I_{\upbeta}C_F}}{0.175283003265127} = \frac{\lambda_{I_{\upbeta}T_r}}{0.507677682409403} = \frac{\lambda_{I_{\upbeta}M_2}}{0.63070357154901},
\end{equation*}
and
\begin{equation*}
\lambda_{I_{\upbeta}C}\overline{C} + \lambda_{I_{\upbeta}T_r}\overline{T_r} + \lambda_{I_{\upbeta}M_2}\overline{M_2} - d_{I_{\upbeta}}\overline{I_{\upbeta}} = 0.
\end{equation*}
Solving these simultaneously leads to
\begin{align*}
    \lambda_{I_{\upbeta} C} &= 1.33 \times 10^{-11} ~\left(\mathrm{g/cell}\right)\mathrm{day^{-1}},\\
    \lambda_{I_{\upbeta} T_r} &= 7.68 \times 10^{-11} ~\left(\mathrm{g/cell}\right)\mathrm{day^{-1}}, \\
    \lambda_{I_{\upbeta}M_2} &= 9.54 \times 10^{-11} ~\left(\mathrm{g/cell}\right)\mathrm{day^{-1}}.
\end{align*}
Consequently, considering \eqref{ibetaqssa}, we have that
\begin{equation*}
    I_\upbeta(0) = \frac{1}{d_{I_{\upbeta}}}\left(\lambda_{I_{\upbeta}C}C(0) + \lambda_{I_{\upbeta}T_r}T_r(0) + \lambda_{I_{\upbeta}M_2}M_2(0)\right) = 9.32 \times 10^{-7} \mathrm{~g/cm^3}.
\end{equation*}
\subsubsection{Estimates for $I_{10}$}
Amongst $48$ different cell lines tested, it was found in \cite{Gastl1993supp} that cancer IL-10 production was maximised in cell lines derived from colon carcinomas. As such, we assume that at steady state, cancer production of IL-10 is equal to half of that by $M_2$ macrophages. We assume that the enhancement factor of IL-2 for IL-10 production by Tregs is similar in CRC to that of inflammatory bowel disease and use the estimate of $\lambda_{I_{10}I_2} = 3$ that was used in \cite{Lo2016supp}. This, in conjunction with values from \cite{Cui2023supp}, and considering \eqref{il10eqn} at steady state leads to the equations
\begin{equation*}
    \frac{\lambda_{I_{10}C}}{1} = \frac{\lambda_{I_{10}M_2}}{2},
\end{equation*}
and
\begin{equation*}
\frac{\lambda_{I_{10}M_2}}{1} = \frac{\lambda_{I_{10}T_{r}}\left(1 + \frac{\lambda_{I_{10}I_{2}}}{2}\right)}{0.472157630570674},
\end{equation*}
and
\begin{equation*}
\lambda_{I_{10}C}\overline{C} + \lambda_{I_{10}M_2}\overline{M_2} + \lambda_{I_{10}T_{r}}\left(1 + \frac{\lambda_{I_{10}I_{2}}}{2}\right)\overline{T_r} - d_{I_{10}}\overline{I_{10}} = 0.
\end{equation*}
Solving these simultaneously leads to
\begin{align*}
    \lambda_{I_{10} C} &= 1.94 \times 10^{-17} ~\left(\mathrm{g/cell}\right)\mathrm{day^{-1}},\\
    \lambda_{I_{10} M_2} &= 3.89 \times 10^{-17} ~\left(\mathrm{g/cell}\right)\mathrm{day^{-1}},\\
    \lambda_{I_{10} T_r} &= 7.34 \times 10^{-18} ~\left(\mathrm{g/cell}\right)\mathrm{day^{-1}}.
\end{align*}
\subsection{Parameters for DCs, Macrophages, and NK Cells}
\subsubsection{Estimates for $D_0$ and $D$ \label{DCestsection}}
Adding \eqref{D0eqn} and \eqref{Deqn} at steady state leads to
\begin{equation*}
    \mathcal{A}_{D_0} - \frac{\lambda_{D_0K} \overline{D_0}\overline{K}}{2} - d_{D_0}\overline{D_0} - \lambda_{DD^\mathrm{LN}}\overline{D} - d_{D}\overline{D} =0.
\end{equation*}
We assume that HMGB1 is the most potent inducer of DC maturation, and as such, at steady state, we assume that
\begin{equation*}
\frac{\lambda_{DH}}{2 \times 10} = \frac{\lambda_{DS}}{2 \times 1}.
\end{equation*}
In \cite{Piccioli2002supp}, it was also shown that the percentage of immature DCs that were lysed as a result of NK cells is roughly linear in the ratio of NK cells to immature DCs. When a 1:1 ratio of activated NK cells to immature DCs is present, after 24 hours, roughly $35.5 \%$ of immature DCs are lysed, whereas if a 5:1 ratio is present, $85.5\%$ of immature DCs are lysed. At steady state, the ratio of NK cells to immature DCs is $\approx 2.39:1$, corresponding to an approximate $52.85\%$ being lysed. However, if we consider only immature DC loss due to degradation, after $24$ hours, only $1 - e^{-d_{D_0}} \approx 3.54\%$ is lost to it. Thus, we assume at steady state that
\begin{equation*}
    \frac{\lambda_{D_0K} \overline{D_0}\overline{K}}{2 \times 0.5285} = \frac{d_{D_0}\overline{D_0}}{0.0354} \implies \lambda_{D_0K} = 2.21 \times 10^{-7} ~\left(\mathrm{cell/cm^3}\right)^{-1}\mathrm{day^{-1}}.
\end{equation*}
Considering \eqref{Deqn} at steady state leads to
\begin{equation*}
    \frac{\lambda_{DH}\overline{D_0}}{2} + \frac{\lambda_{DS}\overline{D_0}}{2} - \lambda_{DD^\mathrm{LN}}\overline{D} - d_{D}\overline{D} =0.
\end{equation*}
Finally, it was found in \cite{Verdijk2009supp} that only a limited number of DCs migrate up to the TDLN, with at most 4\% of DCs reaching the TDLN in melanoma patients when DCs were injected intradermally. We assume at steady state that this holds, too, for MSI-H/dMMR CRC. Taking into account that only $e^{-d_D\tau_m}$ of mature DCs that leave the TS survive their migration to the TDLN, we have that
\begin{equation*}
    \frac{\lambda_{DD^\mathrm{LN}}}{0.04e^{d_D\tau_m}} = \frac{d_D}{1-0.04e^{d_D\tau_m}}.
\end{equation*}
Solving these simultaneously leads to
\begin{align*}
    \mathcal{A}_{D_0} &= 9.89 \times 10^{5} ~\left(\mathrm{cell/cm^{3}}\right)\mathrm{day^{-1}}, \\
    \lambda_{DH} &= 1.98 \times 10^{-1} ~\mathrm{day^{-1}}, \\
    \lambda_{DS} &= 1.98 \times 10^{-2} ~\mathrm{day^{-1}}, \\
    \lambda_{DD^\mathrm{LN}} &= 1.68 \times 10^{-2} ~\mathrm{day^{-1}}.
\end{align*}
To estimate the steady states and initial conditions for $D^\mathrm{LN}$, we consider \eqref{DLNeqn} at steady state, which leads to
\begin{equation*}
    \frac{V_\mathrm{TS}}{V_\mathrm{LN}}\lambda_{DD^\mathrm{LN}}e^{-d_D \tau_m}\overline{D} - d_{D}\overline{D^\mathrm{LN}} = 0 \implies \overline{D^\mathrm{LN}} = \frac{V_\mathrm{TS}\lambda_{DD^\mathrm{LN}}e^{-d_D \tau_m}\overline{D}}{V_\mathrm{LN}d_D} = 6.04 \times 10^{6} ~\mathrm{cell/cm^3}.
\end{equation*}
We set the initial condition for $D^\mathrm{LN}$ to be such that $D^\mathrm{LN}(0)/\overline{D^\mathrm{LN}} = D(0)/\overline{D} \implies D^\mathrm{LN}(0) = 1.05 \times 10^{7} ~\mathrm{cell/cm^3}$.
\subsubsection{Estimates for $M_0$, $M_1$, and $M_2$}
To estimate the macrophage production constants, we consider \eqref{M0eqn}, \eqref{M1eqn}, \eqref{M2eqn} at steady state, and use the data from \cite{Cui2023supp}. We assume that the magnitude of response to a specific cytokine is proportional to its corresponding macrophage polarisation rate, where the response is defined as the Euclidean distance between the centroid vectors of cytokine-treated macrophages and phosphate-buffered saline (PBS)-treated macrophages. \\~\\
Adding \eqref{M0eqn}, \eqref{M1eqn}, and \eqref{M2eqn} at steady state leads to
\begin{equation*}
    \mathcal{A}_{M_0} - d_{M_0}\overline{M_0} - d_{M_1}\overline{M_1}-d_{M_2}\overline{M_2} = 0 \implies \mathcal{A}_{M_0} = 1.00 \times 10^{6} ~\left(\mathrm{cell/cm^{3}}\right)\mathrm{day^{-1}}.     
\end{equation*}
Using values from \cite{Cui2023supp}, and considering \eqref{M0eqn} and \eqref{M1eqn} at steady state leads to the equations
\begin{align*}
    &\mathcal{A}_{M_0} - \frac{\lambda_{M_1 I_\upalpha}\overline{M_0}}{2} - \frac{\lambda_{M_1 I_\upgamma}\overline{M_0}}{2} - \frac{\lambda_{M_2 I_{10}}\overline{M_0}}{2} - \frac{\lambda_{M_2 I_\upbeta}\overline{M_0}}{2} - d_{M_0}\overline{M_0} = 0,\\
    & \frac{\lambda_{M_1 I_\upalpha}\overline{M_0}}{2} + \frac{\lambda_{M_1 I_\upgamma}\overline{M_0}}{2} + \frac{\lambda_{MI_{\upgamma}}\overline{M_2}}{2} + \frac{\lambda_{MI_{\upalpha}}\overline{M_2}}{2} - \frac{\lambda_{MI_{\upbeta}}\overline{M_1}}{2} - d_{M_1}\overline{M_1} = 0, \\
    &\frac{\lambda_{M_1 I_\upalpha}}{2 \times 10.77} = \frac{\lambda_{M_1 I_\upgamma}}{2 \times 12.54} = \frac{\lambda_{M_2 I_{10}}}{2 \times 6.81} = \frac{\lambda_{M_2I_\upbeta}}{2 \times 7.63}.      
\end{align*}
We assume that IFN-$\upgamma$ repolarises M2 macrophages to the M1 phenotype slightly more potently than TNF. Hence, at steady state
\begin{equation*}
    \frac{\lambda_{MI_{\upgamma}}}{2 \times 6} = \frac{\lambda_{MI_{\upalpha}}}{2 \times 5}.
\end{equation*}
At steady state, we also assume that
\begin{equation*}
    \frac{1}{\overline{M_2}}\left(\frac{\lambda_{MI_\upgamma}}{2} + \frac{\lambda_{MI_{\upalpha}}}{2}\right) = \frac{1}{\overline{M_1}}\frac{\lambda_{MI_\upbeta}}{2}.
\end{equation*}
Solving these simultaneously leads to
\begin{align*}
\lambda_{M_1I_\upalpha} & = 6.92 \times 10^{-1} \mathrm{~day^{-1}}, \\
\lambda_{M_1I_\upgamma} & = 8.06 \times 10^{-1} \mathrm{~day^{-1}}, \\
\lambda_{M_2I_{10}} & = 4.38 \times 10^{-1} \mathrm{~day^{-1}}, \\
\lambda_{M_2I_\upbeta} & = 4.90 \times 10^{-1} \mathrm{~day^{-1}}, \\
\lambda_{MI_{\upgamma}} & = 1.71 \times 10^{-2} \mathrm{~day^{-1}}, \\
\lambda_{MI_{\upalpha}} & = 1.43 \times 10^{-2} \mathrm{~day^{-1}}, \\
\lambda_{MI_{\upbeta}} & = 8.11 \times 10^{-3} \mathrm{~day^{-1}}.
\end{align*}
\subsubsection{Estimates for $K_0$ and $K$}
To estimate NK cell production parameters, we do a similar process to macrophages. Adding \eqref{K0eqn} and \eqref{Keqn} at steady state leads to
\begin{equation*}
    \mathcal{A}_{K_0} - d_{K_0}\overline{K_0} - d_{K}\overline{K}= 0 \implies \mathcal{A}_{K_0} = 3.67 \times 10^{5} ~\left(\mathrm{cell/cm^{3}}\right)\mathrm{day^{-1}}.
\end{equation*}
Considering \eqref{Keqn} at steady state leads to
\begin{equation*}
    \frac{1}{2}\left(\frac{\lambda_{KI_2}\overline{K_0}}{2} + \frac{\lambda_{KD_0}\overline{K_0}}{2} + \frac{\lambda_{KD}\overline{K_0}}{2}\right) - d_{K}\overline{K} = 0.
\end{equation*}
We assume that mature DCs are more potent activators of NK cells than immature DCs, so that at steady state
\begin{equation*}
    \frac{\lambda_{KD}}{2 \times 5} = \frac{\lambda_{KD_0}}{2 \times 1}.
\end{equation*}
We finally assume that DC-mediated NK-cell activation is twice as potent as cytokine-induced activation at steady state, so that
\begin{equation*}
    \frac{\lambda_{KD_0}/2 + \lambda_{KD}/2}{2} = \frac{\lambda_{KI_2}/2}{1}.
\end{equation*}
Solving these simultaneously leads to
\begin{align*}
    \lambda_{KI_2} &= 9.24 \times 10^{-1} \mathrm{~day^{-1}}, \\
    \lambda_{KD_0} &= 3.08 \times 10^{-1} \mathrm{~day^{-1}}, \\
    \lambda_{KD} &= 1.54 \mathrm{~day^{-1}}.
\end{align*}
\subsection{T Cell Parameters and Estimates}
\subsubsection{Estimates for $T_0^8$, $T_A^8$, $T_8$, and $T_\mathrm{ex}$}
Considering \eqref{naivecd8eqn} at steady state leads to
\begin{equation*}
    \mathcal{A}_{T_0^8} - \overline{R^8} - d_{T_0^8}\overline{T_0^8} = 0,
\end{equation*}
and in particular,
\begin{equation*}
    \overline{R^8} = \frac{\lambda_{T_0^8 T_A^8}e^{-d_{T_0^8}\tau_8^\mathrm{act}}\overline{D^\mathrm{LN}}\overline{T_0^8}}{4}.
\end{equation*}
Considering \eqref{TA8n8maxeqn} at steady state leads to
\begin{equation*}
    \frac{2^{n^8_\mathrm{max}}e^{-d_{T_0^8}\tau_{T_A^8}} \overline{R^8}}{4} - \lambda_{T_A^8T_8}\overline{T_A^8} - d_{T_8} \overline{T_A^8} = 0.
\end{equation*}
We first consider the case where no pembrolizumab is present. Considering \eqref{t8eqn} and \eqref{Texeqn} at steady state leads to
\begin{align*}
    \frac{V_\mathrm{LN}}{V_\mathrm{TS}}\lambda_{T_A^8T_8}e^{-d_{T_8} \tau_a}\overline{T_A^8} + \frac{\lambda_{T_8I_2}\overline{T_8}}{4}- \frac{\lambda_{T_8C}\overline{T_8}}{2} - \frac{d_{T_8}\overline{T_8}}{2}&=0, \\
    \frac{\lambda_{T_8C}\overline{T_8}}{2} - \frac{d_{T_\mathrm{ex}}\overline{T_\mathrm{ex}}}{2}&=0.
\end{align*}
We assume that at steady state, $95\%$ of positive $T_8$ growth is due to $T_A^8$ migration to the TS, and the other $5\%$ is due to IL-2-induced proliferation. Thus, we have that
\begin{equation*}
    \frac{V_\mathrm{LN}}{V_\mathrm{TS}}\frac{\lambda_{T_A^8T_8}e^{-d_{T_8} \tau_a}\overline{T_A^8}}{0.95} = \frac{\lambda_{T_8I_2}\overline{T_8}/4}{0.05}.
\end{equation*}
To determine $\lambda_{T_\mathrm{ex}A_1}$, we assume that when pembrolizumab is present, at steady state, $20\%$ of exhausted CD8+ T cells are reinvigorated. That is, we assume that
\begin{equation*}
\frac{\lambda_{T_\mathrm{ex}A_1}\overline{T_\mathrm{ex}}/2}{0.2}=\frac{d_{T_\mathrm{ex}}\overline{T_\mathrm{ex}}/2}{0.8}.
\end{equation*}
Solving these equations simultaneously leads to
\begin{align*}
\mathcal{A}_{T_0^8} &= 3.88 \times 10^{5} ~\left(\mathrm{cell/cm^{3}}\right)\mathrm{day^{-1}}, \\
\lambda_{T_0^8 T_A^8} &= 1.12 \times 10^{-10} ~\left(\mathrm{cell/cm^3}\right)^{-1}\mathrm{day^{-1}}, \\
\overline{R^8} &= 1.90 \times 10^{3} ~\left(\mathrm{cell/cm^{3}}\right)\mathrm{day^{-1}}, \\
\lambda_{T_A^8T_8} &= 4.75 \times 10^{-1} ~\mathrm{day}^{-1}, \\
\lambda_{T_8I_2} &= 1.61 \times 10^{-3} ~\mathrm{day}^{-1}, \\
\lambda_{T_8C} &= 7.08 \times 10^{-3} ~\mathrm{day}^{-1}, \\
\lambda_{T_\mathrm{ex}A_1} &= 2.25 \times 10^{-3} ~\mathrm{day}^{-1}.
\end{align*}
\subsubsection{Estimates for $T_0^4$, $T_A^1$, and $T_1$}
Considering \eqref{naivecd4eqn} at steady state leads to
\begin{equation*}
    \mathcal{A}_{T_0^4} - \overline{R^1} - d_{T_0^4}\overline{T_0^4} = 0,
\end{equation*}
where
\begin{equation*}
    \overline{R^1} = \frac{\lambda_{T_0^4 T_A^1} e^{-d_{T_0^4}\tau_4^\mathrm{act}}\overline{D^\mathrm{LN}}\overline{T_0^4}}{4}.
\end{equation*}
Considering \eqref{TA1n1maxeqn} at steady state leads to
\begin{equation*}
    \frac{2^{n^1_\mathrm{max}}e^{-d_{T_0^4}\tau_{T_A^1}} \overline{R^1}}{4} - \lambda_{T_A^1T_1}\overline{T_A^1}-d_{T_1}\overline{T_A^1}=0.
\end{equation*}
We assume, like for CD8+ T cells, that at steady state, $95\%$ of positive $T_1$ growth is due to $T_A^1$ migration to the TS, and the other $5\%$ is due to IL-2-induced proliferation. Thus, we have that
\begin{equation*}
    \frac{V_\mathrm{LN}}{V_\mathrm{TS}}\frac{\lambda_{T_A^1T_1}e^{-d_{T_1} \tau_a}\overline{T_A^1}}{0.95} = \frac{\lambda_{T_1I_2}\overline{T_1}/4}{0.05}.
\end{equation*}
Based on murine data from \cite{Tan2025supp}, we assume that at steady state, $20\%$ of Th1 cells are converted to Tregs. That is, we assume that
\begin{equation*}
\frac{\lambda_{T_1T_r}\overline{T_1}/2}{0.2} = \frac{d_{T_1}\overline{T_1}}{0.8}.
\end{equation*}
Finally, considering \eqref{th1eqn} at steady state leads to
\begin{align*}
    \frac{V_\mathrm{LN}}{V_\mathrm{TS}}\lambda_{T_A^1 T_1}e^{-d_{T_1} \tau_a}\overline{T_A^1} + \frac{\lambda_{T_1I_2}\overline{T_1}}{4} - \frac{\lambda_{T_1T_r}\overline{T_1}}{2}- d_{T_1}\overline{T_1} &= 0.
\end{align*}
Solving these equations simultaneously leads to
\begin{align*}
\mathcal{A}_{T_0^4} &= 1.77 \times 10^{5} ~\left(\mathrm{cell/cm^{3}}\right)\mathrm{day^{-1}}, \\
\lambda_{T_0^4 T_A^1} &= 4.05 \times 10^{-10} ~\left(\mathrm{cell/cm^3}\right)^{-1}\mathrm{day^{-1}}, \\
\overline{R^1} &= 2.48 \times 10^{3} ~\left(\mathrm{cell/cm^{3}}\right)\mathrm{day^{-1}}, \\
\lambda_{T_A^1 T_1} &= 2.66 \times 10^{-2} ~\mathrm{day}^{-1}, \\
\lambda_{T_1I_2} &= 2.00 \times 10^{-3} ~\mathrm{day}^{-1}, \\
\lambda_{T_1T_r} &= 4.00 \times 10^{-3} ~\mathrm{day}^{-1}.
\end{align*}
\subsubsection{Estimates for $T_0^r$, $T_A^r$, and $T_r$}
Considering \eqref{naivetregeqn} at steady state leads to
\begin{equation*}
    \mathcal{A}_{T_0^r} - \overline{R^r} - d_{T_0^r}\overline{T_0^r} = 0,
\end{equation*}
where
\begin{equation*}
    \overline{R^r} = \lambda_{T_0^r T_A^r} e^{-d_{T_0^r}\tau_r^\mathrm{act}}\overline{D^\mathrm{LN}}\overline{T_0^r}.
\end{equation*}
Considering \eqref{TArnrmaxeqn} at steady state leads to
\begin{equation*}
    2^{n^r_\mathrm{max}}e^{-d_{T_0^r}\tau_{T_A^r}} \overline{R^r} - \lambda_{T_A^rT_r}\overline{T_A^r}-d_{T_r}\overline{T_A^r}=0.
\end{equation*}
Finally, considering \eqref{tregeqn} at steady state leads to
\begin{align*}
    \frac{V_\mathrm{LN}}{V_\mathrm{TS}}\lambda_{T_A^r T_r}e^{-d_{T_r} \tau_a}\overline{T_A^r} + \frac{\lambda_{T_1T_r}\overline{T_1}}{2} - d_{T_r}\overline{T_r} &= 0.
\end{align*}
Solving these equations simultaneously leads to
\begin{align*}
\mathcal{A}_{T_0^r} &= 4.43 \times 10^{4} ~\left(\mathrm{cell/cm^{3}}\right)\mathrm{day^{-1}}, \\
\lambda_{T_0^r T_A^r} &= 4.24 \times 10^{-8} ~\left(\mathrm{cell/cm^3}\right)^{-1}\mathrm{day^{-1}}, \\
\overline{R^r} &= 4.39 \times 10^{4} ~\left(\mathrm{cell/cm^{3}}\right)\mathrm{day^{-1}}, \\
\lambda_{T_A^r T_r} &= 3.51 ~\mathrm{day}^{-1}.
\end{align*}
\subsection{Cancer Cell Parameters}
\subsubsection{Estimates for $C$}
Considering \eqref{cancereqn} at steady state leads to
\begin{equation*}
    \lambda_{C}\left(1-\frac{\overline{C}}{C_0}\right)-\frac{\lambda_{CT_8}}{4}\overline{T_8} -\frac{\lambda_{CK}}{4}\overline{K}-\frac{\lambda_{CI_\upalpha}}{2} - \frac{\lambda_{CI_\upgamma}}{2} = 0. 
\end{equation*}
We assume that CD8+ T cells and NK cells kill cancer cells with similar potency, so we approximate
\begin{equation*}
\lambda_{CK}/{4} = \lambda_{CT_8}/4 \implies \lambda_{CK} = \lambda_{CT_8}.
\end{equation*}
We also assume that the rate at which TNF induces tumour necroptosis is larger than that for IFN-$\upgamma$, so we approximate
\begin{equation*}
\frac{\lambda_{CI_\upalpha}}{2 \times 5} = \frac{\lambda_{CI_\upgamma}}{2}.
\end{equation*}
Solving these simultaneously leads to
\begin{align*}
    \lambda_{CK} &= \lambda_{CT_8}, \\
    \lambda_{CI_\upalpha} &= \left(\frac{5}{3} - \frac{3.31 \times 10^{9}}{60 C_0}\right)\lambda_C - \frac{8.33 \times 10^8}{400}\lambda_{CT_8}, \\
    \lambda_{CI_\upgamma} &= \left(\frac{1}{3} - \frac{3.31 \times 10^{7}}{3C_0}\right)\lambda_C - \frac{8.33 \times 10^7}{200}\lambda_{CT_8}.
\end{align*}
\subsubsection{Estimates for $N_c$}
Considering \eqref{necroticcelleqn} at steady state leads to the equation
\begin{equation*}
\frac{\lambda_{CI_\upalpha}\overline{C}}{2} + \frac{\lambda_{CI_\upgamma}\overline{C}}{2} - d_{N_c}\overline{N_c} = 0.
\end{equation*}
This leads to
\begin{equation*}
    d_{N_c} = \frac{1}{\overline{N_c}}\left[\left(1 - \frac{3.31 \times 10^{7}}{3C_0}\right)\lambda_{C} - \frac{2.499 \times 10^6}{2}\lambda_{CT_8}\right].
\end{equation*}
\subsubsection{Fitting $\lambda_C$, $\lambda_{CT_8}$, and $C_0$ \label{fittingsection}}
We fit $\lambda_C$, $\lambda_{CT_8}$, and $C_0$ by choosing the values such that the steady-state values of $C$ and $N_c$ are reached at $155$ days, in particular ensuring that $C$ and $N_c$ reach steady state at exactly $155$ days. Furthermore, we expect monotonicity in the growth of the total cancer population ($C+N_c$) as the cancer progresses without treatment, and so we aim to minimise
\begin{equation}
\operatorname{Objective} = \operatorname{max}\left(\frac{|C(155)-\overline{C}|}{\overline{C}}, \frac{|N_c(155)-\overline{N_c}|}{\overline{N_c}}, \frac{|C(155) + N_c(155) -(\overline{C} + \overline{N_c})|}{\overline{C} + \overline{N_c}} \right), \label{objectivefn}
\end{equation}
subject to
\begin{equation}
\underset{t\in [0,155]} \max(C(t)+N_c(t)) \leq \overline{C}+\overline{N_c}. \label{objectivecons}
\end{equation}
We perform a parameter sweep to minimise \eqref{objectivefn} subject to \eqref{objectivecons}, and set the parameter space to be $\lambda_C \in (0~\mathrm{day}^{-1},2~\mathrm{day}^{-1}]$, $\lambda_{CT_8} \in (0~\left(\mathrm{cell/cm^{3}}\right)^{-1}\mathrm{day}^{-1},1 \times 10^{-6} ~\left(\mathrm{cell/cm^{3}}\right)^{-1}\mathrm{day}^{-1}]$, and $C_0 \in (8 \times 10^{7}~\mathrm{cell/cm^3}, 10^{11} ~\mathrm{cell/cm^3}]$, ensuring that all model parameters are positive.     
The optimal values of $\lambda_C$, $\lambda_{CT_8}$, and $C_0$ were found to be
\begin{align*}
    \lambda_C &= 1.77 \times 10^{-1} ~\mathrm{day}^{-1}, \\
    \lambda_{CT_8}&= 2.58 \times 10^{-8} ~\left(\mathrm{cell/cm^{3}}\right)^{-1}\mathrm{day^{-1}}, \\
    C_0 &= 8.15 \times 10^{7} ~\mathrm{cell/cm^3},
    \intertext{which implies that}
    \lambda_{CK} &= 2.58 \times 10^{-8} ~\left(\mathrm{cell/cm^{3}}\right)^{-1}\mathrm{day^{-1}}, \\
    \lambda_{CI_\upalpha}&= 1.21 \times 10^{-1} ~ \mathrm{day}^{-1}, \\
    \lambda_{CI_\upgamma} &= 2.43 \times 10^{-2} ~\mathrm{day^{-1}}, \\
    d_{N_c} &= 6.55 \times 10^{-1} \mathrm{~day^{-1}}.
\end{align*}
\subsection{Estimates for Immune Checkpoint-Associated Components in the TS\label{tsicissinitappendix}}
\subsubsection{Estimate for $\lambda_{Q}$}
The dissociation rate of the PD-1/PD-L1 complex was found to be $1.44 \mathrm{~sec^{-1}}$ in \cite{Cheng2013supp}. Thus, we have that
\begin{equation*}
    \lambda_{Q} = 60 \times 60 \times 24 \times 1.44 \mathrm{~sec^{-1}} = 1.24 \times 10^{5} \mathrm{~day^{-1}}.
\end{equation*}
\subsubsection{Estimate for $\lambda_{P_DP_L}$}
The formation rate of the PD-1/PD-L1 complex was found to be $1.84 \times 10^{5} ~\mathrm{M^{-1}sec^{-1}}$ in \cite{Cheng2013supp}. To convert this to units of $\mathrm{(molec/cm^3)^{-1}day^{-1}}$, we recall that $1 \mathrm{~M} = 1 \mathrm{~mol/L} = 10^{-3} \mathrm{~mol/cm^3} = 6.022 \times 10^{20}\mathrm{~molec/cm^3}$. As such,
\begin{equation*}
   \lambda_{P_DP_L} = 60 \times 60 \times 24 \times 1.84 \times 10^{5} \times (6.022 \times 10^{20})^{-1}= 2.64 \times 10^{-11} \mathrm{~(molec/cm^3)^{-1}day^{-1}}.
\end{equation*}
\subsubsection{Estimates for Synthesis Rates and Steady States}
We note that estimating parameters, steady states, and initial conditions for PD-1, PD-L1, and the PD-1/PD-L1 complex in the TS is more involved than the previous estimations and requires more information.\\~\\
We first denote $\rho_{P_D^{T_8}}$, $\rho_{P_D^{T_1}}$, and $\rho_{P_D^{K}}$ as the number of PD-1 molecules expressed on the surface of CD8+ T cells, Th1 cells, and activated NK cells in the TS, respectively. To determine these parameters, we used the baseline data collected in \cite{Pluim2019supp} on $5$ advanced cancer patients before their pembrolizumab infusions. The net number of PD-1 molecules on the surface of CD4+ T cells was $2053~\mathrm{molec/cell}$, and so we set $\rho_{P_D^{T_1}} = 2.05 \times 10^{3} ~\mathrm{molec/cell}$. The net number of PD-1 molecules on the surface of CD8+ T cells was $2761~\mathrm{molec/cell}$, and so we set $\rho_{P_D^{T_8}} = 2.76 \times 10^{3}~ \mathrm{molec/cell}$. Despite the net number of PD-1 molecules on the surface of NK cells being below the lower limit of quantification in \cite{Pluim2019supp}, NK cells substantially express PD-1 \cite{Liu2017supp} in CRC, and so we set $\rho_{P_D^{K}} = \rho_{P_D^{T_8}}/5 = 5.52 \times 10^{2}~ \mathrm{molec/cell}$.\\~\\
We next denote $\rho_{P_LX}$ as the number of PD-L1 molecules expressed on $X \in \mathcal{X}$, recalling that $\mathcal{X}=\set{C, D, T_8, T_1, T_r, M_2}$. It was found in \cite{Cheng2013supp} that the PD-L1 expression on activated CD3+ PD-L1+ T cells was $9282 ~\mathrm{molec/cell}$, whilst the PD-L1 expression on mature DCs was $80,372 ~\mathrm{molec/cell}$. However, amongst advanced CRC patients, only $22.4\%$ of CD4+ T cells were PD-L1+, and only $16.1\%$ of CD8+ T cells were PD-L1+ \cite{Saito2021supp}. Moreover, only $22\%$ of colonic DCs were PD-L1+ in \cite{Moreira2021supp}. We thus assume that $\rho_{P_L{T_1}} = \rho_{P_L{T_r}} = 2.08 \times 10^{3} ~\mathrm{molec/cell}$, $\rho_{P_L{T_8}} = 1.49 \times 10^{3} ~\mathrm{molec/cell}$, and $\rho_{P_LD} = 1.77 \times 10^{4} ~\mathrm{molec/cell}$. In their quantitative systems pharmacology model of colorectal cancer, Anbari et al.\ estimated the baseline numbers of PD-L1 molecules per cancer cell and per APC to be $180,000 ~\mathrm{molec/cell}$ and $266,666 ~\mathrm{molec/cell}$, respectively \cite{Anbari2023supp}. This makes sense, noting that PD-L1 expression in macrophages is stronger and more continuous than that in cancer cells \cite{Saito2022supp}. As such, we set $\rho_{P_LC} = 1.8 \times 10^{5} \mathrm{~molec/cell}$ and $\rho_{P_L{M_2}} = 2.67 \times 10^{5} \mathrm{~molec/cell}$. \\~\\
Considering \eqref{Q8eqn}~--~\eqref{PLeqn} at steady state in the absence of pembrolizumab leads to
\begin{align*}
    \lambda_{P_D^{T_8}}\overline{T_8} - d_{P_D}\overline{P_D^{T_8}} &=0, \\
    \lambda_{P_D^{T_1}}\overline{T_1} - d_{P_D}\overline{P_D^{T_1}} &=0, \\
    \lambda_{P_D^{K}}\overline{K} - d_{P_D}\overline{P_D^{K}} &=0, \\
    \sum_{X \in \mathcal{X}}\lambda_{P_L X}\overline{X} - d_{P_L}\overline{P_L} &=0, \\
    \overline{Q^{T_8}} - \frac{\lambda_{P_DP_L}}{\lambda_Q}\overline{P_D^{T_8}}\overline{P_L}&=0, \\
    \overline{Q^{T_1}} - \frac{\lambda_{P_DP_L}}{\lambda_Q}\overline{P_D^{T_1}}\overline{P_L}&=0, \\
    \overline{Q^{K}} - \frac{\lambda_{P_DP_L}}{\lambda_Q}\overline{P_D^{K}}\overline{P_L}&=0.
    \intertext{By considering the total number of PD-1 receptors expressed on each PD-1-expressing cell at steady state, we expect in the absence of pembrolizumab that}
    \overline{P_D^{T_8}} + \overline{Q^{T_8}} &=\rho_{P_D^{T_8}}\overline{T_8}, \\
    \overline{P_D^{T_1}} + \overline{Q^{T_1}} &=\rho_{P_D^{T_1}}\overline{T_1}, \\
    \overline{P_D^{K}} + \overline{Q^{K}} &=\rho_{P_D^{K}}\overline{K}.
    \intertext{We can also consider the total number of PD-L1 ligands at steady state so that}
    \overline{P_L} + \overline{Q^{T_8}} + \overline{Q^{T_1}} + \overline{Q^{K}} &= \sum_{X \in \mathcal{X}} \rho_{P_L X}\overline{X}.
    \intertext{Finally, we expect the synthesis rates of PD-1 and PD-L1 to be proportional to the total number of PD-1 and PD-L1 molecules expressed per PD-1- and PD-L1-expressing cell, respectively, so that}
    \frac{\lambda_{P_D^{T_8}}}{\rho_{P_D^{T_8}}} =     \frac{\lambda_{P_D^{T_1}}}{\rho_{P_D^{T_1}}} &=     \frac{\lambda_{P_D^{K}}}{\rho_{P_D^{K}}}, \\
    \frac{\lambda_{P_LC}}{\rho_{P_LC}} = \frac{\lambda_{P_LD}}{\rho_{P_LD}} = \frac{\lambda_{P_LT_8}}{\rho_{P_LT_8}} = \frac{\lambda_{P_LT_1}}{\rho_{P_LT_1}} = \frac{\lambda_{P_LT_r}}{\rho_{P_LT_r}} &= \frac{\lambda_{P_LM_2}}{\rho_{P_LM_2}}.
\end{align*}
Solving these simultaneously and ensuring all model parameters are positive leads to
\begin{align*}
    \lambda_{P_D^{T_8}} &= 9.26 \times 10^{2} ~\left(\mathrm{molec/cell}\right)\mathrm{day^{-1}}, \\
    \lambda_{P_D^{T_1}} &= 6.88 \times 10^{2} ~\left(\mathrm{molec/cell}\right)\mathrm{day^{-1}}, \\
    \lambda_{P_D^{K}} &= 1.85 \times 10^{2} ~\left(\mathrm{molec/cell}\right)\mathrm{day^{-1}}, \\
    \lambda_{P_LC} &= 2.50 \times 10^{5} ~\left(\mathrm{molec/cell}\right)\mathrm{day^{-1}}, \\
    \lambda_{P_LD} &= 2.46 \times 10^{4} ~\left(\mathrm{molec/cell}\right)\mathrm{day^{-1}}, \\
    \lambda_{P_LT_8} &= 2.07 \times 10^{3} ~\left(\mathrm{molec/cell}\right)\mathrm{day^{-1}}, \\
    \lambda_{P_LT_1} &= 2.89 \times 10^{3} ~\left(\mathrm{molec/cell}\right)\mathrm{day^{-1}}, \\
    \lambda_{P_LT_r} &= 2.89 \times 10^{3} ~\left(\mathrm{molec/cell}\right)\mathrm{day^{-1}}, \\
    \lambda_{P_LM_2} &= 3.71 \times 10^{5} ~\left(\mathrm{molec/cell}\right)\mathrm{day^{-1}}.
\end{align*}
This leads to
\begin{align*}
    \overline{P_D^{T_8}} &= 4.91 \times 10^{8} \mathrm{~molec/cm^3}, \\
    \overline{P_D^{T_1}} &= 1.48 \times 10^{8} \mathrm{~molec/cm^3}, \\
    \overline{P_D^{K}} &= 2.66 \times 10^{9} \mathrm{~molec/cm^3}, \\
    \overline{P_L} &= 6.39 \times 10^{12} \mathrm{~molec/cm^3}, \\
    \overline{Q^{T_8}} &= 6.68 \times 10^{5} \mathrm{~molec/cm^3}, \\
    \overline{Q^{T_1}} &= 2.02 \times 10^{5} \mathrm{~molec/cm^3}, \\
    \overline{Q^{K}} &= 3.62 \times 10^{6} \mathrm{~molec/cm^3}.
\end{align*}
\subsubsection{Estimates for Initial Conditions}
To determine the relevant initial conditions, we can simply consider the total number of PD-1 receptors on each PD-1-expressing cell and the total number of PD-L1 ligands in the absence of pembrolizumab, so that
\begin{align*}
P_D^{T_8}(0) + Q^{T_8}(0) &=\rho_{P_D^{T_8}}T_8(0), \\
P_D^{T_1}(0) + Q^{T_1}(0) &=\rho_{P_D^{T_1}}T_1(0), \\
P_D^{K}(0) + Q^{K}(0) &=\rho_{P_D^{K}}K(0), \\
P_L(0) + Q^{T_8}(0) + Q^{T_1}(0) + Q^{K}(0) &= \sum_{X \in \mathcal{X}} \rho_{P_L X}X(0).
\intertext{We can also consider \eqref{Q8eqn}~--~\eqref{QKeqn} initially, so that}
    Q^{T_8}(0) - \frac{\lambda_{P_DP_L}}{\lambda_Q}P_D^{T_8}(0)P_L(0)&=0, \\
    Q^{T_1}(0) - \frac{\lambda_{P_DP_L}}{\lambda_Q}P_D^{T_1}(0)P_L(0)&=0, \\
    Q^{K}(0) - \frac{\lambda_{P_DP_L}}{\lambda_Q}P_D^{K}(0)P_L(0)&=0.
\end{align*}
Solving these simultaneously leads to
\begin{align*}
   P_D^{T_8}(0) &= 6.70 \times 10^{8} \mathrm{~molec/cm^3}, \\
    P_D^{T_1}(0) &= 2.13 \times 10^{8} \mathrm{~molec/cm^3}, \\
    P_D^{K}(0) &= 2.87 \times 10^{9} \mathrm{~molec/cm^3}, \\
    P_L(0) &= 3.57 \times 10^{12} \mathrm{~molec/cm^3}, \\
    Q^{T_8}(0) &= 5.09 \times 10^{5} \mathrm{~molec/cm^3}, \\
    Q^{T_1}(0) &= 1.62 \times 10^{5} \mathrm{~molec/cm^3}, \\
    Q^{K}(0) &= 2.18 \times 10^{6} \mathrm{~molec/cm^3}.
\end{align*}
We note that excluding bound PD-1 receptors when considering the total number of PD-1 receptors on PD-1-expressing cells does not affect the parameter estimates, steady states, or initial conditions at this level of precision, since the number of unbound PD-1 receptors is several orders of magnitude larger than the number of bound PD-1 receptors on PD-1-expressing cells. Furthermore, this also applies when considering the total number of PD-L1 ligands.
\subsection{Estimates for Immune Checkpoint-Associated Components in the TDLN\label{tdlnicissinitappendix}}
\subsubsection{Estimates for Synthesis Rates and Steady States}
For simplicity, we assume that the total number of PD-1 receptors and PD-L1 ligands on cells in the TDLN is equal to the number on the corresponding cells in the TS. Thus, denoting $\rho_{P_D^{8\mathrm{LN}}}$ and $\rho_{P_D^{1\mathrm{LN}}}$ as the number of PD-1 molecules expressed on the surface of CD8+ T cells and Th1 cells in the TDLN, respectively, we have that $\rho_{P_D^{8\mathrm{LN}}} = \rho_{P_D^{T_8}}$ and $\rho_{P_D^{1\mathrm{LN}}} = \rho_{P_D^{T_1}}$. Similarly, we have that $\rho_{P_L^\mathrm{LN}D^\mathrm{LN}} = \rho_{P_LD}$, $\rho_{P_L^\mathrm{LN}T_A^8} = \rho_{P_LT_8}$, $\rho_{P_L^\mathrm{LN}T_A^1} = \rho_{P_LT_1}$, and $\rho_{P_L^\mathrm{LN}T_A^r} = \rho_{P_LT_r}$, where $\rho_{P_L^\mathrm{LN}D^\mathrm{LN}}$, $\rho_{P_L^\mathrm{LN}T_A^8}$, $\rho_{P_L^\mathrm{LN}T_A^1}$, and $\rho_{P_L^\mathrm{LN}T_A^r}$ denote the number of PD-L1 ligands expressed on the surfaces of mature DCs, effector CD8$^+$ T cells, effector Th1 cells, and effector Tregs in the TDLN, respectively. We recall that the set of PD-L1-expressing cells in the TDLN is $\mathcal{Y} = \set{D^\mathrm{LN}, T_A^8, T_A^1, T_A^r}$. The procedure for estimating parameters, steady states, and initial conditions for PD-1, PD-L1, and the PD-1/PD-L1 complex in the TDLN is the same as in the TS. Considering \eqref{PD8LNeqn}~--~\eqref{PD1LNeqn} and \eqref{PLLNeqn}~--~\eqref{Q1LNeqn} at steady state in the absence of pembrolizumab, and making the same assumptions for estimation as in the TS, we obtain
\begin{align*}
    \lambda_{P_D^{8\mathrm{LN}}}\overline{T_A^8} - d_{P_D}\overline{P_D^{8\mathrm{LN}}} &=0, \\
    \lambda_{P_D^{1\mathrm{LN}}}\overline{T_A^1} - d_{P_D}\overline{P_D^{1\mathrm{LN}}} &=0, \\
    \sum_{Y \in \mathcal{Y}}\lambda_{P_L^\mathrm{LN}Y}\overline{Y} - d_{P_L}\overline{P_L^\mathrm{LN}} &=0, \\
    \overline{Q^{8\mathrm{LN}}} - \frac{\lambda_{P_DP_L}}{\lambda_Q}\overline{P_D^{8\mathrm{LN}}}\overline{P_L^\mathrm{LN}}&=0, \\
    \overline{Q^{1\mathrm{LN}}} - \frac{\lambda_{P_DP_L}}{\lambda_Q}\overline{P_D^{1\mathrm{LN}}}\overline{P_L^\mathrm{LN}}&=0, \\
    \overline{P_D^{8\mathrm{LN}}} + \overline{Q^{8\mathrm{LN}}} &=\rho_{P_D^{8\mathrm{LN}}}\overline{T_A^8}, \\
    \overline{P_D^{1\mathrm{LN}}} + \overline{Q^{1\mathrm{LN}}} &=\rho_{P_D^{1\mathrm{LN}}}\overline{T_A^1}, \\
    \overline{P_L^\mathrm{LN}} + \overline{Q^{8\mathrm{LN}}} + \overline{Q^{1\mathrm{LN}}} &= \sum_{Y \in \mathcal{Y}} \rho_{P_L^\mathrm{LN}Y}\overline{Y}, \\
    \frac{\lambda_{P_D^{8\mathrm{LN}}}}{\rho_{P_D^{8\mathrm{LN}}}} &=     \frac{\lambda_{P_D^{1\mathrm{LN}}}}{\rho_{P_D^{1\mathrm{LN}}}}, \\
    \frac{\lambda_{P_L^\mathrm{LN}D^\mathrm{LN}}}{\rho_{P_L^\mathrm{LN}D^\mathrm{LN}}} = \frac{\lambda_{P_L^\mathrm{LN}T_A^8}}{\rho_{P_L^\mathrm{LN}T_A^8}} = \frac{\lambda_{P_L^\mathrm{LN}T_A^1}}{\rho_{P_L^\mathrm{LN}T_A^1}} &= \frac{\lambda_{P_L^\mathrm{LN}T_A^r}}{\rho_{P_L^\mathrm{LN}T_A^r}}.  
\end{align*}
Solving these simultaneously and ensuring all model parameters are positive leads to
\begin{align*}
    \lambda_{P_D^{8\mathrm{LN}}} &= 9.27 \times 10^{2} ~\left(\mathrm{molec/cell}\right)\mathrm{day^{-1}}, \\
    \lambda_{P_D^{1\mathrm{LN}}} &= 6.89 \times 10^{2} ~\left(\mathrm{molec/cell}\right)\mathrm{day^{-1}}, \\
    \lambda_{P_L^\mathrm{LN}D^\mathrm{LN}} &= 2.46 \times 10^{4} ~\left(\mathrm{molec/cell}\right)\mathrm{day^{-1}}, \\
    \lambda_{P_L^\mathrm{LN}T_A^8} &= 2.07 \times 10^{3} ~\left(\mathrm{molec/cell}\right)\mathrm{day^{-1}}, \\
    \lambda_{P_L^\mathrm{LN}T_A^1} &= 2.89 \times 10^{3} ~\left(\mathrm{molec/cell}\right)\mathrm{day^{-1}}, \\
    \lambda_{P_L^\mathrm{LN}T_A^r} &= 2.89 \times 10^{3} ~\left(\mathrm{molec/cell}\right)\mathrm{day^{-1}}.
\end{align*}
This leads to
\begin{align*}
    \overline{P_D^{8\mathrm{LN}}} &= 2.37 \times 10^{9} \mathrm{~molec/cm^3}, \\
    \overline{P_D^{1\mathrm{LN}}} &= 1.59 \times 10^{10} \mathrm{~molec/cm^3}, \\
    \overline{P_L^\mathrm{LN}} &= 1.26 \times 10^{11} \mathrm{~molec/cm^3}, \\
    \overline{Q^{8\mathrm{LN}}} &= 6.36 \times 10^{4} \mathrm{~molec/cm^3}, \\
    \overline{Q^{1\mathrm{LN}}} &= 4.27 \times 10^{5} \mathrm{~molec/cm^3}.
\end{align*}
We note again that excluding bound PD-1 receptors when considering the total number of PD-1 receptors on PD-1-expressing cells does not affect the parameter estimates, steady states, or initial conditions at this level of precision, since the number of unbound PD-1 receptors is several orders of magnitude larger than the number of bound PD-1 receptors on PD-1-expressing cells. Furthermore, this also applies when considering the total number of PD-L1 ligands.
\subsection{Estimates for Initial Conditions of T Cells and Immune Checkpoint-Associated Components in the TDLN \label{initcondTDLNestsec}}
To determine the relevant immune checkpoint initial conditions, we can simply consider the total number of PD-1 receptors on each PD-1-expressing cell and the total number of PD-L1 ligands in the absence of pembrolizumab, so that
\begin{align*}
P_D^{8\mathrm{LN}}(0) + Q^{8\mathrm{LN}}(0) &=\rho_{P_D^{8\mathrm{LN}}}T_A^8(0), \\
P_D^{1\mathrm{LN}}(0) + Q^{1\mathrm{LN}}(0) &=\rho_{P_D^{1\mathrm{LN}}}T_A^1(0), \\
P_L^\mathrm{LN}(0) + Q^{8\mathrm{LN}}(0) + Q^{1\mathrm{LN}}(0) &= \sum_{Y \in \mathcal{Y}} \rho_{P_L^\mathrm{LN}Y}Y(0).
\intertext{We can also consider \eqref{Q8LNeqn} and \eqref{Q1LNeqn} initially, so that}
    Q^{8\mathrm{LN}}(0) - \frac{\lambda_{P_DP_L}}{\lambda_Q}P_D^{8\mathrm{LN}}(0)P_L^\mathrm{LN}(0)&=0, \\
    Q^{1\mathrm{LN}}(0) - \frac{\lambda_{P_DP_L}}{\lambda_Q}P_D^{1\mathrm{LN}}(0)P_L^\mathrm{LN}(0)&=0.
\end{align*}
Furthermore, we assume that the initial rate of change of all T cell populations is zero. Considering \eqref{naivecd8eqn} and \eqref{TA8n8maxeqn} initially, we have that
\begin{align*}
    \mathcal{A}_{T_0^8} - R^8(0) - d_{T_0^8}T_0^8(0) &= 0, \\
    \frac{2^{n^8_\mathrm{max}}e^{-d_{T_0^8}\tau_{T_A^8}} R^8(0)}{{\left(1+\tau_{T_A^8}T_A^r(0)/K_{T_A^8 T_A^r}\right)\left(1+\tau_{T_A^8}Q^{8\mathrm{LN}}(0)/K_{T_A^8 Q^{8\mathrm{LN}}}\right)}} - \lambda_{T_A^8T_8}T_A^8(0) - d_{T_8}T_A^8(0) &= 0,
\end{align*}
where
\begin{equation*}
    R^8(0) = \frac{\lambda_{T_0^8 T_A^8}e^{-d_{T_0^8}\tau_8^\mathrm{act}}D^\mathrm{LN}(0)T_0^8(0)}{\left(1+\tau_8^\mathrm{act}T_A^r(0)/K_{T_0^8T_A^r}\right)\left(1+\tau_8^\mathrm{act}Q^{8\mathrm{LN}}(0)/K_{T_0^8Q^{8\mathrm{LN}}}\right)}.
\end{equation*}
Similarly, considering \eqref{naivecd4eqn} and \eqref{TA1n1maxeqn} initially, we have that
\begin{align*}
    \mathcal{A}_{T_0^4} - R^1(0) - d_{T_0^4}T_0^4(0) &= 0, \\
    \frac{2^{n^1_\mathrm{max}}e^{-d_{T_0^4}\tau_{T_A^1}} R^1(0)}{{\left(1+\tau_{T_A^1}T_A^r(0)/K_{T_A^1 T_A^r}\right)\left(1+\tau_{T_A^1}Q^{1\mathrm{LN}}(0)/K_{T_A^1 Q^{1\mathrm{LN}}}\right)}} - \lambda_{T_A^1T_1}T_A^1(0) - d_{T_1}T_A^1(0) &= 0,
\end{align*}
where
\begin{equation*}
    R^1(0) = \frac{\lambda_{T_0^4 T_A^1}e^{-d_{T_0^4}\tau_4^\mathrm{act}}D^\mathrm{LN}(0)T_0^4(0)}{\left(1+\tau_4^\mathrm{act}T_A^r(0)/K_{T_0^4 T_A^r}\right)\left(1+\tau_4^\mathrm{act}Q^{1\mathrm{LN}}(0)/K_{T_0^4Q^{1\mathrm{LN}}}\right)}.
\end{equation*}
Finally, considering \eqref{naivetregeqn} and \eqref{TArnrmaxeqn} initially, we have that
\begin{align*}
    \mathcal{A}_{T_0^r} - R^r(0) - d_{T_0^r}T_0^r(0) &= 0, \\
    2^{n^r_\mathrm{max}}e^{-d_{T_0^r}\tau_{T_A^r}} R^r(0) - \lambda_{T_A^rT_r}T_A^r(0)-d_{T_r}T_A^r(0)&=0,
\end{align*}
where
\begin{equation*}
    R^r(0) = \lambda_{T_0^r T_A^r} e^{-d_{T_0^r}\tau_r^\mathrm{act}}D^\mathrm{LN}(0)T_0^r(0).
\end{equation*}
Solving these simultaneously leads to
\begin{align*}
    T_0^8(0) &= 1.20 \times 10^{7} \mathrm{~cell/cm^3}, \\
    T_A^8(0) &= 1.11 \times 10^{6} \mathrm{~cell/cm^3}, \\
    T_0^4(0) &= 4.40 \times 10^{6} \mathrm{~cell/cm^3}, \\
    T_A^1(0) &= 1.01 \times 10^{7} \mathrm{~cell/cm^3}, \\
    T_0^r(0) &= 9.95 \times 10^{4} \mathrm{~cell/cm^3}, \\
    T_A^r(0) &= 7.84 \times 10^{5} \mathrm{~cell/cm^3}, \\
   P_D^{8\mathrm{LN}}(0) &= 3.06 \times 10^{9} \mathrm{~molec/cm^3}, \\
    P_D^{1\mathrm{LN}}(0) &= 2.07 \times 10^{10} \mathrm{~molec/cm^3}, \\
    P_L^\mathrm{LN}(0) &= 2.10 \times 10^{11} \mathrm{~molec/cm^3}, \\
    Q^{8\mathrm{LN}}(0) &= 1.37 \times 10^{5} \mathrm{~molec/cm^3}, \\
    Q^{1\mathrm{LN}}(0) &= 9.23 \times 10^{5} \mathrm{~molec/cm^3}.
\end{align*}
\subsection{Estimates for $A_1$ and $A_1^\mathrm{LN}$}
\subsubsection{Estimate for $f_\mathrm{pembro}$}
To determine $f_\mathrm{pembro}$, we use the formula
\begin{align}
    f_\mathrm{pembro} &= \frac{C_\mathrm{max,ss}(\xi_\mathrm{pembro})-C_\mathrm{min,ss}(\xi_\mathrm{pembro})}{\xi_\mathrm{pembro}}, \label{fA1pembroeqn}
\end{align}
where $C_\mathrm{max,ss}$ and $C_\mathrm{min,ss}$ correspond to the maximum and minimum serum concentrations of pembrolizumab at steady state, respectively, after a dose $\xi_\mathrm{pembro}$ of pembrolizumab is administered. \\~\\
For pembrolizumab, the mean values of $C_\mathrm{min,ss}$ and $C_\mathrm{max,ss}$ were approximately $32.6 \mathrm{~\mu g/mL}$ and $85.8 \mathrm{~\mu g/mL}$ for Treatment 1, and $22.4 \mathrm{~\mu g/mL}$ and $147.7 \mathrm{~\mu g/mL}$ for Treatment 2 \cite{Lala2020supp}. This results in $f_\mathrm{pembro} \approx 2.90 \times 10^{-7} ~\left(\mathrm{g/cm^3}\right)/\mathrm{mg}$ of pembrolizumab administered for all doses. To convert this into units of $\left(\mathrm{molec/cm^3}\right)/\mathrm{mg}$, we note that the molecular mass of pembrolizumab is approximately $149,000 \mathrm{~g/mol}$ \cite{fda_pembrolizumabsupp}, which corresponds to $f_\mathrm{pembro} \approx 1.17 \times 10^{12} ~\left(\mathrm{molec/cm^3}\right)/\mathrm{mg}$.
\subsection{Estimates for PD-1/pembrolizumab Complex on Cells}
\subsubsection{Estimate for $\lambda_{Q_A}$}
The dissociation rate of the PD-1/pembrolizumab complex was measured using biolayer interferometry to be $2.6 \mathrm{~day^{-1}}$ in \cite{Li2021supp}. Thus, we take $\lambda_{Q_A} = 2.6 \mathrm{~day^{-1}}$.
\subsubsection{Fitting $\lambda_{P_DA_1}$}
We estimate $\lambda_{P_DA_1}$ by fitting it to target engagement (TE) at trough for a triweekly regimen, specifically PD-1 receptor saturation by pembrolizumab, based on data from the 03TLC9 study \cite{tga_pembrolizumabsupp}. For example, the TE of pembrolizumab on CD8+ T cells in the TS, which we denote by $\operatorname{TE}^{T_8}$, is mathematically defined as
\begin{equation*}
\operatorname{TE}^{T_8} := \frac{Q_A^{T_8}}{P_D^{T_8} + Q_A^{T_8} + Q^{T_8}} \times 100\%,
\end{equation*}
which is the percentage of all PD-1 receptors on CD8+ T cells at the TS that are bound to pembrolizumab. TEs of pembrolizumab on other cells in the TS and TDLN are defined and notated similarly, with minimal deviation across all cell types. We define the overall TE as the average TE across all cell types in the TS and TDLN. The median TE at trough for a triweekly regimen at various doses is shown in \autoref{TEtable}, noting that we assume a patient mass of $80 \mathrm{~kg}$.
\begin{table}[H]
    \centering
    \begin{tabular}{|c|c|}
    \hline
    \textbf{Dose (mg/kg)} & \textbf{Median TE} (\%) \\
    \hline
    0.1 & 59 \\
    0.2 & 80 \\
    0.5 & 92 \\
    1   & 96 \\
    2   & 98 \\
    5   & 99 \\
    \hline
    \end{tabular}
    \caption{\label{TEtable}Median TE at trough for triweekly pembrolizumab regimens at various doses.}
\end{table}
Noting that it takes approximately 19 weeks for a triweekly pembrolizumab regimen to reach steady-state concentrations \cite{Longoria2016supp}, we consider 14 treatment cycles, corresponding to 294 days, to ensure that PD-1/pembrolizumab complex steady-state concentrations are achieved. We define $f(\lambda_{P_DA_1}, \xi_\mathrm{pembro})$ as the overall TE at trough, in this case at 294 days, for a triweekly regimen with a dosage of $\xi_\mathrm{pembro}$ mg/kg predicted by the model with a PD-1/pembrolizumab formation rate of $\lambda_{P_DA_1}$. To estimate the best value of $\lambda_{P_DA_1}$, we minimise the sum of squares of the differences at trough between $f(\lambda_{P_DA_1}, \xi_\mathrm{pembro})$ and the true value based on the data from \autoref{TEtable}. Thus, assuming pembrolizumab infusions every 3 weeks from $t=0 \mathrm{days}$ up until $294 \mathrm{~days}$, we aim to minimise
\begin{equation}
\begin{split}
\mathrm{Objective} &= \left(f(\lambda_{P_DA_1}, 0.1) - 59\right)^2 + \left(f(\lambda_{P_DA_1}, 0.2) - 80\right)^2 + \left(f(\lambda_{P_DA_1}, 0.5) - 92\right)^2 \\
&+ \left(f(\lambda_{P_DA_1}, 1) - 96\right)^2 + \left(f(\lambda_{P_DA_1}, 2) - 98\right)^2 + \left(f(\lambda_{P_DA_1}, 5) - 99\right)^2.
\end{split}\label{lambdapda1func}
\end{equation}
We perform a parameter sweep to minimise \eqref{lambdapda1func} and set the parameter space to be $\lambda_{P_DA_1} \in (0~\left(\mathrm{molec/cm^3}\right)^{-1}\mathrm{day}^{-1},10^{-12}~\left(\mathrm{molec/cm^3}\right)^{-1}\mathrm{day}^{-1}]$. Solving this, the optimal value of $\lambda_{P_DA_1}$ was found to be
\begin{equation*}
    \lambda_{P_DA_1} = 4.69 \times 10^{-13} ~\left(\mathrm{molec/cm^3}\right)^{-1}\mathrm{day}^{-1}.
\end{equation*}
\subsubsection{Estimate for $d_{Q_A}$}
The internalisation rate of the PD-1/pembrolizumab complex was estimated to be $0.43 \mathrm{~day^{-1}}$ in \cite{Li2021supp}, and so we estimate $d_{Q_A} = 0.43 \mathrm{~day^{-1}}$.
\putbib[References.bib]
\end{bibunit}
\end{document}